\documentclass[a4paper,fleqn,10pt]{cas-sc}

\usepackage{import}
\usepackage[numbers]{natbib}
\usepackage{upgreek}
\usepackage{enumitem}
\usepackage{tabularx}
\usepackage{ragged2e}
\usepackage{multicol}
\usepackage{lscape}
\usepackage{xcolor}
\usepackage{soul}
\usepackage{amsmath}
\usepackage[utf8]{inputenc}
\usepackage[T1]{fontenc}
\usepackage{lmodern}
\usepackage{siunitx}
\usepackage{textcomp}

\def\tsc#1{\csdef{#1}{\textsc{\lowercase{#1}}\xspace}}
\tsc{WGM}
\tsc{QE}
\tsc{EP}
\tsc{PMS}
\tsc{BEC}
\tsc{DE}

\begin{document}
\let\WriteBookmarks\relax
\def\floatpagepagefraction{1}
\def\textpagefraction{.001}

\shorttitle{A Comprehensive Review of Lunar-based Manufacturing and Construction}

\shortauthors{Mohammad Azami et~al.}


\title [mode = title]{A Comprehensive Review of Lunar-based Manufacturing and Construction}  




%
\author[1]{Mohammad Azami}[type=editor,
                        orcid=0000-0002-7086-8994]



\ead{mohammad.azami@mail.concordia.ca}



\affiliation[1]{organization={Department of Electrical and Computer Engineering, Concordia University},
    city={Montréal},
    province={Québec,},
    country={Canada}}

\author[2]{Zahra Kazemi}
\affiliation[2]{organization={Institute for Aerospace Studies, University of Toronto},
    city={Toronto},
    addressline={Dufferin Street 4925}, 
    postcode={M3H 5T6}, 
    province={Ontario,},
    country={Canada}}

\author[3]{Sare Moazen}[%
   ]


\affiliation[3]{organization={CREPEC, Department of Mechanical Engineering, École de technologie supérieure},
    city={Montréal},
    province={Québec,},
    country={Canada}}

\author%
[3]
{Martine Dubé}

\author%
[4]
{Marie-Josée Potvin}

\affiliation[4]{organization={Space Science and Technology, Canadian Space Agency},
    city={Saint-Hubert},
    country={Canada}}

\author%
[1]
{Krzysztof Skonieczny}
\cormark[1]
\ead{krzysztof.skonieczny@concordia.ca}
\ead[URL]{https://www.concordia.ca/faculty/krzysztof-skonieczny.html}

\cortext[cor1]{Corresponding author}



\begin{abstract}
As humankind prepares to establish outposts and infrastructure on the Moon, the ability to manufacture parts and buildings on-site is crucial. While transporting raw materials from Earth can be costly and time-consuming, in-situ resource utilization (ISRU) presents an attractive alternative. This review paper aims to provide a thorough examination of the current state and future potential of Lunar-based manufacturing and construction (LBMC), with a particular focus on the prospect of utilizing in-situ resources and additive manufacturing. The paper analyzes existing research on LBMC from various perspectives, including different manufacturing techniques and compositions, the potential of ISRU for LBMC, characterization of built parts and structures, the role of energy sources and efficiency, the impact of low-gravity and vacuum conditions, and the feasibility of using artificial intelligence, automation, and robotics. By synthesizing these findings, this review offers valuable insights into the challenges and opportunities that lie ahead for LBMC.


\end{abstract}


\begin{highlights}
\item Provides a critical review of the current state of Lunar-based manufacturing and construction (LBMC).
\item Recognizes the advantages and challenges associated with different potential manufacturing techniques for LBMC.
\item Recommends future research directions for LBMC.
\end{highlights}


\maketitle

\setcounter{secnumdepth}{4}
\renewcommand\theparagraph{\arabic{paragraph}}
\newcommand{\subsubsubsection}[1]{\paragraph{#1}\mbox{}\\}

\section*{Abbreviations}
\scriptsize
\begin{multicols}{2}
\begin{itemize}[itemsep=0pt,parsep=0pt]
    \item AM: Additive Manufacturing
    \item AI: Artificial Intelligence
    \item BJ: Binder Jetting
    \item CCC: Cement Contour Crafting
    \item CF: Carbon Fiber
    \item CMC: Ceramic Matrix Composite
    \item DCM: Dichloromethane
    \item DIC: Digital Image Correlation
    \item DIW: Direct Ink Writing
    \item DED: Directed Energy Deposition
    \item EB-DED: Electron-Beam Directed Energy Deposition
    \item EBM: Electron Beam Melting
    \item EB-PBF: Electron Beam Powder Bed Fusion
    \item EDS: Energy-Dispersive X-Ray Spectroscopy
    \item ESA: European Space Agency
    \item EMR: Extrusion of Molten Regolith
    \item FDM: Fused Deposition Modeling
    \item FFF: Fused Filament Fabrication
    \item FGF: Fused Granulate Fabrication
    \item HIP: Hot Isostatic Pressing
    \item ISRU: In-Situ Resource Utilization
    \item L-DED: Laser Directed Energy Deposition
    \item L-PBF: Laser Powder Bed Fusion
    \item LS: Laser Sintering
    \item LBMC: Lunar-Based Manufacturing and Construction
    \item LRS: Lunar Regolith Simulant
    \item LSCC: Lunar Soil Characterization Consortium
    \item ME: Materials Extrusion
    \item MOC: Magnesium Oxychloride
    \item MRE: Molten Regolith Electrolysis
    \item MRS: Martian Regolith Simulant
    \item MMC: Metal Matrix Composite
    \item MWS: Microwave Sintering
    \item MWS-PBF: Microwave Sintering Powder Bed Fusion
    \item PA: Polyamide
    \item PE: Polyethylene
    \item PEEK: Polyether ether ketone
    \item PEKK: Polyether ketone ketone
    \item PI: Polyimide
    \item PSD: Particle Size Distribution
    \item PMC: Polymer Matrix Composite
    \item PLGA: Polylactic-co-glycolic Acid
    \item PBF: Powder Bed Fusion
    \item PCA: Process Control Agent
    \item PMA: Polymer Micro-Agglomerates
    \item SEM: Scanning Electron Microscopy
    \item SLM: Selective Laser Melting
    \item SLS: Selective Laser Sintering
    \item SMWS: Selective Microwave Sintering
    \item SSLS: Selective Solar Light Sintering
    \item SS: Solar Sintering
    \item SS-PBF: Solar Sintering Powder Bed Fusion
    \item SPS: Spark Plasma Sintering
    \item SLA: Stereolithography Apparatus
    \item TES: Thermal Energy Storage
    \item TMS: Technological Maturity Stage
    \item WAAM: Wire Arc Additive Manufacturing
    \item XRD: X-Ray Diffraction
\end{itemize}
\end{multicols}

\normalsize


\section{Motivation and Purpose of the Study}
Humanity is preparing to return to the Moon with the Artemis missions. Already, there is a flurry of activities, including several robotic and orbiter missions from various countries \cite{Hill2024Astronomy}. Presently, robotic missions are self-contained, and no repair or help is possible. While there have been some recent successes, there have also been several failures \cite{Foust2023SpaceNews}, underscoring the difficulty of operating on the Moon. Moreover, all these missions have aimed to deploy small rovers on the Moon, rather than large ones, given the high cost and difficulty of bringing a lander into the vicinity of the Moon.

However, bringing humans back to the Moon for more than a few hours at a time will require substantial installations, many of which will be crucial for human survival and will need methods for repair and maintenance. Planning to bring every conceivable spare part would be prohibitive. Therefore, means to fabricate on-site what is required to develop, maintain, and repair systems are essential. Different construction and manufacturing techniques are explored for in-situ repair or fabrication of available/required structures, usually categorized into traditional and additive manufacturing (AM) techniques. Moreover, utilizing in situ resources, such as Lunar regolith, is an excellent strategy for minimizing the materials transported to the Moon. Considering the challenges of the Lunar surface, Lunar-Based Manufacturing and Construction (LBMC) encounters limitations that should be addressed in all related research.

In that view, although traditional techniques are an integral part of research on LBMC, AM, also known as 3D printing, is an interesting solution as it is highly versatile and allows the printing of parts of complex shapes without prior knowledge of these parts before the start of the mission. Nowadays, there are several techniques for AM, each bringing its own set of constraints, difficulties, and possibilities to establish an outpost on the Moon. This paper aims to enumerate the distinct challenges of working in an extreme environment such as the Moon, review the state of the art for each traditional or AM technique, and discuss the possibilities of each one of them.

\section{Impact of the Lunar Environment}
Fabricating components on the Moon presents several environmental challenges. The temperatures oscillate between very hot and very cold, with the Lunar day spanning 29.5 Earth days, resulting in extended periods of extreme temperatures. The Moon's atmosphere is extremely thin, approaching a vacuum-like state, and leading to high radiation levels, posing significant challenges in selecting suitable materials and equipment. Additionally, the Moon's gravity, approximately one-sixth of Earth's, may impact adhesion and other gravity-dependent processes. Lunar dust, pervasive and consisting of extremely small, abrasive particles, affects the operations of all systems. Finally, as operating any system requires power, the only available source on the Moon is currently solar power. Evaluating the power requirements of manufacturing and construction systems on the Moon is crucial, as well as exploring potential methods for meeting these power needs. \label{impact}

\subsection{Vacuum Condition}
The atmosphere on the Moon is very thin, $3\text{×}10^{\text{-15}}$ bar \cite{Stern1999Geophysics}.  This near vacuum has consequences on the materials used on the Moon and, therefore, on the design of any piece of equipment operating on the Moon.

When polymers are exposed to a near vacuum, volatile particles leak out of the polymer at a fast rate, a phenomenon called outgassing \cite{Jiao2019IOP}.  At a pressure of one atmosphere, the phenomenon also happens, but more slowly and the volatile particles dissipate in the air.  In an environment such as the Moon, the volatile particles will tend to condense on nearby surfaces, which could happen to be a sensor, in effect coating it with a thin layer of polymer, and potentially preventing that sensor from working as planned. For this reason, the space industry abides by rules to ensure the polymers used in extraterrestrial environments will not be a threat to the correct operation of space hardware. The first measure is the Total Mass Loss (TML), which must be below 0.1\%.  This measure is the total amount of material lost in a vacuum during a specified period of time. The second measure is the Collected Volatile Condensed Mass (CVCM), which must be below 0.01\%.  This measure is the total amount of material collected on a probe in the same environment where the material sample has been deposited and submitted to a vacuum for a specified amount of time \cite{ASTME595}.

Several organizations, private and public, maintain lists of materials and their corresponding TML and CVCM, such as the NASA list \cite{NASA_Outgassing}. The number of polymers respecting the two criteria is quite restricted.  Therefore, when planning for 3D printing on the Moon, one must first validate the suitability of the proposed material.  This also applies to any lubricant or any other type of material used in a manufacturing process.

Finally, the TML and CVCM norms also apply to any component of machines brought to the Moon, limiting the choices of material. The near vacuum itself also constrains the design of equipment.  Small pockets of air, trapped at the bottom of a socket for a screw, for example, could push the screw out of the piece of equipment, once exposed to space, with a much lower pressure outside the piece of equipment compared to the pressure of the air trapped behind the screw. Hardware assembled on Earth must provide for means of escape for the air trapped in the piece of hardware.

\subsection{Low Gravity}
Although the gravity field of the Moon is not uniform \cite{SLS-SPEC-159}, its average value is $1.6 \, \text{m/s}^2$ \cite{Wertz-2001}, which is approximately $\frac{1}{6}$ of the gravity on Earth.  This has implications for the operations of any machine on the Moon and for manufacturing and construction processes in particular. When designing for a low-gravity environment, one must consider how mechanisms that usually rely on gravity will function in the lower-gravity field. When it comes to 3D printing, for processes such as Fused Deposition Modeling (FDM) or Fused Filament Fabrication (FFF), gravity plays a role in providing a force to bind two successive layers together. However, the flow can be adjusted to slightly over-extrude and thus, create pressure between the two successive layers.  A group of students designing a 3D printer at Polytechnique Montréal in 2020-2021 took the printer and operated it upside down without any seemingly ill effect.  The mechanical properties of the resulting sample were not measured and the lack of gravity could have affected the interlayer bonding \cite{mec8370-22}. Low gravity should be on the mind of the hardware designer in order to compensate for any motion where components would be expected to naturally fall.


\subsection{Energy Sources and Constraints}
While the preceding challenges can be addressed through appropriate material selection and design, a major impediment is the need for power. The primary current source of power on the Moon is solar energy. Manufacturing and construction techniques will require bursts of peak power, necessitating large arrays of solar panels, which could offset the benefits of 3D printing parts on the Moon instead of bringing a large number of diverse spare parts. Small nuclear sources are a viable alternative. NASA has defined a project called the Fission Surface Power Project, which aims to develop a small nuclear fission reactor to generate electricity on the Moon and inform future designs for Mars. They are currently concluding the initial phase of this project \cite{NASA2024}.

In 2022, NASA awarded three contracts to commercial partners—Lockheed Martin, Westinghouse, and IX (a collaboration between Intuitive Machines and X-Energy)—to develop initial reactor designs and associated power systems, including estimated costs and development timelines. This project is critical for establishing a sustained human presence on the Lunar surface for at least 10 years, as the reactor will operate independently of the Sun, especially during the 14-and-a-half Earth days of Lunar night. The design requirements were kept open and flexible to encourage innovative approaches. However, NASA specified that the reactor should weigh under six metric tons and generate 40 kilowatts of electrical power, enough to support Lunar habitats, rovers, backup grids, and science experiments. The reactor must also operate for a decade without human intervention, prioritizing safety and minimal radiation exposure \cite{NASA2024, BBC2024}.

Phase 2 will develop the final reactor design for Lunar demonstration. The goal is to deliver a reactor to the launch pad in the early 2030s, with a one-year demonstration on the Moon followed by nine operational years. Additionally, NASA has awarded contracts to develop Brayton power converters to improve the efficiency of converting thermal power from nuclear fission into electricity \cite{NASA2024}.

The UK Space Agency has also announced new funding to support the demonstration of a modular nuclear reactor for Lunar missions. Following an initial study in 2022, Rolls-Royce is leading this initiative in collaboration with UK industries and academic institutions. With extensive experience in designing, manufacturing, and supporting nuclear reactors for Royal Navy submarines, Rolls-Royce is leveraging its expertise in compact nuclear reactor technology to propel advancements in space exploration and Lunar infrastructure \cite{BBC2024}.

NASA has also introduced a new direct energy conversion concept known as the Nuclear Thermionic Avalanche Cell (NTAC), coupled with the Metallic Junction Thermoelectric (MJ-TE) generator, both of which are patented NASA technologies. These advancements offer high specific power, ranging from milliwatts to megawatts, making them versatile for various space applications such as propulsion, spacecraft, satellites, rovers, ISRU, habitats, and other mission systems. NTAC has shown superior performance over radioisotope thermoelectric generators (RTG) by up to two orders of magnitude based on experimental and theoretical assessments, while the MJ-TE generator achieves three to four times higher efficiency for the same mass. In a conceptual scenario, a mobile power station equipped with four NTAC units, each capable of delivering 100 kWe or 200 kWe, could efficiently provide electrical power at mission sites. This configuration is significantly lighter and offers substantially more power compared to conventional fission reactors, underscoring the transformative potential of NTAC and MJ-TE technologies in powering diverse space missions. Ongoing prototyping tests aim to further validate their performance and adaptability in practical space environments \cite{Moses2024portable}.

However, nuclear energy carries the risk of emitting radiation, and considering a human settlement is a key scenario, these small nuclear sources could pose an added risk to astronauts. The international community is discussing the establishment of power grids on the Moon, which would facilitate activities such as Lunar-based manufacturing\cite{Csank2022}.

\subsection{Large Temperature Gradients} \label{fluctuation}

One of the primary environmental concerns on the Moon is the extreme temperature gradients, and research on determining the Lunar surface and subsurface temperature has been ongoing \cite{paige2010diviner, fang2014high, malla2015determination, wei2016inversions, wei2016thermal, fisher2017evidence, williams2017global, williams2019seasonal, zhang2014Lunar}. The large temperature fluctuations on the Moon are governed by its rotational periods, lack of atmosphere, and thermo-physical properties of regolith. The Lunar day-night cycle presents a sharp contrast to that of Earth. A full Lunar day, defined as the time it takes for the Sun to return to the same position in the sky, lasts approximately 29.5 Earth days. Lunar days and nights both have durations of about half that period, lasting approximately 14.75 Earth days each. Surface temperatures on the Moon fluctuate around 280\textdegree{C} in a full day/night cycle. It can soar to over +100\textdegree{C}, particularly in equatorial regions, and plummet to around -173\textdegree{C} during the days and nights, respectively \cite{SLS-SPEC-159}.

The most significant contributor to the Moon's thermal environment is solar radiation during Lunar days. As the Moon lacks any atmospheric layer to attenuate or trap the energy, heat transfer on the Moon is dominated by radiation. The different input and output sources are direct solar radiation, reflected sunlight from nearby surfaces, heat conduction to or from objects on the surface, and the flow of heat beneath the surface.
Heat flow measurements were conducted during Apollo 15, and Apollo 17 missions by placing probes at different levels inside the soil and thermocouples on the surface, providing temperature fluctuations as a function of depth. A significant difference in mean temperatures, the average temperature over a complete diurnal Lunar cycle, was reported below the Lunar surface. According to the measurements, the mean temperature at a depth of 35 cm was $45$\textdegree{C} and $40$\textdegree{C} higher than those at the surface, which is attributed to the notably low thermal conductivity of Lunar regolith within its uppermost 1 to 2 cm. Interestingly, no fluctuation due to the Lunar diurnal cycle below 80 cm of the surface was observed, and the measured fluctuations were due to the heat flow from the Lunar crust. This radiogenic heating, similar to the Earth, is a consequence of radioactive decay within its crust. Elements like Thorium (Th), Potassium (K), and Uranium (U) have been reported to be found in the Moon's crust. The distribution of radiogenic elements across the Lunar surface exhibits spatial heterogeneity, influencing localized heat generation.

Last but not least, when objects like vehicles or natural surface features (e.g., rocks, craters) cast shadows significant temperature differences can be created as the shaded regions will cool down rapidly because they are not receiving direct sunlight and the surrounding surface does not conduct enough heat to warm them.


\section{In-situ Resource Utilization: Lunar Regolith}


\subsection{Characteristics: Mineralogy \& Morphology}
The Moon's surface displays distinct regions known as Lunar highlands and Lunar maria. The highlands appear bright from Earth, while the maria are observed as darker regions. The Lunar regolith, a layer of fragmented, unconsolidated material covering the Moon's surface, is the result of billions of years of meteoroid bombardment. The Lunar regolith is formed by meteoroid impacts that fragment exposed Lunar bedrock. Initially, the resulting regolith is fresh and young. However, when this newly formed regolith remains exposed on the Lunar surface, it undergoes progressive modification by micrometeoroid impacts and high-energy solar and cosmic charged particles. This ongoing modification process, known as maturation, leads to the development of mature regolith over extended periods. Two dominant processes shape the regolith: mechanical breakdown through impacts, gradually reducing particle sizes, and the fusion of particles with impact-formed glass, creating aggregates known as agglutinates, which are especially unique to Lunar soil. Lunar soil comprises a loose, clastic material derived from the mechanical disintegration of basaltic and anorthositic rocks. The Lunar science community mostly uses "Lunar soil" as a synonym for "Lunar regolith," although sometimes the term is used based on its definition in the "Lunar Sourcebook." According to this book, "Although "Lunar soil" is lexicographically synonymous with "Lunar regolith", "Lunar soil" usually refers to the finer-grained fraction of the unconsolidated material (regolith) at the Lunar surface. In this discussion, we define Lunar soil as the subcentimeter fraction of the Lunar regolith" \cite{heiken1991Lunar}. The same terms are used in this section.

In a recent study, Cannon \cite{cannon2023Lunar} proposed a classification system for Lunar soil specifically designed for space resource applications, such as constructing Lunar infrastructure or extracting metals. This system classifies soil using two main criteria: bulk chemistry and average particle size. The primary classification is based on the overall elemental composition of the regolith, particularly focusing on its total iron content (FeO). The regolith is divided into three main categories: Highland (H; FeO < 8.6 wt\%), Mare (M; FeO > 15.9 wt\%), and Intermediate (I; $8.6 \leq \text{FeO} \leq 15.9$ \, \text{wt}\%). These classifications are established based on the analysis of lunar samples, as outlined by Korotev and Irving \cite{korotev2021lunar}. The second criterion relates to particle size classification. It categorizes particles as coarse (C), medium (M), or fine (F) based on their \( D_{50} \) (the particle size at which 50\% of the cumulative mass of particles is smaller and 50\% is larger) values: particles with \( D_{50} > 125 \, \mu \text{m} \) are classified as coarse-grained, those with \( D_{50} < 53 \, \mu \text{m} \) as fine-grained, and particles with \( 53 \leq D_{50} \leq 125 \, \mu \text{m} \) as medium-grained.
Cannon's proposed categorization divides the soil into nine primary groups, which can be further specified with tags indicating additional properties like glass content or concentrations of particular elements. Testing the system demonstrated a strong correlation between orbital soil classification maps and measurements from samples returned from the Apollo 17 landing site \cite{cannon2023Lunar}.


According to \cite{heiken1991Lunar}, the median thickness of regolith is approximately 10-15 m in highlands and 3-5 m in maria regions, respectively. Rocks are found relatively rarely on the Moon's surface; it is estimated that below 1\% of the Moon's surface is covered by rocks with a diameter of approximately 1 meter or larger \cite{heiken1991Lunar}.

    
    

Lunar regolith consists of five primary particle types: mineral fragments, pristine crystalline rock fragments, breccia fragments, volcanic and impact-generated glasses, and agglutinates. The particles in Lunar regolith exhibit a wide range of sizes and shapes, reflecting the mechanical breakdown and fusion of particles caused by impact processes \cite{heiken1991Lunar}. These particles are generally somewhat elongated, displaying subangular to angular characteristics, but they can range from spherical to highly angular shapes. The absence of mechanisms such as wind results in very abrasive particles characterized by jagged edges. Moreover, electrostatic and electromagnetic phenomena lead to a tendency for the soil grains to repel one another when disturbed, contributing to the pervasive and abrasive nature of the dust in the Lunar environment. "Dust" is usually referred to the soil particles smaller than $20$ $\mu$m \cite{cannon2023Lunar}. These agglutinates are characterized by their highly angular and irregular shapes and vary in size from tens of microns to a few millimeters, with an average size of approximately 175 microns. Notably, agglutinates contain a significant amount of nanophase metallic Fe$^0$ in their glass, resulting from the reduction of Fe-silicates in the Lunar soils.

The particle size distribution in Lunar soil plays a crucial role in determining its properties, including strength, compressibility, optical characteristics, thermal properties, and seismic properties. It is important to note that the grain size distribution can vary at different regions on the Moon, between different Apollo landing sites, and even within a single site, based on factors like regolith thickness and proximity to impact craters. More mature Lunar soil tends to become finer on average, more uniformly sized, and contain more agglutinates \cite{mckay1974grain}.
The median particle size of Apollo 11, Apollo 12, Apollo 14, and Apollo 15 were reported between 48-115 $\mu$m, 42-94 $\mu$m, 75-802 $\mu$m, and 51-108 $\mu$m, respectively. The mean particle size of soil from Apollo 16 and 17 was measured between 101-268 $\mu$m and 42-166 $\mu$m \cite{mckay1991workshop}. The mean grain size of analyzed soils ranges from 40 $\mu$m to 800 $\mu$m, with an average between 60 and 80 $\mu$m. The median particle size ranges from 40 to 130 $\mu$m, with an average of 70 $\mu$m, meaning that about half of the soil, by weight, consists of particles smaller than the human eye can discern. A fraction of the soil, roughly 10\% to 20\%, is finer than 20 $\mu$m \cite{roberts2019cross, carrier2003particle, cannon2023Lunar}.


The mean grain size of analyzed soils ranges from 40 $\mu$m to 800 $\mu$m, with an average between 60 and 80 $\mu$m. The median particle size ranges from 40 to 130 $\mu$m, with an average of 70 $\mu$m, meaning that about half of the soil, by weight, consists of particles smaller than the human eye can discern. A fraction of the soil, roughly 10\% to 20\%, is finer than 20 $\mu$m \cite{roberts2019cross, carrier2003particle}.

    
    

Table \ref{tab:Rcomposition} shows the chemical composition of the Lunar regolith from different Apollo missions \cite{heiken1991Lunar} \cite{mckay1994jsc}. It is evident that the first and second most abundant elements on the Moon's surface material are Oxygen and Silicon, with around 61 mol\% (approximately 42 wt\%) and around 16-17 mol\% (approximately 21 wt\%) of the atoms, respectively \cite{heiken1991Lunar}. It is worth mentioning that while all cations are chemically bonded with oxygen, they do not exist in straightforward oxide forms, as one might deduce from the provided compositional data in Table \ref{tab:Rcomposition}. Instead, they can be found in complex silicates or oxide minerals such as anorthite, albite, plagioclase, pyroxene, diopside, etc.

\begin{table}
\caption{Chemical composition (wt\%) of Lunar regolith from Apollo 12, Apollo 14, Apollo 15, Apollo 16, and Apollo 17 \cite{heiken1991Lunar, mckay1994jsc}}.
    \centering
    \begin{tabular}{llllll} 
    \toprule
         Oxide (wt\%)&  Apollo 12&  Apollo 14&  Apollo 15&  Apollo 16& Apollo 17\\ 

    \midrule
         SiO$_2$&  42.2&  46.3&  48.1&  46.8& 45.1\\ 
         TiO$_2$&  7.8&  3&  1.7&  1.2& 0.54\\ 
         Al$_2$O$_3$&  13.6&  12.9&  17.4&  14.6& 27.3\\ 
         FeO&  15.3&  15.1&  10.4&  14.3& 5.1\\ 
         MnO&  0.2&  0.22&  0.14&  0.19& 0.3\\ 
         MgO&  7.8&  9.3&  9.4&  11.5& 5.7\\ 
         CaO&  11.9&  10.7&  10.7&  10.8& 15.7\\ 
         Na$_2$O&  0.47&  0.54&  0.7&  0.39& 0.46\\ 
         K$_2$O&  0.16&  0.31&  0.55&  0.21& 0.17\\ 
 P$_2$O$_5$& 0.05& 0.4& 0.51& 0.18&0.11\\ 
 \bottomrule
    \end{tabular}
    \label{tab:Rcomposition}
\end{table}
 \label{regolith}

\subsection{Simulants}
The study and analysis of the Lunar regolith is of great importance for the success of future Lunar exploration missions. The limited availability of Lunar regolith on Earth poses significant challenges for developing and testing technologies for Lunar missions. Therefore, Lunar regolith simulants, synthetically produced materials that mimic the properties of the Lunar regolith, have been developed by different teams around the world. Conducting experiments with Lunar regolith simulants is the only path for the development of future Lunar technologies. Lunar simulants are supposed to have similar mineralogy, particle size distribution, chemical composition, and/or engineering properties as the Lunar regolith. 
A list of different Lunar regolith simulants developed in the last 33 years, as well as their reported usage are listed chronologically in Table \ref{tbl1}. Even though many Lunar regolith simulants have been developed worldwide, it is well known that a simulant developed for one purpose may not be satisfactory for another. For example, in many Lunar regolith simulants such as JSC-1, BP-1, and GRC-1, the bulk material properties were prioritized over mineralogical accuracy by modifying composition and particle size distribution until the desired physical properties were achieved \cite{taylor2016evaluations, long2023geomechanical}. Mineralogical precision is of great importance in applications such as regolith sintering, electrolysis, extraction purposes, additive manufacturing, and any other application that requires regolith processing. Long-Fox et al. designed Lunar mare and highlands simulants based on mineralogy accuracy in which bulk properties naturally emerge from the density and the geometry of the mineral constituents \cite{long2023geomechanical}.

\begin{table}[width=.9\linewidth,cols=4,pos=h]
\caption{Lunar regolith simulants developed over the last 33 years and their applications.}\label{tbl1}
\begin{tabular*}{\tblwidth}{@{} LLLL@{} }
\toprule
Simulant & Type & Reported usage & Year\\
\midrule
MLS-1 \cite{weiblen1990preparation}  & Mare, High-Ti & General & 1990 \\
MLS-2 \cite{weiblen1990preparation} & Highlands & General & 1990 \\
JSC-1/1A \cite{mckay1993new} & Mare, Low-Ti & Geotechnical & 1993 \\
FJS-1 \cite{kanamori1998properties} & Mare, Low-Ti & Geotechnical & 1998 \\
OB-1 \cite{richard2007ob} & Highlands & Geotechnical & 2007 \\
NU-LHT series \cite{martin2022} & Highlands & Geotechnical & 2007 - 2022 \\
CAS-1 \cite{zheng2009cas} & Mare, Low-Ti & General & 2008 \\
GCA-1 \cite{taylor2008jurassic} & Mare, Low-Ti & Geotechnical & 2008 \\
NAO-1 \cite{li2009nao} & Highlands & General & 2008 \\
CHENOBI \cite{chenobi} & Highlands & General & 2009 \\
CLRS-1 \cite{planetary} & Mare, Low-Ti & N/A & 2009 \\
CLRS-2 \cite{planetary} & Highlands, High-Ti & N/A & 2009 \\
GRC-1 \cite{oravec2010design} & N/A & Geotechnical, Roving vehicle & 2010 \\
CUG-1 \cite{he2010Lunar} & Mare & Geotechnical & 2010 \\
BP-1 \cite{suescun2015geotechnical} & Mare, Low-Ti & Geotechnical & 2010 \\
OPRL2N \cite{offplanet} & Mare & Geotechnical & 2010 \\
OPRL2NT \cite{offplanet} & Mare, High-Ti & General use & 2010 \\
OPRH2N, OPRH3N, OPRH4N \cite{offplanet} & Highlands & Geotechnical & 2010 \\
CSM-CL-S \cite{taylor2016evaluations} & Mare & Geotechnical & 2010 \\
TJ-1 \cite{jiang2012properties} & Mare, Low-Ti & Geotechnical & 2012 \\
DNA-1 \cite{cesaretti2014building} & Mare, Low-Ti & General & 2014 \\
KOHLS-1 \cite{yoo2014kau} & Mare, Low-Ti & Geotechnical & 2014 \\
KLS-1 \cite{ryu2018development} & Mare, Low-Ti & Chemical, Geotechnical & 2017 \\
OPRFLCROSS2 \cite{roux2019unique} & Polar Volatile & Ice simulant & 2019 \\
GreenSpar \cite{gruener2020greenland} & Highlands & Geotechnical & 2020 \\
EAC-1A \cite{engelschion2020eac} & Mare, Low-Ti & Geotechnical & 2020 \\
LSS-ISAC-1 \cite{venugopal2020study} & Highlands & Geotechnical, Rover mobility & 2020 \\
LMS-1 \cite{long2023geomechanical} & Mare, Low-Ti & Geomechanical, Geochemical, Mineralogy & 2021 \\
LHS-1 \cite{long2023geomechanical} & Highlands & Geomechanical, Geochemical, Mineralogy & 2021 \\
CSM-LHT-1 \cite{martin2022} & Highlands & General & 2021 \\
CSM-LMT-1 \cite{martin2022} & Mare, Low to Medium-Ti & Genaral & 2021 \\
CSM-LHT-1G \cite{sibille2023vacuum} & Highlands  & Geotechnical & 2023 \\

\bottomrule
\end{tabular*}
\end{table}

\subsection{Extraction of Oxygen and Metals from Regolith}
As will be presented in Section 4.1 and Table~\ref{SpaceGrade}, metals are an important material for manufacturing in space, as on Earth. Once extracted, metals can be used in conventional manufacturing in space, or in additive manufacturing, as will be discussed further in subsequent sections. The extraction of metals from Lunar regolith may sometimes be done alongside the extraction of oxygen, another crucial resource.

\subsubsection{Extraction Processes}

In general, metals exhibit higher mechanical strength compared to regolith. While transporting raw materials to the Moon is costly, researchers have explored the potential of extracting metals from Lunar regolith. Regolith contains various oxides, including silicon oxide, and different metallic oxides such as aluminum oxide, iron oxide, magnesium oxide, titanium oxide, and calcium oxide \cite{papike1982Lunar}. However, regolith is a complex mixture of minerals that contain impure oxides (e.g., ilmenite) and silicates (e.g., anorthite) rather than pure oxides, making metal and oxygen extraction challenging. However, prolonged human presence on the Moon necessitates in-situ production of oxygen \cite{taylor1993oxygen}. Oxygen is essential for both human life support (breathing) and propulsion, given that liquid oxygen is utilized in many bipropellant rockets \cite{bowersox1992processes}. Alongside ice water, Lunar regolith has the potential to be a primary resource for oxygen extraction on the Moon. Some oxygen extraction technologies simultaneously yield metals, rendering metal extraction more economically viable. According to NASA's SP-509 document \cite{mckay1992space}, iron is the most convenient metal to produce from Lunar regolith. The resulting metals have the potential to be utilized as raw materials for Lunar-based manufacturing \cite{lomax2020proving}.

The chemical reduction of metallic oxides is a process that involves using a reducing agent such as hydrogen, known as \textbf{hydrogen reduction}, or carbon compounds like carbon monoxide, known as \textbf{carbothermal reduction}. Extracting oxygen from iron oxide is more energy-efficient compared to most other metal oxides. Ilmenite ($\mathrm{FeTiO_3}$), found in concentrations of 10-25\% by weight in iron-rich mare basalts, contains 31.6\% titanium and 36.8\% iron. According to Apollo 17 findings, these basalts provide a sufficient source of iron for various space applications \cite{mckay1992space}. One of the most frequently proposed techniques for metal and oxygen extraction is the thermal reduction of ilmenite at 1000\textdegree{C} using hydrogen derived from electrolyzed water in a fluidized bed \cite{ellery2020sustainable, mckay1992space, bowersox1992processes}:

\[ \text{FeTiO}_3 + \text{H}_2 \rightarrow \text{Fe} + \text{TiO}_2 + \text{H}_2\text{O} \]

Raising the temperature to approximately 1600\textdegree{C}, the melting point of iron, causes iron to be released through liquation. To obtain oxygen and recycle hydrogen, water must be split into its components via electrolysis \cite{ellery2020sustainable}:

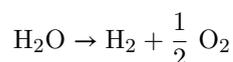
\[ \text{H\(_2\)O} \rightarrow \text{H\(_2\)} + \frac{1}{2}\text{ O\(_2\)} \]

To extract oxygen and metals from Lunar regolith using the carbothermal process, reactions involve metal oxides such as FeO, Fe$_2$O$_3$, TiO$_2$ (for partial reduction), and SiO$_2$. At temperatures exceeding 1600\textdegree{C}, these oxides react with methane (CH$_4$), leading to the formation of metallic elements (M), carbon monoxide (CO), and hydrogen gas (H$_2$) \cite{gustafson2009oxygen, gustafson20102010}:

\[ \text{MO}_x + x \, \text{CH}_4 \rightarrow \text{M} + x \, \text{CO} + 2x \, \text{H}_2  \]

The generated CO and H$_2$ can further react to produce methane and water \cite{gustafson2009oxygen, gustafson20102010}:

 \[ x \, \text{CO} + 3x \, \text{H}_2 \rightarrow x \, \text{CH}_4 + x \, \text{H}_2\text{O}  \]

The water generated can subsequently undergo electrolysis to produce hydrogen and oxygen, by the water electrolysis equation mentioned earlier  \cite{gustafson2009oxygen, gustafson20102010}.

\textbf{Carbothermal reduction.} The Precursor In-situ Lunar Oxygen Testbed (PILOT) project, part of NASA's Outpost Precursor Testbed for ISRU and Modular Architecture (OPTIMA) program, initially involved Lockheed Martin as the prime contractor and later ORBITEC, focusing on oxygen production methods from Lunar regolith. ORBITEC specialized in carbothermal reduction using methane and solar energy, while Lockheed Martin concentrated on hydrogen reduction. The project's goal was to integrate ISRU technology for excavation, oxygen extraction, storage, and distribution. ORBITEC developed the Carbothermal Regolith Reduction Module, which was tested using both CO$_2$ lasers and concentrated solar energy, demonstrating the capability to extract 1 metric ton of oxygen per year from Lunar regolith, particularly under conditions found at Lunar polar locations. Initial validation tests were conducted using a 440 W CO$_2$ laser to simulate concentrated solar energy, confirming the functionality of the regolith handling system, the carbothermal reduction reactor, the gas handling system, and the processed regolith removal system. Through extensive carbothermal reduction experiments utilizing direct energy heating, ORBITEC achieved oxygen yields up to 28 wt\%. However, these high yields typically require a carbon-rich environment, which can cause a carbon cap to form over the melt region, impeding thermal energy transfer and slowing the reduction process. Despite this challenge, the regolith samples processed had very low carbon content, ranging from 0 to 0.03 wt\%, corresponding to a maximum carbon loss of only 1.7 kg per metric ton of oxygen produced. This minimal carbon presence highlights the effectiveness of the reduction method despite the challenges. The integration of the Solar Energy Module with the Carbothermal Regolith Reduction Module marked a critical advancement in the project. This module includes seven solar concentrators, all mounted on a frame connected to a commercial two-axis solar tracking unit. Solar energy is delivered to the carbothermal reduction reactor via fiber optic cables, each designed to provide about 140 W of power, achieving processing temperatures up to 1900\textdegree{C}. While the system was initially intended to deliver around 1 kW, the actual power delivered has ranged between 500-700 W, with recent upgrades to the inlet optics potentially boosting this to 800-900 W. A series of experiments using JSC-1A Lunar regolith simulant and concentrated solar energy demonstrated the system's efficiency. In one test, conducted for 80 minutes with approximately 700 W of solar power, the processed simulant produced a melt with a diameter of 53.3 mm and a mass of 77.6 g, achieving an oxygen yield of 11.3 wt\%. Subsequent experiments with 550-600 W of power resulted in maximum melt masses of 33 g and oxygen yields of 2.6 wt\%, highlighting the crucial role of power levels in optimizing the oxygen extraction process. Following previous phases, carbothermal reduction experiments were conducted using tephra collected from a Lunar analog site on Mauna Kea, Hawaii, managed by PISCES. Situated at an altitude of 9,000 ft, this site closely mimics the terrain and soil composition found on the Moon. Initial single-batch tests utilizing concentrated solar energy revealed that the Mauna Kea tephra produced a maximum melt mass of 22 grams and yielded 7.6 wt\% oxygen by weight with 580 W of solar power. Subsequent multiple-batch tests involved processing the resulting gas through NASA Kennedy Space Center's Gas Clean-Up Module and feeding it into a microchannel methanation reactor. These tests demonstrated that with a hydrogen ratio of 3.7:1, sustained operation without coking was achievable, underscoring the system's potential for reliable and efficient Lunar operations \cite{gustafson2009oxygen, gustafson2010demonstrating}.

Townsend et al. \cite{townsend2010performance} reported on the performance of regolith feed systems, employing mechanical auger and pneumatic systems, during the OPTIMA tests on Mauna Kea, Hawaii. These tests demonstrated the feasibility of producing oxygen at a rate equivalent to 660 kg/year in a Lunar Outpost scenario. The auger-based regolith feed system effectively met all operational objectives by consistently feeding 8 kg batches of tephra from the input hopper into the reactor and expelling them to the surface. Despite encountering challenges such as high temperatures and abrasive particles, the system exhibited robust performance during initial field testing while future lifetime testing was recommended to assess wear rates and ensure long-term reliability. Terrestrial and reduced gravity experiments validated the efficacy of dense-flow pneumatic transfer methods for conveying Lunar regolith simulants like NU-LHT-2M and tephra up to a height of 5 ft (1.5 m). These experiments highlighted operational challenges related to variable gravity levels impacting material compaction and pneumatic transfer efficiency. The integration of cyclone separators in pneumatic systems proved crucial for delivering regolith to specified locations and filtering conveyance gas for reuse. These findings informed the development of NASA's second-generation ROxygen oxygen production system, aimed at achieving a minimum annual production of one metric ton of oxygen from Lunar regolith simulants, with an emphasis on reliability, durability, and minimal maintenance \cite{townsend2010performance}.

NASA recently conducted a project using carbothermal reduction to extract both oxygen and metals from Lunar soil simulants, replicating Lunar surface conditions. The Carbothermal Reduction Demonstration (CaRD) team utilized the Dirty Thermal Vacuum Chamber, a specialized spherical chamber designed for testing unclean samples. Inside this chamber, a high-powered laser simulated solar energy concentration to melt Lunar soil simulant within a carbothermal reactor developed by Sierra Space Corp. This reactor, crucial for heating and oxygen extraction, mirrors techniques long used on Earth to produce materials such as solar panels and steel through the generation of carbon monoxide or dioxide at high temperatures. Following heating, carbon monoxide was detected using the Mass Spectrometer Observing Lunar Operations (MSolo) device. This technology will be utilized in upcoming Lunar missions aimed at exploring Lunar water ice and other potential resources. The successful testing of the carbothermal reactor in a vacuum environment during the CaRD project raised its technical readiness level to six, indicating it has a fully functional prototype and is prepared for space testing. The test results showed significant improvements in the efficiency of oxygen extraction. The 2010 field demonstration \cite{gustafson2009oxygen, gustafson2010demonstrating} achieved an oxygen extraction rate of 1.45 grams per kilowatt-hour (g/KWh) delivered to the reactor. In comparison, the CaRD ambient tests showed an extraction rate of 13.42 g/KWh, while the CaRD vacuum tests achieved rates of 11.53 g/KWh, 15.79 g/KWh, and 10.77 g/KWh in subsequent tests. These results indicate that the CaRD system is substantially more efficient in extracting oxygen from Lunar regolith compared to the previous field demonstration, highlighting the potential for scaled-up operations to produce several times its own weight in oxygen per year of operation \cite{paz2023carbothermal}.

\textbf{Molten salt electrolysis.} Molten salt electrolysis is a technique employed to extract metals and oxygen from Lunar regolith through high-temperature electrolysis. In this method, a suitable molten salt electrolyte, typically composed of compounds like calcium chloride ($\mathrm{CaCl_2}$) or lithium fluoride (LiF), is heated to typically over 700\textdegree{C} to achieve a molten state. The Lunar regolith itself is not directly heated to such temperatures. Instead, regolith serves as the source of metal oxides that dissolve into the molten salt electrolyte. Subsequently, an electric current is applied to the molten mixture, prompting metal ions from the regolith to migrate to the cathode and undergo reduction to their elemental forms, while oxygen ions move to the anode and are released as oxygen gas. This process facilitates the extraction of metals such as aluminum, iron, and titanium, as well as the recovery of oxygen \cite{schwandt2012production, lomax2020proving}. Lomax et al. \cite{lomax2020proving} conducted a study that presents a proof-of-concept demonstrating the molten salt electrolysis (Metalysis-FFC) process's ability to extract oxygen and generate metallic alloys as by-products from Lunar regolith simulants. The research revealed the potential for achieving near-complete (96\%) oxygen recovery from regolith through molten salt electrolysis at approximately 950\textdegree{C}. The analysis of the produced alloys reveals three primary categories: an Al/Fe alloy, often accompanied by Si; a Fe/Si alloy, sometimes containing Ti and/or Al; and a Ca/Si/Al alloy, sporadically incorporating Mg.

\textbf{Molten regolith electrolysis.} Molten regolith electrolysis (MRE) is another method for extracting metals and oxygen directly from Lunar regolith. Unlike molten salt electrolysis, which requires a separate molten salt electrolyte, molten regolith electrolysis employs the Lunar regolith itself as both the source of metal oxides and the electrolyte once it reaches temperatures typically between 1600-1800\textdegree{C}. The process begins by heating the Lunar regolith until it melts, allowing metal oxides within the regolith to dissolve into the molten mass. Electrolysis then proceeds, with an electric current applied directly to the molten regolith \cite{ellery2020sustainable, guerrero2023system, burke2024modeling}. This method is tailored specifically for Lunar resource utilization, eliminating the need for additional electrolytes and potentially enhancing efficiency and simplicity in sustainable resource extraction on the Moon. However, achieving and maintaining such high temperatures poses significant challenges, particularly in terms of energy consumption. Guerrero-Gonzalez and Zabel \cite{guerrero2023system} carried out a study to compare the proficiency of three different production plants that extract metals and oxygen out of Lunar regolith at the Lunar South Pole. They analyzed hydrogen reduction of ilmenite and carbonylation for the production of low-carbon steel, electrolysis of molten regolith to extract ferrosilicon alloys, and electrolysis of molten salt with vacuum distillation to produce aluminum-silicon alloys. The primary objective of the research was to create detailed models for various ISRU production plants, under a consistent set of assumptions, enabling a quantitative comparison to facilitate high-level decision-making regarding the crucial process parameters. The study found that as the production plants grew in size, they demonstrated a slight economy of scale, indicating more efficient extraction of metals and oxygen. The study revealed that ferrosilicon alloy extraction using molten regolith electrolysis is the most efficient ISRU process. It requires 6776 kg of equipment and 311.34 kW of solar power to produce 25 tons of metal and 23.9 tons of oxygen annually, with a low mass payback ratio of 0.14 kg of hardware per kg of product. Low-carbon steel production is effective when the ilmenite concentration in the regolith is above 7.5 wt\%, achieving a mass payback ratio of less than 0.73 kg of hardware per kg of product. In contrast, aluminum-silicon alloy production using the FFC-Cambridge process is less efficient due to the substantial amount of molten salt required; this process would need a reduction in salt usage by at least five times to be competitive \cite{guerrero2023system}.

MRE holds significant promise in achieving NASA's Lunar Surface Innovation Initiative (LSII) goal of producing 10 tons of oxygen annually from Lunar regolith during the Artemis program. This straightforward, single-step process requires minimal consumables and yields both oxygen and metals simultaneously. However, previous small-scale demonstrations faced challenges due to high temperatures generated by external heaters, which created a corrosive interface between the molten regolith and the containment material in the reactor, thus limiting its lifespan. To address this, the Gaseous Lunar Oxygen from Regolith Electrolysis (GaLORE) project, funded by NASA's Space Technology Mission Directorate (STMD), advanced a 'cold-walled' reactor design. This innovative design melts regolith in a localized area between two electrodes, forming a thermal insulation barrier to protect the containment wall from the corrosive melt \cite{grossman2021galore}. Grossman et al. \cite{grossman2021galore} used thermal modeling of the proposed cold-walled reactor to develop parameters for a feasible reactor shape and size, targeting minimal energy consumption. The design of heater devices considers limitations in electrical power, the wide range of Lunar regolith compositions, the limited availability of replacement metallic parts, and the low thermal conductivity of regolith in a vacuum \cite{grossman2021galore}.

Schreiner et al. \cite{schreiner2016parametric} developed a parametric sizing model for an MRE reactor to simultaneously extract oxygen and molten metals from Lunar regolith. This model is based on regolith material property models from their earlier publication \cite{schreiner2016thermophysical}, validated based on data obtained from Apollo mission samples and regolith simulants. The required oxygen extraction rate determines the processing requirements and reactor current. A multiphysics simulation was employed to create a comprehensive database of reactor designs and performance characteristics for MRE systems. This effort led to the development of a novel design methodology that utilizes this database to parametrically design MRE reactors capable of meeting specific regolith processing rates and reactor current requirements. The study presented predictions for the mass and power consumption of MRE reactors across various oxygen yield levels, analyzing the impacts of operating temperature, regolith type, and design flexibility. The model forecasts oxygen extraction efficiencies ranging from approximately 0.15 to 0.375 kilograms of oxygen per kilogram of regolith, with higher efficiencies observed at elevated temperatures. For Highland regolith, increasing temperatures enhance current efficiency from 72.5\% to 82\%, while for Mare regolith, efficiencies start at 57.5\% and rise to 70\% at higher temperatures. Higher operating temperatures generally lead to reduced reactor mass and sometimes lower power requirements, although these effects are heavily influenced by the specific characteristics of the regolith type. Regolith composition is a critical factor impacting MRE reactor design and performance. At operating temperatures around 2000~K, reactors optimized for Lunar Highland are 33\% lighter and require 60\% less power compared to those designed for Lunar Mare, while maintaining robust performance levels even when processing Mare regolith as opposed to Highland regolith \cite{schreiner2016parametric}.

\textbf{Acid leaching.} Acid leaching is a chemical process that can be used to extract metals from Lunar regolith by dissolving them in an acidic solution. This method involves mixing the regolith with acids like sulfuric acid (H\(_2\)SO\(_4\)), hydrofluoric acid (HF), or hydrochloric acid (HCl), which react with the metal oxides in the regolith to form soluble metal salts. The dissolved metals in the resulting solution can then be recovered through precipitation or electrochemical methods, including electrowinning or electrorefining. One advantage of acid leaching is its ability to selectively dissolve specific metals, making it a versatile option for extracting valuable elements such as iron, aluminum, and titanium from Lunar regolith \cite{ellery2020sustainable, ellery2018towards}. However, despite the potential for being conducted at lower temperatures compared to molten regolith or molten salt electrolysis techniques in most cases, thereby reducing energy requirements, acid leaching faces significant challenges on the Moon. These include the need for large quantities of acids and the management of toxic and corrosive by-products. Moreover, the extreme temperature variations on the Lunar surface complicate the use of liquid acids. During the Lunar nights, acids can freeze without adequate heating systems, while during the Lunar days, temperatures exceeding 120\textdegree{C} can cause most acids to boil. For example, HCl boils at around 108\textdegree{C}, which is lower than the maximum daytime temperature on the Moon. This temperature difference presents challenges in handling and storing the acid. The risk of acid loss through evaporation, coupled with potential hazards from toxic and corrosive fumes, can impair the efficiency of the leaching process if not effectively managed. Another significant challenge is the continuous sourcing or in-situ production of acids essential for Lunar operations. For instance, sulfuric acid can be synthesized from sulfur dioxide (SO\textsubscript{2}) and water vapor, while water (H\textsubscript{2}O) can be electrolyzed to yield hydrogen gas (H\textsubscript{2}) and oxygen gas (O\textsubscript{2}). Chlorine gas (Cl\textsubscript{2}), vital for HCl production, is present in Lunar regolith at trace levels, typically less than 0.1\% by weight. However, producing acids on the Moon itself demands substantial energy, infrastructure, and resource investments. Therefore, refining and repurposing acids after the leaching process is crucial to minimize waste and enhance efficiency in Lunar resource utilization. Superacids, such as solid acid catalysts like HF--SbF\(_5\), CF\(_3\)SO\(_3\)H, and FSO\(_3\)H, can conduct complex reactions at low temperatures. However, their effective use necessitates a continuous provision of reagents from Earth \cite{ellery2020sustainable}.

A method advocated by Thibodeau et al. \cite{Thibodeau2024silica} involves extracting silica and alumina from Lunar regolith, particularly targeting anorthosite (CaAl\(_2\)Si\(_2\)O\(_8\)). HCl is utilized as the leaching agent, aligning with a circular industrial model for in-situ Lunar resource utilization. The process, conducted at a laboratory scale, includes beneficiation, hot HCl leaching, centrifugal separation, HCl sparging, and double calcination. This process successfully yielded silica and precipitated aluminum chloride hexahydrate (AlCl\(_3 \textbf{.}6\)H\(_2\)O), which was subsequently calcined to produce alumina. These materials find applications in neutral refractory ceramics, fused silica glass, and thermal insulation fibers. Furthermore, alumina can undergo further processing using electrochemical methods such as the Metalysis FFC process to obtain pure aluminum metal, while silica can be processed into high-purity silicon via molten rock electrolysis for early-stage photovoltaic cell manufacturing. The feasibility of HCl leaching for Lunar resource extraction with minimal reliance on Earth-sourced reagents is highlighted, offering a sustainable pathway toward establishing Lunar manufacturing capabilities. The authors suggested that future research should prioritize automating and optimizing these processes, with a specific focus on physical beneficiation to improve material quality and operational efficiency. This strategy aims to reduce waste and promote a circular industrial ecosystem essential for sustainable Lunar industrialization \cite{Thibodeau2024silica}.

\textbf{Bioleaching.} Biological techniques for extracting metals from Lunar regolith involve using microorganisms' metabolic capabilities to dissolve and recover metals. This process, known as bioleaching or biomining, employs bacteria or fungi that naturally produce acids or enzymes capable of breaking down minerals in the regolith. By exposing the regolith to these microorganisms in controlled environments, metals like iron, aluminum, titanium, and others can be solubilized and extracted \cite{dusengemungu2021overview, tichy1998bioleaching, volger2020mining}. This approach offers several advantages over traditional extraction methods, potentially including reduced energy consumption and the ability to operate under conditions that may be inhospitable for humans, among others. Volger et al. \cite{volger2020mining} introduced a method for biologically extracting iron from Lunar and Martian regolith through the utilization of microorganisms. This process integrates leaching, precipitation or accumulation, and magnetic extraction steps to enhance iron concentration. Four distinct organisms were examined for their potential contributions to iron leaching and accumulation. The study emphasizes the significance of process yields as a crucial metric for evaluating the efficacy of microbial processes in space applications \cite{volger2020mining}.

\textbf{Sublimation.} Shaw et al. \cite{shaw2023metal} conducted a study modeling the kinetics of vacuum sublimation for the volatile major oxides in Lunar regolith, deriving evolution rates for Fe, Na, and K. Due to the Moon's high vacuum environment, the research investigates the sub-liquidus operation of sublimation. Their findings demonstrate the kinetic feasibility of sublimation for Na and K under sub-liquidus conditions, while thermodynamically viable Fe sublimation is constrained by kinetics.

\textbf{Summary.} Despite the numerous papers published on this topic, it appears that it is still in its early stages of development. Further research is essential to achieve a more efficient and cost-effective process. There is an extensive review paper by Ellery et al. \cite{ellery2020sustainable} that provides detailed coverage of this topic. For further details, please refer to it.

\subsubsection{Physical Separation}

Physical separation methods are used to beneficiate Lunar regolith by increasing the concentration of valuable minerals before proceeding to subsequent ISRU stages, including enrichment and extraction processes. Gravitational, magnetic, and electrostatic separation techniques are the primary methods used for this purpose. Gravitational separation relies on density differences to segregate components: heavier particles settle more rapidly under gravity, while lighter particles remain suspended or are carried away. This method effectively separates minerals based on their varying densities. Magnetic separation exploits the magnetic properties of minerals, using magnets or electromagnets to separate magnetic minerals (such as iron-containing minerals) from non-magnetic ones in Lunar regolith. Electrostatic separation utilizes differences in particle conductivity and surface charge to separate minerals. Charged particles respond to an electric field by attracting or repelling, enabling separation based on their electrostatic properties \cite{rasera2020beneficiation}.

\textbf{Magnetic separation} is a commonly used method in mineral processing for removing unwanted iron from minerals and for separating roasted hematite, magnetite, and siderite from silica \cite{wills2015wills}. This method holds significant promise for extraterrestrial applications as well. Both paramagnetic and ferromagnetic techniques exist in the composition of Lunar regolith, enabling the applicability of magnetic beneficiation of regolith \cite{rasera2020beneficiation}. Magnetic separation is explored for enriching various target materials, including ilmenite \cite{williams1979mining, taylor1992magnetic, berggren2011Lunar}, anorthite \cite{oder1989magnetic}, and iron oxide \cite{berggren2011Lunar}, among others. However, the electromagnetic properties of Lunar minerals have not yet been extensively studied through empirical means. Additionally, it is crucial to determine these properties using Lunar-origin samples due to the significant chemical differences between Lunar and terrestrial minerals \cite{rasera2020beneficiation}.

As an example, the Lunar Soil Particle Separator (LSPS) introduces an innovative method for beneficiating Lunar regolith. By enhancing the concentration of ilmenite or other iron-oxide-bearing materials in Lunar soils, the LSPS significantly boosts ISRU oxygen yield. This technological advancement holds the potential to reduce the size of hydrogen reduction reactors and greatly decrease the energy input required for soil heating. Research utilizing grade N-50 neodymium-iron-boron permanent magnets has demonstrated successful enrichment of magnetic and paramagnetic materials from binary mixtures and Lunar simulants. Berggren et al. \cite{berggren2011Lunar} reported achieving notable enrichment levels: iron oxide increased from 10 wt\% to 87 wt\% with two passes and ilmenite from 10 wt\% to 75 wt\% with three passes in quartz sand. However, results with JSC-1A simulant showed minimal improvement, while with NU-LHT-2M simulant, iron oxide-rich minerals increased from 4 wt\% in the feed to 10.5 wt\% in the concentrate stream \cite{berggren2011Lunar, rasera2020beneficiation}.

\textbf{Electrostatic separation} is an effective technique for differentiating minerals with varying electrical conductivities or charge characteristics. For example, conductive particles are drawn to the electrode, while non-conductive particles remain unaffected and can be separated accordingly. Electrostatic separation is particularly advantageous for processing Lunar regolith, as it enables the efficient segregation of valuable minerals like ilmenite and other metal oxides, which are essential for ISRU \cite{rasera2020beneficiation, fraas1970factors, captain2007tribocharging, trigwell2006electrostatic}. Additionally, this method is useful for size sorting, allowing for the separation of regolith particles based on size \cite{williams1979mining, agosto1984electrostatic, adachi2017particle}.

In the available literature, electrostatic separation is primarily investigated for enriching ilmenite from Lunar regolith \cite{agosto1984electrostatic, williams1979mining, agosto1983electrostatic, agosto1985electrostatic, li1999dry, quinn2013evaluation, trigwell2009use, trigwell2013electrostatic}. Besides, in the literature, three methods have been utilized for beneficiating Lunar regolith through the use of electrostatic forces: conductive induction via slide separators \cite{agosto1984electrostatic, agosto1983electrostatic, agosto1985electrostatic, agosto1992part}, tribocharging with parallel plate separators \cite{trigwell2006electrostatic, trigwell2009use, trigwell2013electrostatic, li1999dry, quinn2013evaluation}, and the electrostatic traveling wave technique \cite{adachi2016sampling, adachi2017electrostatic, adachi2017particle, kawamoto2018improvement}.

For comprehensive coverage of physical separation techniques, refer to the extensive review paper by Rasera et al. \cite{rasera2020beneficiation}.\label{Extraction}

\section{Building and Manufacturing Materials and Processes}


\subsection{Space-Grade Materials Selection}

 Space-grade materials are carefully selected and specially engineered to endure the harsh conditions of outer space, encompassing factors such as the vacuum, radiation, temperature fluctuations, and microgravity. These materials find application in manufacturing spacecraft, satellites, space stations, spacesuits, and a variety of other equipment intended for space missions. The selection of space-grade materials hinges on their durability, ability to withstand radiation, and consistent performance in the demanding space environment. These materials encompass a range of options, including aluminum and titanium alloys, space-grade plastics, fiber-reinforced composites, ceramics, fabrics, electronics, lubricants, specialized coatings, radiation shielding coatings, and thermal insulation materials, among others \cite{pater2007advanced, toor2018space, fayazbakhsh2010materials, soboyejo2016review, schmidt2005ceramic, rawal2001metal,voevodin2005nanocomposite, krishnamurthy1995polymers, grossman2003space, rinaldi2018additive}. 

For the purpose of this discussion, our primary emphasis centers on materials selection for in-space manufacturing (ISM). We have already delved into the influence of Lunar conditions in Section \ref{impact}. In addition to the application-specific requirements, there are general requirements for materials selection for ISM, specifically LBMC. When considering LBMC, it is advantageous for the fabrication technology and materials to be compatible with Lunar regolith. This is in line with the goal of maximizing the use of Lunar regolith for LBMC to reduce launch expenses. To put it differently, these materials should ideally be suitable for processing from regolith or for integration with regolith-based materials.

It is essential that these materials show minimal outgassing tendencies to minimize the risk of contamination and have sufficient thermal stability to endure the Moon's extreme temperature fluctuations. Given the absence of Earth's protective atmosphere, materials must also demonstrate resistance to the solar and cosmic radiation found on the Lunar surface, if they are not intended to be protected by a shield. Mechanical strength is another essential criterion to endure the Lunar environment. On the other hand, given the fact that the Lunar gravity force and the weight of objects on the Moon are one-sixth that of Earth, the compressive strength required for Lunar construction is expected to be one-sixth of what is needed for a comparable structure on Earth, as mentioned in Montes et al.'s evaluation \cite{montes2015evaluation}, typically ranging from 25-40 MPa \cite{kosmatka2002design}. Due to the abrasive nature of regolith, materials should be selected or treated to minimize dust accumulation and damage. Mass and volume considerations are of utmost importance, as launching heavy payloads of raw materials from Earth can be cost-prohibitive. Materials should be chosen in a manner that streamlines the fabrication process and reduces reliance on various pieces of equipment, given the limited accessibility to such equipment. Considerations like recyclability, longevity, and durability in the Lunar environment round out the primary requirements for materials selection \cite{pater2007advanced, toor2018space, fayazbakhsh2010materials, soboyejo2016review, schmidt2005ceramic, rawal2001metal,voevodin2005nanocomposite, krishnamurthy1995polymers, grossman2003space, rinaldi2018additive}.

Table \ref{SpaceGrade} lists the most commonly used materials for space applications.

\begin{landscape}
\begin{table}[htp]
\centering
\caption{A Compilation of Commonly Utilized Space-Grade Materials \cite{pater2007advanced, toor2018space, fayazbakhsh2010materials, soboyejo2016review, schmidt2005ceramic, rawal2001metal,voevodin2005nanocomposite, krishnamurthy1995polymers, grossman2003space, rinaldi2018additive}}  \label{SpaceGrade}
\small  
\begin{tabularx}{\linewidth}{l l X}
\toprule
\textbf{Category} & \textbf{Material} & \textbf{Description} \\
\midrule
\multirow{6}{*}{Metals} & Aluminum Alloys & Lightweight, good corrosion resistance, widely used in space applications. \\
& Titanium and Titanium Alloys & High strength-to-weight ratio, corrosion-resistant. \\
& Stainless Steel & Provides strength and corrosion resistance in space applications. \\
& Invar Alloy & Low thermal expansion, ideal for precision components. \\
& Copper Alloys & Excellent electrical conductivity, used in electrical components. \\
& Gold and Silver & High electrical conductivity, oxidation-resistant, for electrical applications. \\
\midrule
\multirow{7}{*}{Ceramics} & Alumina (Aluminum Oxide) & High strength, electrical insulation, and thermal resistance, used in insulators and structural components. \\
& Silicon Carbide (SiC) & High thermal stability, high strength, and corrosion resistance, used as heat shields and optical components. \\
& Zirconia (Zirconium Oxide) & High strength, thermal resilience, and chemical stability, utilized in bearings, sensors, and cutting tools. \\
& Boron Nitride (BN) & Excellent thermal conductivity and electrical insulating properties, used in heat sinks and as electrical insulators. \\
& Sapphire & Single-crystal alumina with optical clarity and scratch resistance, used in optical windows, lenses, and sensors. \\
& Piezoelectric Ceramics & Used for precise spacecraft control and measurement. \\
& Ceramic Thermal Insulation & Specialized materials for spacecraft protection from extreme temperatures. \\

\midrule
\multirow{10}{*}{Polymers} & Kapton (Polyimide) & Resistant to radiation, used in spacecraft circuits, insulation, and protective layers. \\
& Polyethylene & Lightweight with low outgassing properties, used for insulation and protection in space missions. \\
& Polyether Ether Ketone (PEEK) & High-performance polymer with mechanical and thermal qualities for spacecraft components. \\
& Polysulfone & Resistance to radiation and high temperatures, used in specialized components. \\
& Teflon (Polytetrafluoroethylene) & Low friction, non-stick properties, chemical resistance. Used in coatings and insulation. \\
& Polyimides & Heat-resistant polymers for high-temperature space applications. \\
& Polyethylene Terephthalate (PET) & Lightweight and durable, used in various space equipment. \\
& Fluorosilicone Rubber & Resistant to extreme temperatures, used for seals and gaskets. \\
& Silicone Rubber & Highly flexible and used for sealing and gasket functions. \\
& Nylon & Employed for strength and durability in space applications. \\
\midrule
\multirow{7}{*}{Composites} & Carbon Fiber Composites & High strength-to-weight ratio, used in structural spacecraft components. \\
& Glass Fiber Composites & Affordable and strong, used in secondary spacecraft structures. \\
& Aramid Fiber Composites & High tensile strength and impact resistance for spacecraft reinforcement. \\
& Fiber Metal Laminates (FMLs) & Balance lightweight composites with durable metals in spacecraft components. \\
& Carbon-Carbon Composites & Exceptional thermal properties for heat shields and rocket nozzles. \\
& Foam Core Composites & Provide thermal insulation, especially in cryogenic fuel tanks. \\
& Ceramic Matrix Composites & Used in components like heat shields and high-temperature structural parts. \\
\bottomrule
\end{tabularx}
\end{table}
\end{landscape}


\subsection{Lunar-Based Additive Manufacturing}


\subsubsection{Powder Bed Fusion}

Powder Bed Fusion (PBF) is a major AM category in which a fine layer of powdered material, evenly distributed across the build platform, is exposed to a high-power energy source, such as a laser beam, electron beam, microwave, or solar radiation to melt and fuse powder particles layer-upon-layer selectively. PBF is commonly implemented through various methods, including selective laser sintering (SLS), selective laser melting (SLM), electron beam melting (EBM), selective solar light sintering (SSLS), and selective microwave sintering (SMWS). Researchers have found some of these methods applicable to challenging extraterrestrial environments, particularly for printing Lunar regolith materials.


\subsubsubsection{Laser Powder Bed Fusion (L-PBF)}

L-PBF-based technologies, including SLM and SLS, rely on laser energy to fully melt/sinter material particles and fuse them together \cite{abedi2019high}. These processes are particularly notable for their ability to work with materials with high melting points, such as powdered ceramics and multiphase soil. Thanks to this capability, these methods have been proposed for the direct fabrication of fine geometries from Lunar regolith with no need for additives. Considering that laser processing parameters directly influence the properties and characteristics of the printed materials, researchers have worked to precisely optimize the processing parameters to control and improve the outcome of the print.


Fateri et al. studied the feasibility of laser melting technology for printing Lunar regolith \cite{fateri2013experimental, fateri2015process}. To minimize particle size variation and ensure uniform energy absorption within the powder bed, JSC-1A regolith simulant was sieved through a $63$ $\upmu$m mesh. Optimal process settings were determined to be $50$ W for laser power and a scan speed of $50$ mm/s, which yielded acceptable microscopic results with uniform surface properties. The layer thickness was set between $100$ $\upmu$m and $300$ $\upmu$m. They printed various complex geometries at both millimeter and micrometer scales. The authors investigated the morphology and hardness of the manufactured objects using the Berkovich method, which yielded a surface hardness of $1245$ HV. The measured surface roughness averaged at $\textrm{Ra}$ = $1.5$ $\upmu$m. They reported promising outcomes in terms of structural integrity, geometrical accuracy, and density for the manufactured parts.


Laser processing parameters were optimized in another research by Goulas et al. \cite{goulas20163d}. They conducted experiments within an inert argon environment. The as-received JSC-1A regolith simulant was filtered through a $125$ $\upmu$m mesh. Their optimized parameters consisted of a $50$ W laser power, a scanning speed of $210$ mm/s, a laser beam spot size of $300$ $\upmu$m, and a $210$ $\upmu$m hatching space (with a $35$\% overlap) on each layer, resulting in a laser energy density of $1.011 \, \text{J/mm}^2$. These settings produced macroscopically flawless specimens. Subsequent analysis involved SEM, X-ray fluorescence, and Vickers micro-hardness testing on the laser-processed specimens. The printed material exhibited a relative porosity of $40.8$\% and a material hardness of $670 \text{±} 11$ HV. Notably, they achieved a dimensional accuracy of $99.8$\%. In another study \cite{goulas2019mechanical}, Goulas et al. documented that an input energy of $0.92 \, \text{J/mm}^2$ yielded favorable morphological characteristics and mechanical properties. For this study, they set the laser scanning speed at $320$ mm/s, layer thickness at $150$ $\upmu$m, and hatch spacing ranging from $170$ to $250$ $\upmu$m to produce configurations with fine features without failures. The printed objects displayed a porosity ranging from $44$\% to $49$\%. The authors reported a maximum compressive strength of $4.2 \text{±} 0.1$ MPa, an elastic modulus of $287.3 \text{±} 6.6$ MPa, and an average hardness value of $657 \text{±} 14$ HV. The authors concluded that the achieved strength value is sufficient for manufacturing structures or relevant replacement parts on the Moon, particularly considering the absence of storms. However, the Moon's lower gravity might negatively affect PBF-based printing processes. They also implemented the SLM method to 3D print Martian regolith \cite{goulas2017assessing}.

The processability of regolith from other Lunar regions has also been studied. In an investigation carried out by Caprio et al. \cite{caprio2020determining}, the L-PBF of NU-LHT-2M highlands regolith simulant was analyzed. They investigated the feasibility and limitations of the process and thoroughly examined the optimal base plate and energy conditions. They reported a maximum compressive strength of $31.4$ MPa, well above the properties achieved through alternative methods. The corresponding microhardness and porosity were measured at $680$ HV and $37$\%, respectively. These properties were attained using specific process parameters, including a laser power of $50$ W and a scan speed of $225$ mm/s. The study discussed atmospheric and gravitational conditions through a simplified theoretical framework, analyzing particle settling and free-fall times \cite{caprio2020determining}. However, these findings are purely theoretical and lack experimental validation at this point. Additionally, the study did not assess how these conditions affect the properties and qualities of the print; a more thorough evaluation is necessary. These aspects are fundamental for process stability and print quality due to the Rayleigh and Mie scattering effect \cite{steen2010laser}. Rayleigh scattering involves the dispersion of the laser beam in multiple directions by fine particles smaller than the laser wavelength, leading to inefficient energy transfer and inconsistent energy absorption in the powder bed. In the presence of atmospheric gases, gas molecules can further contribute to the dispersion of laser beams. Therefore, Rayleigh scattering is minimal in the vacuum environment. Mie scattering, on the other hand, refers to the phenomenon where a laser beam is scattered by suspended particles whose diameters are comparable to the wavelength of the incident laser. The suspended particulate matter comes from particle ejections near the melt pool and vaporized ejecta \cite{steen2010laser}. The presence of an atmosphere can affect Mie scattering by introducing additional particles like dust, which further disturb the laser beam. Thus, the vacuum conditions on the Moon might mitigate Mie scattering. However, Mie scattering can become more significant in the Moon's low gravity because of the higher concentrations of suspended powder particles. In other laser-based fabrication processes, any step that increases the density of suspended particles, such as the powder deposition step, can increase the Mie effect. For instance, in the Laser Directed Energy Deposition (L-DED) technique, the powder is actively blown from a nozzle, increasing the chances of particles becoming airborne and suspended in the air, thus enhancing the Mie effect. Despite their significance, the scattering effects in laser-based sintering of Lunar regolith have not been addressed in the literature. Caprio et al. studied the vacuum environment and noted substantial particulate outgassing during the fusion of the Lunar regolith simulant. They recommended implementing a strong gas recirculation system to prevent suspended particles from interfering with laser emission \cite{caprio2020determining}. However, such a system adds complexities while in extraterrestrial environments simplicity is preferable. Detailed experiments by Sibille et al. \cite{sibille2023vacuum} evaluated the strength properties and quality of sintered regolith under vacuum conditions. Their findings indicated that the processing temperature for FJS-1 basalt could be reduced by 100\textdegree{C} to achieve similar densification and compressive strength as in air. For the JSC-1A mare simulant, vacuum-sintered specimens showcased higher compressive strength at lower temperatures compared to those sintered in the air (152 MPa vs. 98 MPa). The sintering temperatures were 1100\textdegree{C} in vacuum and 1125\textdegree{C} in air. This reduction in temperature is attributed to the formation of iron and magnesium oxide layers on mineral grains in air, leading to higher processing temperatures than in vacuum or reduced atmospheres. For NU-LHT-2M Lunar highland regolith, test specimens processed at 1300\textdegree{C} and cooled at 1\textdegree{C}/min were fully sintered with 5.3-5.6\% open porosity. In a vacuum, samples with similar porosity were obtained at 1250\textdegree{C} \cite{sibille2023vacuum}.

Mechanical properties have also been reported for L-PBF-printed HIT-L-1 simulant with L-PBF parameters of a $75$ W laser power, a $4$ mm spot diameter, a $1$ mm/s scan speed, a $2$ mm hatch spacing, and a $1.2$ mm layer thickness \cite{wang2023additive}. These settings correspond to a volumetric energy density in the range of $3.5$ to $4.3$ $\textrm{J/mm}^3$. The glass-ceramics were formed at $823$\textdegree{C} during thermal post-treatment, and polycrystalline glass-ceramics with better mechanical properties were achieved at $1100$\textdegree{C}. The average compressive strength, fracture toughness, and Vickers hardness were measured at $50.71$ MPa, $1.49$ $\textrm{MPa}.\textrm{m}^{1/2}$, and $897.91$ HV, respectively. These enhanced mechanical properties, compared to those reported by Goulas et al. \cite{goulas20163d, goulas2019mechanical}, are attributed to the higher energy input, facilitating sufficient melting, improved fusion, and inter-layer bonding. This results in enhanced densification and overall improved mechanical properties. The relative porosity of the print was determined to be $40$\% \cite{wang2023additive}. While Wang et al. were successful in 3D printing Lunar regolith with favorable mechanical characteristics, their proposed processing method raises concerns regarding its energy consumption and sustainability in the Lunar environment. Additionally, their use of a scan speed as slow as $1$ mm/s makes the printing process exceedingly time-consuming.


Laser-based AM technology is capable of printing Lunar soil without any additives. Researchers were able to achieve high microhardness \cite{goulas20163d, goulas2019mechanical, fateri2013experimental, fateri2015process}, as well as compressive strength \cite{wang2023additive, caprio2020determining} for the printed regolith simulant. However, these prints are characterized by remarkably high porosity, primarily attributed to the high melting point of soil minerals that do not fully liquefy during laser processing \cite{azami2023laser, goulas20163d}. The low absorption coefficient of the soil and the problematic absorption of laser energy further contribute to this challenge, resulting in improper flowability and fluidity \cite{azami2023laser}. Moreover, the printed regolith is brittle due to the oxide ceramic elements it contains \cite{goulas20163d, indyk2017structural}. The inclusion of metallic particles demonstrated a considerable enhancement in the quality and characteristics of the final product \cite{azami2023laser, liao2021Lunar}.

In this regard, using the SLM method, Liao et al. \cite{liao2021Lunar} incorporated aluminum alloy into the Lunar regolith and produced a Lunar regolith-AlSi10Mg composite. The CAS-1 Lunar regolith, sieved through a $200$-mesh screen, was mixed with spherical AlSi10Mg powder with a mean particle size of $33.1$$\upmu$m in a 1:1 weight ratio. The process parameters were set as follows: laser power of $200$ W, scanning speed up to $800$ mm/s, layer thickness of $20$$\upmu$m, and a hatching space of $0.1$ or $0.12$ mm. The experiments were conducted within an inert argon gas atmosphere to prevent oxidation. Based on microstructure analysis, two types of defects were observed in the printed composites. Pore defects were predominant at low scan speeds, while lack-of-fusion defects occurred at high scan speeds, with other process parameters held constant. The maximum relative density of the fabricated test specimens reached $92.5$\% for an $800$ mm/s scan speed and a $0.12$ mm hatching space. Correspondingly, compressive strength testing of the Lunar regolith-AlSi10Mg composite yielded a strength of $264$ MPa. During the process, various metallurgical reactions were observed in the molten pool, leading to the loss of Lunar regolith. Aluminum reduced $\mathrm{SiO}_2$, FeO, and $\mathrm{TiO}_2$ in simulant Lunar regolith, while $\mathrm{Al}_2 \mathrm{O}_3$ was transformed into gaseous $\mathrm{Al}_2 \mathrm{O}$ at high temperatures.



L-PBF processes have also found applications for larger-scale projects on the Moon. Recently, a high-power $\mathrm{CO_2}$ laser was implemented for paving on the Moon and the creation of dust-free roads, with a maximum laser spot diameter of $100$ mm \cite{gines2023laser}. Large samples, each 250 \text{mm}\texttimes 250 \text{mm} with interlocking capabilities, were produced by melting the EAC-1A simulant directly on the powder bed using the laser. The process employed laser parameters of $3$ kW and a speed of $5$ mm/min in a single-layer fabrication approach. These printed samples can cover large areas of Lunar soil, forming roads and launchpads. The manufactured samples were analyzed for their mineralogical composition, internal structure, and mechanical properties. To ensure the printed samples could support the weight of heavy exploration equipment, their compressive strength was tested and compared to solar-sintered specimens from the RegoLight project \cite{fateri2019solar}. Compressive strength was measured using 10 \text{mm}\texttimes10 \text{mm}\texttimes 10 \text{mm} cubic samples. The results showed a compressive strength range of $56.19$ to $216.29$ MPa, with an average of $93.97$ MPa, approximately $50$ times higher than the $2.49$ MPa achieved in the solar-sintered project \cite{fateri2019solar}. However, a significant standard deviation of $55.88$ MPa was observed, likely due to internal defects such as porosity and micro-cracks within the samples \cite{gines2023laser}.

ICON and SEArch+ have developed design schematics with concepts for surface-site deployment, construction sequencing, and structural design for large-scale infrastructures for permanent Lunar settlements \cite{yashar2021project}. These designs include landing pads, roadways, habitats, shelters, and blast shields, and their double-protective "Lunar Lantern," all constructed using Lunar regolith and enabled by additive construction technologies. They have investigated various extrusion deposition and sintered-based additive methods. In collaboration with NASA KSC, ICON developed the Laser Vitreous Multi-Material (VMX) transformation process. This process autonomously deposits, compacts, and selectively preheats and sinters thin layers of prefiltered Lunar highland regolith. The system operates with 6-DOF robotic arms inside a Lunar surface simulation chamber that measures 6.4 m \texttimes 8.5 m in area with a bed depth of 0.9 meters \cite{ICON2023}. The Laser VMX vacuum-produced material showcased high performance, with compressive strengths of 344.7 MPa at 25\textdegree\text{C} and 251.66 MPa after one year of Lunar thermal vacuum (TVAC) cycles. Its flexural strength was 37.92 MPa, and the thermal expansion coefficient was $3.6 \text{×} 10^{\text{-6}} \, \text{\textdegree C}^{\text{-1}}$ over a temperature range from -150\textdegree\text{C} to 50\textdegree\text{C} in a vacuum. Simulations of the Laser VMX process are conducted to predict how the laboratory results will scale to actual Lunar surface conditions.



The feasibility of using the SLS method on the Moon to 3D print Lunar regolith has also been studied by Xu et al. \cite{xu20193d}. They conducted experiments to print CLRS-2 Lunar highlands regolith powder and ilmenite powder and investigated the impact of laser manufacturing parameters on the properties of the printed Lunar regolith. More pores were present in the printed CLRS-2 specimens compared to the ilmenite printed specimens, resulting in inferior mechanical properties for CLRS-2. To enhance these properties, CLRS-2 powder and ilmenite powder were pre-pressed at pressures of $10$ MPa and $40$ MPa to become denser. This preprocessing step reduced the porosity of the laser-sintered CLRS-2 but had no effect on ilmenite. Laser-sintered materials that underwent $10$ MPa pre-pressurization displayed much smoother surfaces and fewer porosities than samples without pre-pressurization.

The substantial energy density required for laser-based processes poses challenges for Lunar operations, given that energy production and storage remain issues in extraterrestrial settings. The electrical energy required for laser-based additive manufacturing is a major concern on the Moon. To enhance sustainability for Lunar operations, it is essential to rely on in situ resources for electricity generation to support laser-based additive manufacturing. Solar light, an abundant renewable resource, can be converted into electricity through photovoltaic (PV) cells, which are a standard method for energy conversion. In situ fabrication of solar PV cells has been proposed, as more than 90\% of solar cell materials are available on the Moon (e.g., Si, Fe, $\mathrm{TiO_2}$, Ca, and Al) \cite{ellery2022generating, criswell1998photovoltaics}. Manufactured solar cells consist of a Lunar glass substrate, a doped silicon layer between aluminum electrodes, and a $\mathrm{TiO_2}$ antireflective coating. Single crystal thin-film Si solar cells can potentially achieve efficiencies of around 25\%, though this necessitates the use of advanced deposition and surface coating technologies \cite{blakers2013high}. However, in situ fabrication of solar cells from extraterrestrial sources presents difficulties due to limited controllability, leading to efficiencies as low as 5–10\%  \cite{freundlich2005manufacture, ellery2022generating}. Moreover, these manufactured cells experience long-term degradation from radiation exposure and the high temperatures of the Lunar day \cite{girish2012photovoltaic}, which further reduces their energy conversion efficiency by an additional 30\%. An alternative solar-electric energy generation method proposed by Ellery \cite{ellery2022generating} involves leveraging Lunar resources for solar power production using a Fresnel lens-thermionic conversion system. This system can achieve conversion efficiencies around 15–20\%, with the potential for higher efficiencies through advanced techniques and improvements. Therefore, although solar light is abundant, the conversion of this energy into electricity for laser operations presents certain challenges, making laser-based additive manufacturing highly inefficient in the Lunar setting.

\subsubsubsection{EB-PBF}

Electron-beam PBF (EB-PBF), commonly known as EBM, replaces the laser with an electron beam to selectively melt and fuse powdered materials. EB-PBF operates in a vacuum \cite{sadeghi2019influence}, which is particularly well-suited for applications in space environments. While laser-based printing often results in materials with high porosity due to its sensitivity to the reflectivity of the processed material, impacting the overall mechanical properties of the final product \cite{azami2023laser, goulas20163d}, EBM is less influenced by the absorption characteristics of the material. However, EBM machines are typically more complex and expensive to operate and maintain, making them less accessible for Lunar missions. It also comes with a limitation that prevents its widespread application for LBMC: its applicability is limited to metals and is less effective for non-metallic materials \cite{howell2008site, mclemore2008Lunar}. This limitation can be attributed to the fundamental differences in how electrons interact with various types of materials. In response to this limitation, Howell and colleagues \cite{howell2008site} incorporated aluminum as a binder material to enhance the ductility of the material and improve its fluidity within the melt pool. They successfully printed Lunar soil simulants (JSC-1 and LHT-1M) aluminum composites via the EBM method. There is no information available about the mechanical properties of their prints. Despite the potential shown by Howell et al., there has been a notable absence of subsequent research on the topic. This lack of studies can be due to the limitation of EBM technology for processing soil (non-metallic) material. Another issue is the interaction of electrons with powder particles, which can lead to the emission of powders, particularly in the microgravity Lunar environment. Given these limitations and challenges,  there exists a research gap that must be addressed before this process can be used on the Moon. 

\subsubsubsection{SS-PBF}

Solar sintering PBF (SS-PBF), also known as SSLS, harnesses concentrated sunlight to fuse powdered materials. Taking into account the full potential of resources present on the Lunar surface, including solar light and Lunar regolith, while avoiding the use of additive binders in methods like binder jetting or direct ink writing, and direct sintering without the need for a laser or EB source, like those used in L-PBF or EB-PBF methods, represents a viable approach to on-site manufacturing, thereby reducing reliance on Earth-supplied materials and technology. 

The solar sintering of Lunar regolith was initially limited to sintering a thin layer of Lunar regolith to fabricate launchpads, aiming for surface stabilization and dust mitigation \cite{kayseworks,nakamura2011solar, cardiff2008dust} during take-off and landing of missions, which can cause problems for nearby infrastructures and human activities on the Moon \cite{gaier2007effects, wagner2006apollo, noble2007assessing, khan2008Lunar}. Hintze et al. \cite{hintze2009Lunar} constructed a $1$ $\mathrm{m^2}$ solar concentrator, comprising a Fresnel lens installed on a frame designed to track the sun and collect solar energy for sintering JSC-1 regolith simulant. The highest measured temperature produced by the concentrator was $1350$\textdegree{C}, well above the temperature required to melt the regolith simulant. Their work also discussed the physical and mechanical assessments, including load strength and abrasion resistance, both in the field and for the laboratory-prepared sintered test specimens. Their findings showcased the potential of the proposed method, which uses a simple and cost-effective technology for sintering Lunar regolith and preparing large launch pads.

Meurisse et al. \cite{meurisse2018solar} and Fateri et al. \cite{fateri2019solar}, in their collaborative works, demonstrated the feasibility of printing basic geometries through selective solar sintering of Lunar regolith using mirror-based solar collectors. Their setup included artificial sunlight-Xenon lamps and a water-cooled mirror inclined to 45\textdegree \, onto which the horizontal irradiated light was projected to adjust the light beam orientation and direct it vertically onto the build platform. These two Xenon lamps collectively delivered a concentrated light density of $1.2 \, \text{MW/m}^2$ onto a $20$ mm focal point. The lamps were calibrated to obtain a sharp beam with the smallest focal point. The as-received JSC-2A powder, resembling Lunar mare soil, was dispersed in $100\upmu$m-thick layers and sintered layer upon layer through the selective solar sintering process. Tests were carried out under both ambient conditions and in a vacuum. Bricks were printed with a scanning speed of $47 \, \text{mm/s}$, a layer thickness of $100\upmu$m, a hatch spacing of $15$ mm, and a focal spot of $12$ mm in the ambient condition. However, these printing parameters led to a molten state for the layers and a porous brittle medium for the print in the vacuum environment. The processing parameters were subsequently adjusted for the vacuum condition and set at $1.0 \, \text{MW/m}^2$ for lamp intensity and a scanning speed of $65 \, \text{mm/s}$. Morphological analysis of the printed cubes revealed the presence of open pores, some of which measured up to $0.60$ mm in size, indicating weak interlayer bonding. Additionally, closed porosity with dimensions of up to $0.45$ mm was observed due to the outgassing of minerals at the melting temperature. Solar sintered materials under ambient atmosphere were characterized by compressive strength of $2.49$ MPa and Young's modulus of $0.21$ GPa. For a given set of sintering parameters, smaller sintered components displayed greater strength compared to larger ones. The assessment of vacuum-sintered parts was omitted. It was speculated that the absence of an atmosphere might adversely affect the sintering process due to a substantial release of gases from the highly molten material, causing the fracture of the initial sintered layers or resulting in uneven surface textures. On the other hand, the vacuum environment might enhance mechanical properties by preventing the oxidation of Lunar minerals and facilitating a more uniform distribution of heat within the sintered material, improving material strength and reducing thermal stresses.

Outward Technologies utilized concentrated solar energy to power powder bed fusion (PBF) and fused deposition modeling (FDM) processes \cite{solar2023}. These methods were used to fabricate volumes of 1 \text{m} \text{\texttimes} 0.3 \text{m} \text{\texttimes} 0.1 \text{m} and various lab-scale components from CSM-LHT-1 Lunar highlands regolith simulant. The components included rover parts and structural elements like trusses for bridges, towers, and ramps. The sintered specimens achieved 95\% density. The fused regolith demonstrated a compressive strength of $25$ MPa and Young's modulus of $10.3$ GPa, comparable to M25 concrete. Additionally, its stress-strain curve revealed 20\% more strain at failure and greater toughness than M25 concrete. The flexural strength of the PBF-printed specimens was measured at $3.5$ MPa. This solar-powered PBF process could be suitable for large-scale manufacturing, potentially replacing concrete-type infrastructure on the Moon. However, the solar sintering AM presents certain geometric limitations due to the low precision of its beam focus, making it inappropriate for creating highly complex designs with fine details, acute angles, and smooth surface finish. As a bulk method, solar sintering is a simple and cost-effective technology, ideal for sintering Lunar regolith to prepare large launchpads for dust-free stabilized road and landing pads, as well as for fabricating less detailed parts \cite{hintze2009Lunar, kayseworks, nakamura2011solar, cardiff2008dust}. In contrast, laser-based and electron beam sintering technologies use high-powered energy beams with precise energy delivery in controlled environments, ensuring consistent quality and enabling intricate details with fine resolutions \cite{fateri2013experimental, fateri2015process, goulas20163d}. Notably, laser-based additive manufacturing methods have showcased a promising dimensional accuracy of 99.8\% \cite{goulas20163d}. Long sintering processes can also be challenging, particularly with the sintered parts' thermal stresses. Slow scanning speed and rapid cooling resulted in stress buildup, causing the sintered part to warp during the process. The most effective strategy involved alternating scanning of successive layers along the length and width of the part and sintering a contour before each layer to minimize warping of the part edges.

    
    

\subsubsubsection{MWS-PBF}

The Microwave sintering PBF, also known as SMWS, is a method that involves subjecting powdered material to controlled microwave radiation to induce heating and initiate the sintering process. Microwave heating is currently recognized as a more practical approach for constructing structures and extracting resources on the Moon compared to solar \cite{fateri2019solar, meurisse2018solar} and laser sintering \cite{fateri2015process} methodologies. This preference is due to the volumetric heating characteristic of microwave processes, which consumes approximately $23$\% less energy than laser sintering and in shorter fabrication times \cite{lim2019numerical}. Microwave radiation can penetrate the material deeply (up to $65$ cm), making it an efficient method for subsurface heating \cite{fateri2019localized, taylor2005microwave}. This ensures a more consistent sintering process and minimal temperature gradients across the targeted material, ultimately improving the mechanical properties of the sintered material.

Based on successful microwave sintering of Lunar regolith in previous studies \cite{taylor2010apparatus, taylor2010mineralogical, taylor2005microwave, taylor1988generation, barmatz2013microwave, barmatz2011microwave, barmatz2012microwave}, the SMWS method was suggested for on-site additive manufacturing on the Moon. NASA researchers in \cite{barmatz20143d, howe2013faxing} developed a 3D microwave print head facility using a $2.45$ GHz wave generator to explore the feasibility of additive manufacturing using Lunar and Martian simulants. This involved integrating a temperature-resistant quartz tube vertically through the cavity. The JSC-2A simulant was gradually fed through the tube and heated beyond its melting point to enhance the flow rate. The resulting melted simulant exited the tube to be selectively placed onto a substrate to form walls or pave a road. The researchers understood how to control the sintering and melting of the target material subjected to microwave radiation and determined optimal operating parameters for large-scale manufacturing. Various operating parameters such as power, flow rate, tube diameter, and location were adjusted based on sample size and the estimated temperature dependence of the dielectric constant to optimize performance. Evaluation extended to Lunar highland and mare simulants as well as Mars soil simulants. To verify whether the enhanced heating effect was associated with a chemical interaction with oxygen in the surrounding atmosphere, measurements were repeated on a Lunar JSC-2A simulant with a nitrogen atmosphere surrounding the cavity, yielding consistent results. Consequently, the remaining studies were then performed in Earth's atmosphere. To establish a suitable construction procedure, it is essential to ensure that regolith simulants' particle size and shape distribution align closely with those of authentic Lunar regolith \cite{lim2017extra}. A thorough examination of the necessity for nanophase iron in Lunar simulants is also advised. 

Although promising for additive manufacturing on the Moon, microwave sintering takes a different path in the fabrication process compared to laser and solar sintering methodologies, which are typically used for layerwise sintering and material addition to building up geometries. This divergence is attributed to the challenges associated with fine-tuning and controlling process parameters to achieve the desired material properties and print resolution. The relative lack of attention to this specific method for Lunar additive manufacturing could hold potential for future research and development. Microwave radiation's deep penetration into materials allows for processing within the volume, ensuring an even distribution of heat throughout the substance. Reduced temperature gradients throughout the sintered material minimize the risk of fracture and pore formations. This characteristic makes microwave heating a promising method for thermal post-treatment to alter the microstructure and mechanical properties of pre-sintered materials \cite{kim2021microwave, kim2021microstructural}.

\subsubsection{Material Extrusion}
Material Extrusion (ME) is categorized within ASTM's classification of AM technology, officially defined as a specific additive manufacturing process where the material is deposited through a nozzle or orifice \cite{principles2015iso}. Fused Filament Fabrication (FFF), also known as Fused Deposition Modeling (FDM), is a major subcategory of ME and one of the most widely adopted AM techniques overall. Direct Ink Writing (DIW) is another significant subcategory of ME. Another intriguing category within ME, particularly relevant for AM using regolith, involves the extrusion of molten regolith, typically produced as a by-product during regolith oxygen extraction.

\subsubsubsection{Fused Filament Fabrication}
Fused Filament Fabrication (FFF) is one of the most commonly used AM techniques for manufacturing objects from a filament, usually made of thermoplastic polymers. This method involves a layer-by-layer process in which the 3D printer warms up the thermoplastic material until it loses enough viscosity to be deposited in bead-like patterns to build individual layers \cite{pulipaka2023effect,kazemi2022uncertainty}. FFF is recognized as one of the most cost-effective AM techniques, requiring lower energy levels compared to other AM methods, minimizing the waste of materials, and being very effective in terms of manufacturing time \cite{arif2018performance, ding2019effects, subramaniyan2021state}. Also, it does not involve the usage of any liquid-phase material. It has also been proven to work well in low-gravity or zero-gravity conditions \cite{prater2016summary, werkheiser2015space}. All these features make it a promising method to be used in a Lunar environment.

However, considering the extreme heating/cooling cycles that occur (see \ref{fluctuation}), along with the outgassing requirements \cite{william1993outgassing}, not all materials can be used as feedstock for FFF on the Lunar surface. High-performance thermoplastic polymers like Polyether ether ketone (PEEK) \cite{rival2019influence}, Polyether ketone ketone (PEKK) \cite{siarov2018additive}, and Polyetherimide (PEI) \cite{kern1993simulated}, along with their composites containing glass fiber \cite{liu2022mechanical}, carbon fiber \cite{flanagan2017permeability, kern1993simulated}, and ceramic particles \cite{goyal2008microhardness}, have gained particular attention for in-space manufacturing. The reinforcement part can be added to improve the material's specific properties or make it easier to fabricate. For example, adding chopped carbon fiber can enhance the mechanical strength of the polymers \cite{stepashkin20183d} and simplify the printing process by forming a more uniform heat distribution in the part while printing, reducing the probability of warpage caused by thermal stresses \cite{yang2021effects}. Despite ongoing research in space agencies, including NASA \cite{townsend2022print}, and academia, there remains a research gap in the literature concerning ISRU-based FFF. This method involves incorporating on-site materials, such as Lunar regolith, into a space-grade polymeric binder to reduce the cost of raw material launches and enhance the properties of the final product. When it comes to AM of polymer/regolith, in comparison to methods with the potential to use 100\% ISRU-based materials, such as PBF, it is important to note that this method is not entirely reliant on ISRU and necessitates the use of off-site materials. Nonetheless, the inclusion of polymer in the composition of the fabricated component offers several advantages. The polymer can sustain stresses and transfer them to reinforcing fiber or other agents. The beneficial effects of the polymer on ductility, toughness, and elongation have been extensively validated \cite{azami2023laser, kazemi2022overall, facenda2018literature}.

Gelino et al. \cite{gelino2024selection} studied various regolith/Polylactic Acid (PLA) formulations. They tested mixtures including 70:30, 80:20, and 85:15 weight percent ratios of Lunar mare simulant Black Point-1 (BP-1) with PLA, an 80:20 mixture of Lunar Highlands Simulant-1 (LHS-1) with PLA, and an 80:20 BP-1 with PLA blend that included a flow enhancer additive. These samples were printed under simulated Lunar dirty thermal vacuum conditions (-190\textdegree{C}, $10^{\text{-3}}$ torr). Each formulation underwent several tests to evaluate the achieved mixture ratio, mechanical strength, outgassing products during vacuum printing, and the porosity and density of the printed items. The LHS-1/PLA composite exhibited a favorable combination of properties and was used in the AM of the samples and the subsequent tests. The 80:20 LHS-1 weight percent formulation was selected as the preferred material because the regolith simulant closely represents the south polar Lunar regions and the composite has acceptable mechanical properties. This formulation achieved a flexural modulus of 5.3 GPa and flexural strength of 24 MPa in the 0\textdegree \, orientation. Despite the promising results, PLA is not suitable as a space-grade polymer and does not meet NASA's outgassing standards or the temperature resistance requirements for Lunar conditions. Azami et al. \cite{Azami2024additive} investigated the FFF process involving PEEK/Lunar regolith composites and compared them with pure PEEK and PEEK/carbon fiber blends. They noted that higher solid material content posed challenges during extrusion, leading to increased sample porosity. The incorporation of 20 wt\% carbon fiber improved tensile strength by 8.37\%. Conversely, the introduction of 15 wt\% and 30 wt\% Lunar regolith reduced tensile strength by 14.63\% and 26.78\%, respectively. Additionally, increasing the amount of regolith heightened the brittleness of the fractures and decreased elongation at break.

PEEK is particularly well-suited for use in extraterrestrial environments, especially Lunar conditions, due to its 3D printability, high glass transition temperature Tg (143\textdegree{C}), and operating temperature (250\textdegree{C}). It also offers high tensile strength (around 100 MPa) and low density ($1320 \, \text{kg/m}^3$), meets NASA and ESA's outgassing requirements, high radiation and chemical resistance, and a semi-crystalline structure. These characteristics make it a potential alternative to metals in some applications and a suitable choice for structural and mechanical applications \cite{zanjanijam2020fused, kilroy2008mechanical, rival2020electronic}.

However, printing PEEK can be challenging due to its semi-crystalline nature, high melting point, and high melt viscosity. Printing PEEK in a heated environment is usually recommended, at a minimum of 70 \textdegree{C}, preferably above 150 \textdegree{C}. Given the high printing temperature of PEEK, printing at room temperature increases the cooling rate, causing significant temperature gradients and thermal stresses in part, resulting in warpage, low fabrication accuracy, poor layer adhesion, or even delamination thereafter. Therefore, it is typically printed on a heated bed and in a heated environment, with a low printing speed \cite{zanjanijam2020fused, liu2023material, geng2019effects}. In some studies, the as-fabricated PEEK samples underwent an annealing heat-treatment phase to allow them to recrystallize and improve their mechanical properties \cite{ostberg1987annealing, basgul2020does, yang2017influence}. However, when printing in a vacuum environment, due to the lack of a convection cooling mechanism, the cooling rate is lower compared to a non-vacuum ambient, resulting in lower temperature gradients. Liu et al. \cite{liu2023material} showed that similar or slightly better tensile properties can be expected when printing PEEK in a vacuum, compared to the results obtained with a 150 \textdegree{C} ambient air temperature by Yang et al. \cite{yang2017influence}. This is a promising conclusion for Lunar-based manufacturing, facilitating the printing requirements in a vacuum.

Liu et al. \cite{liu2023material} employed a 100 MPa vacuum environment for printing PEEK samples using FFF. They observed an improvement in the crystallinity of the samples, increasing from 14.9\% to 27.8\%, representing an 86.6\% increase. This rise in crystallinity resulted in a shift in fracture behavior, transitioning from ductile failure to brittle failure, as anticipated. The parts printed in the vacuum environment achieved a relative density of 99.37\%.

One remaining challenge in the FFF of materials, particularly PEEK, is the significant anisotropy in different directions. While the strength in the plane parallel to the layers (X-Y plane) is usually acceptable, parts exhibit lower strength levels in the direction perpendicular to the layers (Z direction) \cite{rinaldi2018additive, arif2018performance, gao2021fused}. This is attributed to the weaker bonding between adjacent layers compared to the materials in the same layer. This presents challenges for specific applications that require high strength in all directions \cite{gao2021fused}. Although research is being conducted to reduce this anisotropy in FFF parts \cite{ghorbani2022eliminating}, the problem persists to some extent. Designers typically consider this limitation when designing for AM.

In a study conducted by Arif et al.~\cite{arif2018performance}, the effect of build orientation on the mechanical properties of PEEK (Victrex\textsuperscript{\textregistered} 450G) samples was investigated. It was found that the tensile strength of the samples with building orientations and raster angles of horizontal/0\textdegree, horizontal/90\textdegree, and vertical/90\textdegree \,is $82.58 \text{±} 1.03 \, \text{MPa}$ (at yield), $72.88 \text{±} 1.92 \, \text{MPa}$ (at break), and $9.99 \text{±} 0.94 \, \text{MPa}$ (at break), respectively, which represents a reduction of more than 87\% from horizontal/0\textdegree \, to vertical/90\textdegree \, samples. In comparison, the tensile strength of molded PEEK is $98 \, \text{MPa}$ (at yield). Furthermore, the horizontal/90\textdegree \, and vertical/90\textdegree \, samples exhibited brittle fracture behavior, with tensile elongation percentages of $2.91 \text{±} 0.14\text{\%}$ and $0.33 \text{±} 0.03\text{\%}$, respectively. In contrast, the horizontal/0\textdegree and molded samples displayed ductile fracture characteristics, with tensile elongation percentages of $110.97 \text{±} 5.31\text{\%}$ and $45\text{\%}$, respectively \cite{arif2018performance}.

Polyetherimide (PEI) is another high-performance thermoplastic polymer that, in contrast to PEEK, lacks a distinct crystalline structure. The presence of imide groups imparts impressive mechanical and thermal properties to PEI. Additionally, the inclusion of ether groups within the polymer's structure introduces chain flexibility, contributing to its favorable printability in the molten state. It should be noted that when compared to PEEK, PEI shows a reduced service temperature range \cite{wu2022characterization, sundarram2015preparation}.

PEI has a $1.27 \, \text{g/cm}^3$ density and a tensile stress (at yield) of 110 MPa. PEI boasts a wide processing temperature range, spanning from 350\textdegree{C} to 425\textdegree{C}, and exhibits a high glass transition temperature, ranging from 217\textdegree{C} to 250\textdegree{C}. Consequently, PEI finds applications in various industries, including aerospace. Similar to PEEK, its outgassing values fall within the allowable range specified by NASA for in-space processing. Furthermore, owing to its favorable melt viscosity, PEI has gained popularity as a material of choice for AM techniques, such as FFF. ULTEM\textsuperscript{\textregistered} is one of the most renowned PEI brands, initially introduced to the market by General Electric in 1982 \cite{wu2022characterization, sundarram2015preparation, caldona2021additively}.

Despite its lower operating temperature compared to PEEK and its high tendency to absorb moisture, which can result in print quality issues \cite{vina2000effect}, PEI offers several advantages over PEEK, particularly in the context of space applications. PEI is often considered easier to work with for 3D printing due to its lower printing temperature, improved layer adhesion at similar ambient temperatures, and lower melt viscosity. This also makes it suitable for open printing environments \cite{abderrafai2023additive}. Another advantage of PEI is its inherent flame resistance. PEI boasts a high Limiting Oxygen Index (LOI) and carries a UL-94 V-0 rating \cite{wu2022characterization}. Additionally, PEI can be transparent, a property that proves advantageous in applications where visibility or light transmission is required \cite{sundarram2015preparation}. Furthermore, PEI tends to be more budget-friendly, although this advantage may be marginal for space-related applications when compared to other incurred costs. It should be noted that PEI and PEEK are miscible, allowing for the blending of the two to create PEI/PEEK filament with desired properties \cite{el2021preparation}.

Similar to PEEK, a significant number of experimental \cite{wu2018multi} and modeling \cite{gilmer2019characterization} research studies have been conducted on the FFF of PEI and its composites. As an example, Gilmer et al. \cite{gilmer2021temperature} utilized finite difference models to concurrently assess heat transfer, polymer diffusion (quantified as the degree of healing, Dh), and the emergence of residual stress during FFF of PEI material. Their findings indicated that the stress development in FFF demonstrates periodic fluctuations over time, influenced by temperature profiles. They concluded that variations in deposition and bed temperatures had a limited impact on stress development, whereas layer deposition time, particularly concerning print speed and duration above Tg, exerted significant influence \cite{gilmer2021temperature}.

The printability of high-performance thermoplastic polymers is significantly dependent on their rheological characteristics. Ajinjeru et al. \cite{ajinjeru2018influence} assessed the impact of introducing carbon fiber (CF) into PEI. They observed an increase in the viscosity of the composites (by 2.5\texttimes \, for 20 wt\% CF and 3\texttimes \, for 30 wt\% CF) and a strengthening of the shear-thinning effect. This noticeable variation in viscosity for PEI composite materials when exposed to varying shear rates and temperatures provides a wide processing range for their use as large-scale FFF feedstock materials. This adaptability can be achieved by adjusting either the screw speed or the processing temperature \cite{ajinjeru2018influence}. In another study conducted by researchers affiliated with NASA \cite{chuang2015additive}, the feasibility of manufacturing large-scale complex aircraft parts using Ultem 9085 and Ultem 1000/10 wt\% chopped carbon fiber was investigated. The study concluded that the addition of just 10\% carbon fiber content significantly increased the viscosity of the materials, making printing more challenging and resulting in higher porosity in the fabricated samples. While the initial Ultem 1000/CF blend showed porosity ranging from 0\% to 15\%, the filaments produced through extrusion and the Ultem 1000 composite vanes printed via FFF exhibited approximately 25\% porosity. This increased porosity can also be attributed to the presence of trapped moisture or outgassing due to degradation at the high printing temperature of 420\textdegree{C} in the FFF process. In contrast, the Ultem 9085 filament produced parts with 5 to 8\% porosity when printed at 375\textdegree{C} \cite{chuang2015additive}.

While incorporating Lunar regolith into PEI can potentially reduce manufacturing costs on the Moon, a research gap exists in exploring its impact on mechanical properties, printing quality, and determining the upper limit for regolith content.

Another promising material for use in space industries and in-space manufacturing is Polyetherketoneketone (PEKK). In comparison to PEEK, PEKK features an additional ketone group in its chemical structure. While PEKK shares many similarities in properties with PEEK, there are a few noteworthy differences. PEKK offers superior thermal stability, boasting a higher glass transition temperature (Tg) at around 162\textdegree{C} in contrast to PEEK, with a Tg of about 143\textdegree{C}. Additionally, PEKK has a lower melt viscosity, making it easier to work with during printing. Furthermore, the printing temperature of PEKK is slightly lower than that of PEEK. However, due to PEEK's earlier introduction to the market and its established track record, PEKK has remained somewhat less popular when compared to PEEK \cite{yuan2018comparison, alqurashi2021polyetherketoneketone, perez2021crystallinity}.

While research on the FFF of PEKK is not as extensive as that on PEEK, a body of research is dedicated to the FFF of PEKK and its composites. Rabinowitz et al. \cite{rabinowitz2023taguchi} employed the Taguchi method to optimize FFF process parameters to maximize the flexural strength of PEKK samples. This study revealed that varying parameters such as raster angle and layer height significantly impacted the flexural strength of carbon fiber-reinforced PEKK. Optimized settings resulted in a flexural strength of $111.3 \text{±} 5.3$ MPa and a flexural modulus of 3.5 GPa. Furthermore, post-processing through pressing significantly improved both the samples' flexural strength and flexural modulus \cite{rabinowitz2023taguchi}.

In two papers published by Rashid et al. \cite{rashed2022effects, rashed2023optimization}, they investigated the impact of various process parameters, including build orientation, infill pattern, number of contours, and raster angle, on the properties of PEKK samples. Their research revealed that build orientation had the most significant influence on tensile properties (similar to the results obtained in \cite{arif2018performance} for PEEK), followed by the number of contours, resulting in variations in Young's modulus and elongation at break. Additionally, the studies found that differences in porosity, rather than crystallinity, were responsible for the variations in mechanical properties among the different sample groups, emphasizing the importance of carefully optimizing FFF process parameters for desired material performance. The research findings also highlighted the significant influence of the number of contours on porosity and its substantial impact on a range of mechanical properties. Specifically, increased contours were associated with reduced porosity and improved mechanical and dynamic mechanical properties \cite{rashed2022effects, rashed2023optimization}.

One notable project is the Marsha initiative by AI SpaceFactory, aimed at creating sustainable habitats for human habitation on Mars using locally sourced Martian materials. The habitat features a vertical, cylindrical design optimized for Martian conditions, ensuring stability, efficient spatial utilization, and protection against the planet's harsh weather and radiation. Marsha utilizes a biopolymer-basalt composite, facilitating its construction using a form of Fused Filament Fabrication (FFF) process controlled by robotic arms \cite{spacefactory_marsha, anderson2021solar, roman20203d}. Marsha achieved first place in NASA's 3D-Printed Habitat Challenge, Phase 3 (2019) \cite{designboom_marsha}. Similarly, AI SpaceFactory's Lina project focuses on developing sustainable habitats for the Moon, employing advanced 3D printing technologies akin to those used in the Marsha project for Mars. Lina aims to construct durable, livable structures capable of withstanding the Lunar environment, using Lunar regolith-based materials to reduce dependence on Earth-sourced resources. LINA will be 3D printed using a high-performance mixture of Lunar regolith and Earth-sourced polymer binder, exploring various configurations to optimize between low reliance on off-site polymers and load-bearing capacity. The project uses BP-1 Lunar regolith simulant and employs a unique 60-degree printing angle to create continuous, vaulted roofs. A prepared regolith berm acts as an inclined print bed, supported by reusable metal tiebacks to anchor initial layers and prevent warping during cooling. As construction progresses, a mobile excavator follows to cover LINA with a protective regolith overburden. Unlike Marsha, designed as a freestanding structure on Mars, LINA is designed to expand from a single unit into a cluster, forming a larger, interconnected Lunar outpost. Its orientation provides natural self-shading, utilizing Lunar topography to shield inhabitants from solar and cosmic radiation hazards \cite{spacefactory_lina}.

Additionally, research has been conducted on printing metals or ceramics using the FFF technique. In this process, after preparing a metallic \cite{wagner2022filament, wagner2022fused} or ceramic \cite{hadian2021material, clemens2023material} filament that contains a polymeric binder, it is printed using the FFF technique. Subsequently, the as-printed samples (known as green bodies) undergo a polymer removal process, typically referred to as debinding, followed by sintering. Ultimately, this results in metallic or ceramic samples \cite{hadian2021material}. This method also holds promise for Lunar-based manufacturing, particularly for printing regolith samples. However, if the process is carried out in a Lunar outdoor environment, there will be limitations in the choice of binder polymer due to outgassing requirements \cite{william1993outgassing}. It is worth mentioning that the use of recyclable polymeric binders is advantageous, as it helps reduce the cost of raw materials delivery \cite{fateri2018feasibility}.

All in all, the most crucial problem for Lunar-based FFF is that carbon is very scarce on the Moon (around 142-226 ppm by mass, based on Apollo 11 data \cite{moore1970total}), and polymers cannot be considered a material that can be developed on-site. This makes the 3D printing/extrusion of polymer-based materials on the Moon challenging. However, there are some points that make on-site polymer manufacturing important. The first case is when a part is needed on-site urgently, and fabrication and transportation from Earth (lead time) can be lengthy. This necessitates enhancing the capabilities of manufacturing, particularly using FFF, on-site. Second, as discussed above, enabling the fabrication of polymer/regolith composite parts through materials extrusion can make the process considerably more economical by greatly reducing the mass launched from Earth. Lastly, FFF offers great potential for fabrication using repurposed thermoplastics. Therefore, conducting research on this topic is crucial.

\subsubsubsection{Fused Granulate Fabrication}

Fused Granulate Fabrication (FGF) is an AM technique that utilizes granular materials as the primary feedstock for object fabrication. Sometimes referred to as 'Fused Granular Fabrication,' this method diverges from traditional filament or resin-based approaches by employing granules or pellets composed of various materials, including plastics, metals, or ceramics. In this process, granules are heated to their melting point and then extruded through a nozzle, closely mirroring the fundamental principle of FFF. The extruded material is deposited layer by layer to form the desired object \cite{kalle2023potential, tagscherer2022mechanical}.

FGF offers several advantages, including material versatility, simplification of filament manufacturing, and the potential for cost savings through the use of readily available granules instead of specially manufactured filaments. Moreover, when compared to filament-based techniques, FGF typically expedites the fabrication process, enabling the rapid production of large-scale components. FGF can achieve higher throughput in mass production scenarios due to its continuous extrusion process \cite{kalle2023potential, tagscherer2022mechanical}.

As indicated in the literature \cite{sarasua1997recycling}, whenever a material undergoes a process involving melting, it can adversely affect its properties, including tensile Young's modulus, strength, and impact strength. This is primarily due to chain scission and crosslinking (when done in the atmosphere) \cite{zaghdoudi2019scission}. One significant advantage of FGF over FFF is its potential to eliminate the melting step associated with filament production. This advantage becomes even more pronounced for in-space manufacturing, where recycling becomes necessary due to the high material delivery costs. This capability allows for greater material recycling over multiple cycles.

However, it is important to note that FFF still holds specific advantages over FGF. FFF often delivers higher print resolution and surface quality because it relies on pre-manufactured filaments with consistent properties. Furthermore, FFF 3D printers are typically less complex because they do not require the additional equipment necessary for granulate processing and extrusion \cite{kalle2023potential, tagscherer2022mechanical}. Another issue for FGF arises when attempting to let pellets or other initial material fall naturally through a funnel, as the lower gravity reduces the efficiency of this process.

The choice between FFF and FGF hinges on specific project requirements. FGF excels in scenarios where material flexibility, cost-efficiency, and rapid manufacturing are critical, particularly for large-scale parts. In contrast, FFF is preferred for applications that require higher print quality and simplicity. In contexts such as in-space manufacturing, where energy resources may be limited, the advantages of FFF, with its streamlined complexity and finer resolution, become particularly valuable \cite{kalle2023potential, tagscherer2022mechanical}.

As a notable research on FGF using space-grade materials, Martin-Perez et al. \cite{martin2022fused} investigated the processability of company scrap LMPAEK containing 40\% short carbon fiber through FGF, providing a sustainable solution for aircraft structural parts. Similar to PEEK and PEKK, Low-melt Polyaryletherketone (LMPAEK) is a high-performance polymer developed by Victrex from the Polyaryletherketone (PAEK) family. It is designed to exhibit thermal, mechanical, tribological behavior, and chemical resistance similar to PEEK but with a less complex manufacturing process due to its lower melting point \cite{Victrex}.

Similar to other fabrication techniques, ISRU is a necessity for space-based FGF. However, ISRU-based FGF has received relatively less attention. Researchers at Polytechnique Montreal conducted a study on FGF using PEI/Lunar regolith simulants under the supervision of the Canadian Space Agency (CSA) \cite{alphonius2023Lunar}. Still, further research needs to be conducted on the FGF of polymer/regolith to assess the real potential of this technique for in-space manufacturing.

\subsubsubsection{Extrusion of Molten Regolith}

One potential option for in-space additive manufacturing involves extruding molten regolith to achieve the desired shape. This approach is of particular interest because molten regolith/ceramic/glass represents one of the few viable options for extrusion without polymer binders. Although accessing the equipment and energy necessary for regolith melting on the Moon poses challenges, these resources could potentially be obtained from specific candidate metal/oxygen extraction processes that require regolith melting. As discussed in Section \ref{Extraction}, metals can be extracted as a by-product of the process. Another possibility is to directly use the molten material created during the oxygen extraction process for additive manufacturing. As an example of existing/ongoing research, Kilncore, a company based in Quebec, Canada, has introduced and patented a technology that combines molten regolith electrolysis and AM, enabling better efficiency than when these processes are carried out separately \cite{missout2022flow, kilncore}.

Lunar regolith has a wide range of compositions and properties (see Section \ref{regolith}). Consequently, there is no single, uniform melting point for it. The melting point of Lunar regolith materials can vary depending on their specific composition and the presence of various minerals. Generally, its melting temperature range is 1100\textdegree{C} to 1380\textdegree{C}. Above 1380\textdegree{C}, the regolith is completely molten \cite{langseth1973heat}. Therefore, finding a temperature suitable for printing would be challenging, and a trade-off should be made between the fabricated part's characteristics and energy consumption. Additionally, the presence of various materials in the molten substance can complicate and make it challenging to control the solidification phase.

In a study carried out at NASA's Kennedy Space Center, Mueller et al. \cite{mueller2014additive} melted and extruded BP-1 basalt regolith and JSC-1A Lunar regolith simulant using a robotic arm to explore the potential of extrusion-based additive manufacturing for Lunar-based construction. They conducted an analysis of flexural strength using a 3-point bending test. Their findings revealed that the strength of the samples surpassed that of typical residential concrete and was comparable to certain types of less durable glass materials \cite{bon2003structural}. Despite these promising results, the researchers concluded that the technology is still in its early stages of development and that further research is necessary to fully realize its potential \cite{mueller2014additive}.


This technology holds great promise for in-space manufacturing. One key advantage is its potential for 100\% ISRU-based fabrication, providing independence from off-site materials. Additionally, as mentioned, it utilizes the molten regolith already produced for oxygen extraction, which helps energy efficiency. However, compared to AM of regolith with a polymeric binder, this method offers less control and requires significantly more power. All in all, limited research has been conducted on this technology, and it is still in its early stages. Conducting more research and assessing the characteristics of the fabricated components will provide additional evidence to evaluate the true potential of this technology.

\subsubsubsection{Direct Ink Writing}

Direct ink writing (DIW) is a process in which material particles are finely dispersed within a liquid solvent, extruded through a printhead, and selectively placed onto a substrate \cite{chen20193d}. The printed materials can solidify through proper drying and sintering post-processing steps. It has been proven that microgravity has a minimal impact on DIW prints \cite{prater20193d, werkheiser20143d}. Furthermore, due to the utilization of chemical reactions, this method has the lowest energy requirements compared to melting and sintering. The DIW method has shown good precision throughout the printing process, resulting in high-quality finished parts.



Contour Crafting (CC) technology was an early attempt by NASA, the US Army Corps, and the University of Southern California to build large-scale domes and walls for the Lunar surface \cite{khoshnevis2006mega, khoshnevis2005Lunar}. CC relies on a low-energy process that enables faster fabrication processes and superior surface finishes compared to other rapid prototyping methods. This technology is compatible with a diverse range of materials, including thermosets, thermoplastics, photopolymers, metal and ceramic pastes mixed with binders like water or sulfur, as well as unconventional rapid prototyping materials like plaster, cement, clay, and concrete \cite{khoshnevis1998innovative}. The CC nozzle barrel can extrude viscous, aggregate-filled substances. Khoshnevis et al. performed preliminary studies on the feasibility of CC technology with sulfur concrete as an alternative binding agent to water, which requires much less energy than competitive construction materials and is 100\% recyclable \cite{khoshnevis2012contour,khoshnevis2016construction}. The sulfur concrete comprised about 80\% regolith and 20\% sulfur. The mixture is pushed inside the hot barrel of the CC nozzle, heated to 130\textdegree{C}, which is the melting point of sulfur. The sulfur concrete structures could achieve a compressive strength of 21 MPa, sufficient for construction purposes \cite{khoshnevis2012contour}. They reported that a 35\% sulfur concentration by mass enables continuous extrusion and high-quality walls. However, challenges arise due to the low melting point of sulfur, which can cause sulfur concrete structures to melt during Lunar days, limiting its applicability in Lunar conditions \cite{khoshnevis2012contour, grugel2008sulfur, rahman2023towards, vaniman1992uses}. Additionally, the strength of sulfur concrete may not be adequate for some high load-bearing sub-structures \cite{khoshnevis2012contour}. The volatility of sulfur in low-pressure environments and outgassing further add complexities to its application on the Moon \cite{grugel2008sulfur, vaniman1992uses}. Studies have shown that $1$ cm of sulfur concrete can deteriorate within five years in a vacuum \cite{grugel2008sulfur}.


Later, Taylor et al. \cite{taylor2018sintering} combined $26$ vol.\% polylactic-co-glycolic acid (PLGA) copolymer with $74$ vol.\% JSC-1A Lunar regolith simulant (sieved to particles smaller than $50$ $\upmu$m), along with organic diluents, to synthesize Lunar regolith ink. They successfully fabricated cylindrical structures consisting of a regular lattice of micro-struts with an approximate diameter of $500$ $\upmu$m. The fabrication process involved the direct extrusion of ink followed by high-temperature sintering cycles in either air or hydrogen atmospheres. The as-printed components underwent thermal treatment cycles to eliminate the polymer binder and sinter the Lunar regolith. The effects of sintering time, temperature, and atmosphere were investigated.


The micro-trusses sintered in an air environment displayed higher relative densities and peak compressive strengths. These improvements are primarily attributed to the liquid-phase sintering, which forms a glassy phase within the structure. Taylor et al. \cite{taylor2018sintering} reported a maximum relative density of $87$\% for the strut when the sintering temperature was raised to $1100$\textdegree{C} in an air atmosphere, with a sintering duration of $24$ hours. Increasing the sintering temperature can improve the degree of sintering and, consequently, the material's strength. However, air-sintered micro-trusses at temperatures above $1110$\textdegree{C} began to melt, leading to the loss of their microarchitecture. Sintering micro-trusses in a hydrogen environment may not necessarily result in a peak strength increase, but it has practical applications in oxygen and water collection for human use on the Moon through a reduction process of the ilmenite ($\mathrm{FeTiO}_3$) mineral found in Lunar regolith to metallic iron by hydrogen \cite{allen1996oxygen, lee2013roxygen, zhao1991reduction, schwandt2012production}. The peak compressive stress was found to be up to $19$ MPa and increases as the strut densifies. The authors compared it to conventional terrestrial clay bricks with a relative density of over $75$\%, which typically exhibit compressive strengths ranging from $10$ to $25$ MPa \cite{brick2007specifications}, and non-sintered chemically bonded D-shaped manufactured blocks (discussed in ~\ref{sec:4.2.4}), which possess a compressive strength of $20$ MPa \cite{cesaretti2014building}. The comparisons show that the material under discussion possesses properties comparable to these established materials and holds potential for practical applications.


The same authors printed soft elastomeric materials via DIW of Lunar and Martian regolith inks, referred to as LRS and MRS, respectively \cite{jakus2017robust}. The ink consisted of $70$-$75$\% simulant powders, $25$-$30$\% PLGA elastomeric binders, and a solvent mixture. Specifically, they used JSC-1A Lunar and JSC MARS-1A simulants, both of which had been sieved to a $325$ $\upmu$m-mesh size. The solvent mixture predominantly consisted of dichloromethane (DCM), a highly volatile solvent, complemented by smaller amounts of other additives to improve the flow characteristics of dissolved PLGA and minimize particle-particle interactions during the printing process. Cylindrical test specimens printed with LRS and MRS inks were subsequently used for compression testing. This study exclusively characterized the as-printed green components without analyzing sintered products. The printed LRS and MRS test specimens displayed an elastic strain of $15$\%-$20$\% for different strain rates. Notably, LRS objects were strained up to $250$\% before failure, whereas MRS specimens typically failed within strain limits ranging from $50$\% to $175$\%. The Young's moduli for 3D-printed LRS and MRS were determined to be $8$-$13$ MPa and $2$-$3$ MPa, respectively. Density assessments of LRS and MRS struts also showed $27$\% and $35$\% porosities, respectively. Comparable elastic characteristics were observed in cyclic compression tests. The LRS and MRS composite could be recycled. Notably, the monomers lactic acid and glycolic acid used in the synthetic PLGA elastomeric binder could be derived from biological sources such as compost and human urine \cite{dietzen1997extraction}. DCM solvent could be directly synthesized through the thermal reaction of chlorine gas and methane. The remaining constituents of the solvent mixture could be removed by ethanol washing of the printed materials \cite{jakus2016hyperelastic}. The resulting solutions might be sequentially distilled to separate and collect the other constituents for recycling purposes \cite{jakus2017robust}. The sustainability and resource efficiency of Jakus et al.'s proposed method is of great importance, particularly in extraterrestrial environments where resource conservation is critical. However, a notable concern arises due to the excessive use of DCM, a neurotoxic substance that can vaporize during the mixing process, emitting hazardous outgas that pose a threat in enclosed spaces \cite{rahman2023towards}.





Furthermore, the overall mechanical characteristics of the print could be improved by investigating alternative materials and structural designs, as demonstrated by Ma et al. \cite{ma20233d}. They introduced geopolymer composite structures inspired by biological architectures, which exhibited impressive strength and toughness through the DIW process \cite{ma20233d}. Geopolymers, formed by combining aluminosilicate with an alkaline activator solution \cite{davidovits1991geopolymers}, displayed notable characteristics, including high compressive strength \cite{ryu2013mechanical}, resistance to freeze-thaw cycles \cite{sun2013chemical, fu2011freeze}, stability at elevated temperatures \cite{kong2010effect}, and the ability to absorb radiation \cite{oz2022radiation, fahri2021study, montes2015evaluation}. These qualities make them potential candidates for manufacturing and construction in severe Lunar environments. Various biomimetic cellular sandwich structures, including honeycomb, triangular, wave, and rectangular configurations, were studied to assess how structural design impacts mechanical performance. To simulate Lunar regolith, a mixture of powders consisting of basalt, plagioclase, forsterite, albite, ilmenite, hematite, silica, alumina, and cristobalite was used \cite{bish2013x, bish2013first}. The simulant also contained $30$\% amorphous phase \cite{isachenkov2022effect}. The ink formulation incorporated finely chopped short carbon fibers, with an average length of $600$ $\upmu$m and a diameter of $5.3$ $\upmu$m. $\mathrm{NaOH}$ ($85$ wt\%), sol-silica ($40$ wt\%), along with quartz sand, were also added to the formulation. The alkali metals on the Moon can be used as the alkaline solution needed for geopolymerization \cite{matta2009sodium}. DIW processes followed by a curing period of $7$ days at ambient temperature and an additional $2$ days in an oven at $60$\textdegree{C} were carried out. The study investigated the mechanical properties of bio-inspired structures, emphasizing their remarkable anisotropic behavior and stable, non-catastrophic failure characteristics, regardless of the compression direction. In the Y-axis direction, compressive strengths varied between $11.3$ and $20.1$ MPa across the different sandwich structures, with maximum fracture strains ranging from $5.8$\% to $10.2$\%. Moving to the Z-axis direction, compressive strengths ranged from $23.8$ to $46.7$ MPa, with corresponding maximum fracture strains varying between $12.1$\% to $13.9$\%. Notably, the Z-axis compressive strength surpassed that of the Y-axis and many other printed materials with different architectures. Despite having microcracks under loading conditions, these patterns still maintain a certain level of structural integrity.




Pilehvar et al. \cite{pilehvar2020utilization} also pursued a sustainable manufacturing process to print Lunar regolith geopolymer using urea from human waste as a chemical additive to achieve good workability of Lunar regolith geopolymers and decrease the water demand for geopolymers. Urea can break hydrogen bonds, thereby reducing the viscosities of aqueous mixtures \cite{usha2002effect}, making it a potential superplasticizer. The authors used DNA-1 Lunar regolith simulant, developed by ESA, as a chemical and mineralogical representation of Lunar mare regolith. A $12$ M ($480$ g/L) sodium hydroxide solution was selected as the alkaline solution. Workability and buildability tests showed an alkaline solution-to-regolith ratio of 0.35:1, and a urea dosage corresponding to $3$\% of the Lunar regolith mass was optimal. The as-printed samples underwent freeze-thaw cycles in which the specimens were first heated at $80 \text{±} 2$\textdegree{C} for $48$ hours, followed by being placed in a freezer at a temperature of $\text{-80} \text{±} 2$\textdegree{C} for the next $48$ hours. The mixture could be pumped to build objects layer by layer with good flowability and retain its shape under external loads. The as-printed specimens displayed an initial compressive strength of $13$ MPa. Although the printed material had lower compressive strength after freeze-thaw cycles, the authors noted that it remained applicable to Lunar infrastructure. However, no further studies have been conducted in their work to assess how this Lunar regolith geopolymer behaves in a Lunar setting. Urea can indeed undergo hydrolysis, breaking down into ammonium ions and producing volatile gases that can evaporate in a low-pressure environment. 


A proper solvent is essential to harness the capabilities of material extrusion techniques for the utilization of in-situ resources. The ideal solvent should be non-toxic, provide sufficient inter-particle bonding within the regolith, and remain resilient in extreme Lunar environmental conditions. As illustrated by Rahman et al. in their study, hydrophilic hydrogel-forming polymers can address these challenges \cite{rahman2023towards}. Even at low concentrations of hydrogel, these polymers can interact with water, forming a network of hydrogen bonds and altering the viscosity and viscoelastic properties of the mixture.




\subsubsection{Stereolithography} \label{SLA}

Stereolithography (SLA) is a form of 3D printing technology used to fabricate parts layer by layer. It operates through photochemical processes, wherein light induces the cross-linking of chemical monomers and oligomers, leading them to form solid-state polymers \cite{hull1984apparatus}. SLA can be either top-down or bottom-up. Both approaches operate on the same fundamental concept but differ in terms of manufacturing direction and occasionally the type of illumination used \cite{melchels2010review,huang2020review}. The advantages of stereolithography lie in its flexibility for fabricating parts with varying geometries and dimensions, its high accuracy, and its rapid production capabilities \cite{corbel2011materials}.

SLA finds utility in fabricating highly precise ceramics or ceramic composites, a technique also known as lithography-based ceramic manufacturing (LCM). This procedure entails the SLA of a polymer-ceramic suspension, typically followed by a debinding-sintering step \cite{griffith1996freeform}. In this process, the initial step involves creating a photocurable polymer/ceramic slurry (suspension). After additive manufacturing of the samples, achieving the desired mechanical characteristics necessitates the removal of the photopolymer from the structure, often accomplished through solvent and/or thermal debinding \cite{altun2021additive, hadian2021material}. Subsequently, the bonding of ceramic particles must be enhanced via a sintering phase.

The ceramic-polymer SLA technique has also been considered a potential fabrication method for space applications. Regolith is a part of this composition to reduce the cost of raw material delivery. Altun et al. \cite{altun2021additive} utilized EAC-1A Lunar regolith simulant for manufacturing components using LCM. Their study involved analyzing the suspension's rheological behavior, thermogravimetric characterization of the as-printed parts, investigating the mechanical and microstructural properties, and computed tomography of the as-sintered samples. The resulting average compressive strength and elasticity modulus were recorded as 5.4 MPa and 403.3 MPa, respectively. Relative density analysis conducted through the Archimedean principle and water displacement indicated a reduction from 67.9\% to 55.9\% after the debinding and sintering processes. The sintering occurred at 1000\textdegree{C}, as reported in the paper, which is at the lower end of the regolith melting zone. Considering the sintered samples' high porosity and, consequently, low strength, higher sintering temperatures could yield better results. Additionally, it was observed that after debinding and sintering, the density was reduced. This further suggests that the sintering process may not have been appropriately selected. While it could facilitate an assessment of the effectiveness of sintering, no mechanical strength data are reported for the as-printed samples. However, given the negative impact of debinding and sintering on density, it could be expected that the mechanical strength would likely be reduced after these processes.

Liu et al. \cite{liu2019digital} utilized SLA technology to print a slurry comprising CLRS-2 Lunar regolith simulant powders mixed with photocurable resins. The regolith simulant underwent ball milling for grinding before being mixed with the photocurable resin. They reported average compressive strength and flexure strength values of the sintered samples as 428.1 MPa and 129.5 MPa, respectively. These enhanced mechanical properties can be attributed to the regolith's particle size reduction. This reduction facilitates particle movement during sintering and enhances density by minimizing pores \cite{rhodes1981agglomerate}.

Isachenkov et al. \cite{isachenkov2022effect} examined the impact of particle size distribution on the curing depth and printability of LHS-1 and LMS-1 Lunar highland and mare regolith simulants in the SLA process. They investigated the effect of wet ball milling under various fraction sizes and durations on the particle size distribution. Their findings revealed a significant influence of the particle size distribution on the photo-curability of the suspension. Another study by Isachenkov et al. \cite{isachenkov2023technical} conducted a comparison of potential additive manufacturing (AM) techniques, considering factors such as energy consumption, required consumables from Earth, and compressive strength as key indicators. This evaluation was based on existing literature. Referencing Liu et al.'s \cite{liu2019digital} reported compressive strength of 428.1 MPa, they ultimately determined that SLA stands out as the most promising approach for the indoor production of small, precise parts.

Wang et al. \cite{wang2023optimized} conducted a study that explores the influence of Lunar environmental conditions on SLA using low-titanium Lunar Regolith Simulant (LRS). Various particle size distributions of regolith were examined, with smaller particles (ground for a longer time) yielding better accuracy. Additionally, they have shown that adjusting solid content controls viscosity. Among the different debinding-sintering conditions, samples debound in nitrogen followed by sintering in the air were found to be preferable due to their combination of organic bonding at low temperatures and rapid degreasing at high temperatures. However, it should be mentioned that due to the limited access to these gases on the Lunar surface, vacuum debinding and sintering are preferable. Also, they showed that there is a trade-off between fabrication convenience and achieving samples with low porosity, high precision, and acceptable mechanical strength. This is achieved by maintaining a balance between not being too big and not being too small. Overall, under the study's optimal conditions, a compressive strength of 386 MPa was achieved for the sintered sample.

Xiao et al. \cite{xiao2023additive} employed SLA to 3D print a paste composed of 80 wt\% Lunar regolith simulant. This composition ratio was determined to achieve a balance between fluidity and retention. The study also delved into the effects of introducing water into the paste. Additionally, the impacts of printing parameters such as layer thickness and building orientation on both performance and microstructure were investigated. In the optimal scenario, a flexural strength of 132.21 MPa and a compressive strength of 444.23 MPa are reported. The density and open porosity were measured at 2.73 $\, \text{g/cm}^3$ and 3.12\%, respectively. The compression strength reported by Wang \cite{wang2023optimized}, Liu \cite{liu2019digital}, and Xiao \cite{xiao2023additive} are well beyond what other available papers have reported and should be independently replicated and verified.

Isachenkov et al. \cite{isachenkov2022characterization} conducted UV-Vis-NIR absorbance spectroscopy on LHS-1 Lunar highland and LMS-1 Lunar mare simulants that were crushed to a particle size of 1 micron. They also performed a photopolymerization assessment on the photocurable slurries of each simulant to evaluate their suitability for the laser-SLA technique. Their findings revealed that due to variations in mineral composition, utilizing mare regolith as source material for this type of additive manufacturing would be considerably more challenging compared to highland regolith. While both materials can be fabricated using the laser-SLA method, the slurry based on mare regolith polymerizes 2.5 times less effectively than the highland regolith-based slurry.

Dou et al. \cite{dou2022ceramic} have introduced an innovative method for creating ceramic paste formulations, specifically focusing on alumina-based paste tailored for stereolithography (SLA) in environments with microgravity and applications in space. They accomplished this by incorporating thickening agents like HE-cellulose or carbomer 940 into the paste to manipulate its rheological properties, resulting in a material that exhibits Bingham pseudoplastic behavior. This effectively addresses the challenge of wall climbing during the printing process in low-gravity conditions. By carefully selecting particle size distribution, slurry compositions, and printing parameters, they achieved relative densities exceeding 99\% and obtained exceptional surface quality, highlighting the potential of SLA for space-based manufacturing. In recognition of the importance of utilizing locally available materials, it is advisable to conduct similar investigations using regolith powder in future research.

Despite the promising outcomes of the mentioned papers that used LCM, there are significant challenges to its wide application in space, especially when regolith is used as ceramic. First of all, SLA-based techniques usually require the use of liquid-phase materials. Given the outgassing requirements imposed by NASA and other space agencies due to the vacuum environment, there are inherent limitations to using these techniques in a vacuum and on outdoor Lunar surfaces. Extreme cooling and heating cycles further limit the use of liquid materials. While using them in a chamber or a space shelter/room with a controlled atmosphere and temperature can be a solution, it imposes other challenges and requirements, such as the extra cost of controlling temperature and atmosphere and the need to make walls that can withstand atmospheric pressure differences between the inside and outside, etc.

We usually aim to use locally available materials for ISM as much as possible. However, in SLA-based manufacturing, the cure depth is inverse to the percentage of solid content (regolith here). This limits how far we can go in terms of regolith percentage. Besides that, in general, choosing smaller particles results in higher density after sintering, while it makes the debinding process harder and more challenging. Gases can be trapped in the structure during debinding and can cause defects in the structure. Further research should be conducted to find the optimal particle size distribution (PSD) for ceramics, and it is challenging, time-consuming, and costly to achieve that specific PSD using Lunar regolith, given the different characteristics of Lunar regolith from various areas on the Moon.

Moreover, in the available literature, the debinding process primarily relies on thermal methods, which essentially "burn out" the polymer. However, given the high value of off-site materials on the Lunar surface, wasting them may not be a wise option. Efforts should be made to recycle as much as possible, as sustainable ISM will be required. Another aspect that should not be neglected is the power requirements for thermal methods, especially when it comes to sintering. Thermal debinding/sintering would require a high level of power. Finally, debinding polymers from large-scale parts is very challenging. It is crucial to precisely maintain a constant gas flow rate during the debinding process to minimize the possibility of forming defects. Controlling this for large-scale parts and ensuring uniform debinding across the part is even more challenging. This would also require significant time and energy and a large furnace, which would be difficult to access on the Moon. All of these factors can impose significant constraints on the design of the part.

\subsubsection{Binder Jetting}\label{sec:4.2.4}

Binder jetting is an additive manufacturing process in which a printhead selectively deposits a pure liquid binder agent onto a very fine layer of powdered material, typically sand, ceramics, metal, or composites. This deposition creates cohesive bonding between powder particles, forming the desired layer following the cross-section of the intended 3D object. The process is repeated with subsequent layers of powdered material being spread and bonded atop the previous ones until the complete 3D object is formed \cite{kunchala20183d, mostafaei2021binder, chavez2020influence}. The resulting green body then goes through post-processing steps, such as curing and sintering cycles, to strengthen the bonds between particles and achieve a dense final product.


Like DIW, binder jetting relies heavily on binders and chemical additives, demanding a consistent supply of these materials from Earth. This reliance stands as a drawback in binder-based AM techniques. Independence from Earth-based resources and utilization of the full potential of Lunar surface resources is a priority. While recycling and reusing binders can support ongoing operations, these approaches are inadequate during periods of growth or expansion. Studies have also been conducted in the context of Lunar-based manufacturing to explore the potential of binder jetting technology for Lunar-related applications, aiming to adapt binder jetting methods to meet the unique challenges and demands of the Lunar environment, such as extreme temperature fluctuations, outgassing, micrometeoroid impacts, and radiation exposure.

An ESA feasibility study \cite{ceccanti20103d} described the use of local regolith to build infrastructure shelters and habitats on the Moon with the D-shaping method. The D-shape binder jetting printer is equipped with a nozzle head housing $300$ nozzles at a $20$ mm inter-axis distance, covering a 6-meter-wide printing platform. This enables the creation of geometric configurations on a substantial scale, making it suitable for architectural and construction purposes. The authors adopted volcanic ashes sourced from Italy's Bolsena volcano, which replicate the key characteristics found in the widely used JSC-1A Lunar regolith simulant. Tests were conducted both in the air and under vacuum conditions. The primary objective was to prevent ink-related issues, boiling or freezing, once sprayed onto the regolith in vacuum conditions and maintain the liquid state long enough to allow the reactions of the reticulation process to occur. The vacuum tests revealed that selecting a proper injection method prevents ink evaporation or freezing. This study primarily focused on the feasibility of the D-shape binder jetting method for Lunar-based manufacturing and construction, and it proved effective both in a vacuum environment and when using a regolith simulant \cite{ceccanti20103d}.


Cesaretti et al. \cite{cesaretti2014building} built upon earlier research by Ceccanti et al. \cite{ceccanti20103d} and successfully printed building blocks for habitat wall structures with a specific topology known as closed foam. They claimed that this particular design choice offers enhanced compressive strength and structural robustness, enabling it to resist meteorite impacts and effectively handle external forces. However, further investigation is required to assess its capability to resist high-velocity meteoroids. Sorel cement, a solution of $23$ wt\% salt ($\mathrm{MgO}$ and $\mathrm{MgCl}_2$) in $77$ wt\% $\mathrm{H}_2 \mathrm{O}$, was used as a binder agent. This binder chemistry was specifically designed to interact with $\mathrm{MgO}$ present in the Lunar regolith. DNA-1 was used as the Lunar regolith simulant based on natural volcanic material. $5$ wt\% additional $\mathrm{MgO}$ was incorporated into the simulant to ensure a sufficient quantity to facilitate the desired chemical reaction with the binder ink. Magnesium compounds could in theory be substituted with $\mathrm{ZnO}$ and $\mathrm{ZnCl}_2$ due to the superior thermal properties of $\mathrm{ZnO}$, though this would be limited by the scarcity of zinc on the Moon. Large-scale blocks with geometries designed for Lunar architecture were printed through the D-shape material depositing method. Tests were carried out in a vacuum environment to verify the feasibility of the chemical process and investigate the ink behavior once sprayed in a vacuum. The mechanical properties of the printed blocks were comprehensively analyzed. They showcased a compression strength of $20.35$ $\mathrm{MPa}$, flexural strength of $7$ $\mathrm{MPa}$, and Young's modulus of $2.35$ $\mathrm{GPa}$. These results were considered satisfactory for unsintered green blocks with $13$\% open porosity. Furthermore, the authors studied thermophysical properties, including linear thermal expansion and freezing resistance. The thermal expansion was found to be relatively low, indicating that the material is less prone to significant dimensional changes in response to temperature variations. This characteristic is crucial for Lunar construction, where extreme daily temperature fluctuations are a challenge. Another critical aspect studied is the printed blocks' freezing resistance. Given the extreme Lunar conditions, the ability of these structures to withstand freezing temperatures is important. 


While addressing the unique challenges posed by vacuum conditions and extreme temperature variations in the Lunar environment, Cesaretti et al. achieved promising mechanical and thermophysical characteristics for their prints that underscore their meticulous approach to ensure the viability of the print for Lunar applications \cite{ceccanti20103d, cesaretti2014building}. However, the reduced atmospheric pressure and lower gravity on the Moon present challenges for precise liquid droplet ejection, a fundamental aspect of the binder jetting process. On Earth, gravity helps with binder deposition, but on the Moon, where gravity is much weaker, advanced technological solutions are needed to precisely control and direct the deposition of liquid binder. It may lead to dimensional and geometrical inaccuracies of the print, a concern that Cesaretti et al. have not addressed.

Furthermore, their approach involves substantial consumption of chemicals and notably water. The high demand for water in Sorel cement and other cementitious binders \cite{sakamoto2020preparation, neves2020characterization}, particularly in the context of extraterrestrial environments, where liquid water is a highly prized and precious resource, might hinder the adoption of their methodology. It becomes imperative to explore innovative approaches that allow for the recovery and reuse of liquid water without compromising the material performance, as demonstrated by Osio et al. \cite{osio2021sintering}. According to their studies, Lunar regolith-based magnesium oxychloride ($\mathrm{MOC}$) cements, which rely on a reaction between $\mathrm{MgO}$ and a $\mathrm{MgCl}_2$ solution, are potential candidates for water recovery in space. This type of cement is notable for forming gel phases that contribute to developing a high-compressive strength binder \cite{walling2016magnesia}. Sintering these materials at temperatures up to $1200$\textdegree{C} results in the complete decomposition of the Lunar regolith-based $\mathrm{MOC}$ cement, releasing $\mathrm{HCl}$ and water that could be captured and reused while maintaining material properties standards. This sintering process also reduces the volume of permeable pores by $7$\%, leading to significant increases in compressive strength, particularly when the material is initially pre-dried. On average, $\mathrm{MOC}$ paste samples display a compressive strength of about $7$ MPa, while sintered specimens reach an average of $20$ MPa.

Cesaretti et al. also did not assess powder and chemical emissions during the printing process, raising concerns about their approach's alignment with sustainability practices. These emissions can arise from various sources, such as powders used in the powder bed due to the unique microgravity environment of the Moon and the outgassing of liquid binders. Such emissions can affect air quality and resources. The choice of materials and the production process can have significant environmental consequences. Future research should comprehensively evaluate the emission rates to address this aspect and enhance the applicability of their work. In this regard, Hayes et al. \cite{hayes2020influence} characterized particle emission rates for powders with varying particle sizes and compositions. The powders studied included Hydroperm gypsum plaster, Lunar regolith simulant (LHS-1), and Zeolite $13$X (a molecular sieve), with a specific focus on Lunar regolith powder. Their findings indicated that variations in particle emissions during the binder jetting process primarily depend on the type of powder used, emphasizing the importance of considering air quality as a design factor for printing processes. Given the Lunar microgravity environment, their study predicted that particulate concentrations at a distance of $20$ meters from the printer can be up to $330$\% higher compared to Earth conditions. There is a need to comprehensively evaluate powder size, morphology, and composition when addressing emission-related concerns for novel powders used in powder-binder jetting technology.




\subsubsection{Directed Energy Deposition}

Directed Energy Deposition (DED) is an additive manufacturing technique that utilizes a high-energy source, such as a laser or electron beam, to precisely melt and fuse material as it is deposited. Typically, this process involves supplying the material in the form of wire or powder feedstock. This method allows for the precise addition of material layer by layer, enabling the creation of complex three-dimensional structures. Directed Energy Deposition finds common use in various industries, including aerospace and automotive, for tasks such as repairing or enhancing existing components or fabricating new parts. Usually, laser-engineered net shaping (LENS) or laser metal deposition (LMD), wire arc additive manufacturing (WAAM), electron beam free-form fabrication (EBF) or electron beam welding (EBW) are categorized within the family of DED \cite{ahn2021directed, najmon2019review}.

The primary distinction between powder-DED and PBF lies in the fact that in PBF, a thin layer of powdered material is deposited and spread over the build plate, with a laser or electron beam selectively melting and fusing the powder to form each layer of the 3D object. In contrast, in Powder-DED, the energy source precisely melts and fuses the material as it's being deposited locally. DED is a suitable method for fabricating large and intricate parts with a somewhat reduced level of precision, making it especially valuable for restoring or enhancing worn or damaged components. On the other hand, PBF is well-known for its capability to manufacture finely detailed components with a high degree of precision, making it ideal for crafting complex shapes \cite{donate2020effect, saboori2019application}.

Among the DED technologies, Wire Arc Additive Manufacturing (WAAM) holds great potential for fabricating large-scale metal components for space industries, such as structural elements of spacecraft. However, WAAM relies on the presence of gases (typically shielding gases) to create the electric arc \cite{da2021effect, romano2023damage}, which seems impossible in a vacuum environment. Therefore, it is not suitable for outdoor additive manufacturing on the Moon. However, if another type of heat source is used, including a laser or electron beam, wire-DED would be feasible for in-space manufacturing, despite the challenges. Electron beam (EB) heat source is usually conducted in a vacuum environment and is primarily suitable for metals and less effective for non-metals, mainly due to the fundamental differences in how electrons interact with different types of materials. The use of an electron beam source for in-space manufacturing will be limited to additive manufacturing of metals, with wire or powder feedstock \cite{howell2008site, NARRA202285, sadeghi2019influence}. Transporting metals to the Moon would be very expensive unless the metal is extracted on the Moon (see Section \ref{Extraction}), a technology that is currently at a low level of readiness. Furthermore, wire-based DED with any kind of energy source is also challenging, as the wire cannot be made directly from regolith and should be metallic. However, incorporating regolith into a metallic matrix and extruding it into wire to reduce material costs could be an interesting research direction, although it can be very energy-consuming (melting and extruding metal/regolith into wire and melting it again for wire-DED). 

The most promising In-Situ Resource Utilization (ISRU)-based DED is likely laser-DED of regolith, which also comes with its challenges. Laser sintering of ceramic materials can be challenging due to ceramics' low absorption of laser light. Ceramics typically do not melt during the process and do not flow to fill gaps. Adding around 20 wt\% metallic content, as suggested by Azami et al. \cite{azami2023laser}, can help increase laser absorption, enabling higher temperatures during printing and resulting in improved material characteristics. Another challenge in powder-DED in a vacuum is that the material cannot be fed using gas flow, necessitating an alternative feeding mechanism.

In their investigation into the feasibility of fabrication of Lunar regolith with laser-based techniques, Balla et al. \cite{balla2012first} employed the LENS technique to fabricate specimens using sieved JSC-1AC simulant with particle sizes falling within the range of 50–150 microns. The process involved a 50 W Nd:YAG laser featuring a 1.65 mm laser spot size. The best results were obtained when using an energy density of $2.12 \, \text{J/mm}^2$ while maintaining a layer thickness of 254 microns and a powder feed rate of 12.35 g/min. This combination of parameters effectively generated the required melt pool, avoided excessive spreading of the liquid pool, and prevented the cracking of solidified parts \cite{balla2012first}.

The researchers subjected the laser-processed components to comprehensive characterization, which included X-ray diffraction, differential scanning calorimetry, field-emission scanning electron microscopy (FE-SEM), and X-ray photoelectron spectroscopy. This thorough analysis aimed to assess the impact of laser processing on the microstructure, constituent phases, and chemical composition of the Lunar regolith simulant. Examination of the samples through FE-SEM revealed a finely uniform structure characterized by nanoscale grains. The material's hardness was determined to be $500 \text{±} 18$ HV. The results indicate that LENS-based laser processing transformed crystalline regolith into structures with a nanocrystalline and/or amorphous character, resulting from complete melting followed by solidification \cite{balla2012first}.

All in all, very limited literature is available on ISRU-based DED. This scarcity of information is primarily attributed to the challenges associated with the adoption of this technology for regolith-based materials, the hurdles in utilizing an electron beam (EB) energy source for non-metals, and the difficulties involved in laser-based fabrication of ceramics. However, if further research is conducted to explore the integration of EB technology with regolith or regolith composites, to reduce porosity in laser-fabricated regolith, or to enhance the extraction of metals from regolith, this technique holds great promise for space industries. This promise is especially significant due to its ability to manufacture large-scale parts with complex geometries and repair damaged components.


\subsection{Non-Additive (Traditional) Manufacturing Techniques}
While additive manufacturing methods hold great potential for future space explorations, traditional manufacturing methods have a long history of success and reliability in producing high-quality components on Earth. Hence, they hold the potential as trusted and dependable options for future Lunar missions. Different manufacturing methods are under consideration for Lunar-based manufacturing and construction including sintering-based, regolith concrete, binder-based, and regolith casting. In some cases, such as regolith concrete and binder-based methods, these works are strongly related to subsequent AM work on Direct Ink Writing and regolith/polymer composite Material Extrusion, respectively, discussed in earlier sections.


\subsubsection{Regolith Casting}
Casting is a manufacturing process where a material is heated above the melting temperature, is then poured into a mold or cavity, then cools down and solidifies. Depending on the cooling rate different materials can be produced that are classified based on their crystallization degree as glass, glass-ceramic, and casts (ceramics). Glass material has been cooled down fast enough to keep the amorphous structure of the liquid phase, while casts have a crystalline structure and glass-ceramic falls between these two. Regolith casting is not an extensively investigated topic and few papers can be found. In 1988, a project by McDonnell Douglas Space System Company, ALCOA/Goldsworthy Engineering, and the Space Studies Institute (SSI) investigated the use of solar thermal power to produce glass fiber and glass samples from Lunar regolith simulant \cite{magoffin1990Lunar}. The team developed a crucible and the required hardware for demonstrating the Lunar glass production concept. Using a 75 kW solar concentrator they were able to melt the low-grade basaltic regolith simulant at $1200$\textdegree{C} and make glass and glass fiber.

Ishikawa et. al. used a microwave heating method to melt a high-Ti mare simulant between $1200\text{-}1500$\textdegree{C} to produce construction materials for Lunar bases. Uniaxial compressive strength of 120 MPa was obtained \cite{ishikawa1992simple}.
In another study by Tucker et al. glass fiber or larger glass rods were fabricated by melting JSC-1 Lunar regolith simulant in a high-temperature furnace between $1450\text{-}1600$\textdegree{C}. The produced glass fiber was then incorporated into sulfur concrete as a reinforcement and results showed a significant improvement of about $50$ \% in flexural strength. The fraction of regolith glass fiber was $1$ wt\% \cite{tucker2006production}.


In a more recent study by Schelppi et al., three mare simulants (JSC-2A, FJS-1, and EAC-1) and LHT-3M highland simulant were melted inside a susceptor-assisted microwave oven to produce glass and mirrors \cite{schleppi2019manufacture}. Susceptor-assisted microwave processing has the advantage of being independent of the composition of regolith, in contrast to regular microwave processing. However, this method is less efficient due to the losses of the susceptor-assisted heating. Reflective properties were characterized after coating the glasses with aluminum and silver. Resulting mirrors with porous and/or smooth surfaces demonstrated the ability to reflect incident solar light in the range of $400$ to $1250$ nm, with reflectivity ranging from 30\% for the least effective sample to 85\% for the most effective. Uncoated samples exhibited lower reflectivity, reflecting less than 7\% of the same incident solar light.

MoonFibre is an ongoing project at RWTH Aachen University focused on developing technology to produce fibers from Lunar regolith. They have successfully spun continuous fibers as thin as 17 \(\mu\)m using various regolith simulants that represent different regions of the Moon. The spinning process involves feeding raw simulant material into a heated bushing equipped with multiple nozzles, from which a rotating winder draws the molten material to form fibers. Since both gravity and atmospheric pressure are crucial for the spinning process, the next challenge is adapting this technology for use in a low-gravity, airless environment.

\subsubsection{Regolith Sintering }
Regolith sintering is an attractive method for Lunar manufacturing and construction as it only requires heat, which can be provided by solar energy and regolith. The sintering procedure starts with the cold molding of a raw powder, forming an un-sintered "green" body. Subsequently, this green body is heated up to the sintering temperature (below the melting point) in a furnace, where it is maintained for a predefined duration before undergoing a controlled cooling process.

The sintering process consists of three different stages. The initial stage involves the formation and growth of necks between particles. By the end of this stage, there is an increase in contact area by approximately 20\%, accompanied by minimal densification, resulting in only a slight compacting of the material, typically a few percent. During this phase, a noticeable reduction in the specific surface area occurs due to surface smoothing. At the same time, grain boundaries between particles maintain their position in the contact plane due to tensile stresses stemming from surface tension. The presence of a more or less continuous network of pore channels along grain edges characterizes the second stage. During this phase, pore channels progressively contract, and grains grow in size. Most of the densification and an increase in contact area primarily occur during this intermediate stage. Continuing the process, pore channels continue to diminish until they eventually close, forming isolated spheroidized pores. This marks the beginning of the final stage of densification. In this concluding stage, the volume fraction of pores approaches zero asymptotically, while in some cases, these closed pores may trap gases, making their removal a challenging task.

Regolith sintering for Lunar construction was first investigated by Simonds in 1973 \cite{simonds1973sintering}. Using a glass powder with a composition close to regolith from Apollo 14 with a size below $37 \mu$m, samples were sintered at $750$\textdegree{C} and $800$\textdegree{C} for 7.5 hours. While a cohesive material was formed at $800$\textdegree{C}, sintering at $750$\textdegree{C} resulted in loose clods. Moreover, the sintering of coarser grains (size distribution between $149 \mu$m and $173 \mu$m) did not form a cohesive structure below $850$\textdegree{C}. Since sintering is driven by surface energy, coarse grains sinter slower than fine ones.
Allen et al. \cite{allen1992sintering} studied the sintering behavior of Basalt Lunar regolith simulant by evaluating its compressive strength. The material sintered at $1000$\textdegree{C} with the initial phase of sintering occurring before 15 minutes with no further significant sintering after 30 minutes. Maximum compressive strength of 26 MPa was achieved at $1100$\textdegree{C}.

Matyas et al. \cite{matyas2011experimental} proposed a thermal mass to survive Lunar nights, to be made by sintering Lunar regolith. For this purpose, they used Lunar JSC-1AF mare and NU-LHT-2M highland simulants. They studied the effect of time, temperature, and atmosphere (air, vacuum, and argon) on sintering and densification. Thermal diffusivity and radiative permittivity were measured as the two properties that provide information on how the material can store heat during the Lunar day and release it during the Lunar night. Results showed that the Lunar mare simulant was successfully densified and sintered at $1200$\textdegree{C} with a porosity content between 3.6\% and 9.9\%, depending on the atmosphere and cooling rate. On the contrary, a higher temperature of $1300$\textdegree{C} was needed for sintering highland simulant, as sintering at $1200$\textdegree{C} resulted in a poorly densified structure with the porosity content of 62.8\% to 89.0\%. Furthermore, it was shown that thermal diffusivity of up to $0.76 \, \text{mm}^2\text{/} \text{s}$ can be achieved upon sintering the regolith, which is more than 100 times higher than the regolith powder thermal diffusivity ($0.0066 \, \text{mm}^2 \text{/} \, \text{s}$). The higher thermal diffusivity indicated a potential for manufacturing a thermal energy storage medium capable of storing the necessary heat for Lunar rovers to endure the harsh cold nights of the Moon.

It is reported that sensible mediums like water and rock are useful for thermal energy storage (TES) applications \cite{li2013performance}. The availability, cost-effectiveness, and simplicity of Lunar rock make it highly advantageous for TES purposes on the Moon. 
Wegeng et al. established the foundational concepts for TES using processed Lunar regolith as the primary energy storage medium. Although untreated Lunar regolith has poor thermal mass properties, it can be enhanced through methods like compacting, sintering, melting, solidifying, and incorporating high-conductivity materials to improve its thermal diffusivity. These methods result in processed regolith with superior thermal mass properties suitable for TESs. They estimated that a reference TES, with 90000 kg of "regolith-derived thermal mass" could store between 16 to 19 GJ of energy, providing heat or power to a Lunar outpost during Lunar nights \cite{wegeng2007thermal}.

The concept of a thermal battery made of sintered regolith that provides heat and power during Lunar nights was investigated by Bonanno and Bernold \cite{bonanno2015exploratory}. ALRS-1 was sintered at $1100$\textdegree{C}, and its heat-storing capacity was investigated by determining specific heat capacity ($C_p$) from DSC measurements. The experiments showed that $C_p$ of sintered regolith fell in the range of the basalt on Earth and demonstrated that a thermal battery could be developed based on sintered regolith.

As previously mentioned, Lunar regolith size distribution varies from a few microns to 800 $\mu$m. Gualtieri and Bandyopadhyay investigated the effect of the initial powder particle size distribution on the porosity and compressive properties of the samples sintered from a mix of JSC-1 A, JSC-1AF, and JSC1-AC Lunar mare regolith simulant \cite{gualtieri2015compressive}. Two sets of samples were sintered at $1200$\textdegree{C} for 20 minutes with the initial powder particle size of higher than $212$$\mu$m for high porosity and between $25$$\mu$m and $212$$\mu$m for low porosity samples. Different powder particle sizes resulted in 99\% and 92\% relative density, and the porosity content significantly affected the mechanical properties. Failure stress increased by almost 77\% from 103.2 MPa to 232 MPa as the porosity decreased to 1\%. The effect of the porosity on the compressive properties of the sintered regolith was studied in other research emphasizing that sintered regolith can meet structural needs in future Lunar explorations \cite{indyk2017structural}.

Hoshino et al. \cite{hoshino2016key} tried to find the optimum temperature and holding time to achieve high strength of sintered FJS-1 Lunar regolith simulant blocks with a diameter of 30 mm and 20 mm thickness. Implementing the optimum parameters achieved in the experiments, the team attempted manufacturing large blocks for construction (100 mm \texttimes 100 mm \texttimes 50 mm). Large blocks showed many cracks and crazing, showing that the time-temperature profile of the small blocks was unsuitable for larger parts. Another optimization was performed to prevent cracking. Results showed the importance of heating rate over holding time in larger parts. Blocks showed compressive and bending strength of 30 MPa and 5 MPa, respectively, which is in the range of the typical concrete on the Earth. Furthermore, the findings from sintering blocks measuring 60 mm in diameter and 50 mm in thickness, conducted under both vacuum and ambient air, demonstrated that the necessary sintering temperature to achieve the same strength in a vacuum environment was around 100 K lower compared to that required under atmospheric conditions. In another study by Song et al. \cite{song2019vacuum} it was shown that porous structures, with low thermal conductivity (0.265 W/mK) for use as a thermal insulator during Lunar missions, can be obtained by sintering CLRS-1 simulant under vacuum conditions. The vacuum-sintered samples showed the formation of macro-pores and a reduction in weight and density over air-sintered samples. The significant impact of the vacuum environment on the properties of the sintered regolith reveals the importance of investigating Lunar application-related research under vacuum conditions. Results of research by Fateri et al. \cite{fateri2019thermal} highlighted the difference between vacuum and ambient sintered Lunar regolith simulant once again; pressing regolith powder before sintering in vacuum did not influence the density of the final part, while it reduced the density in ambient sintered parts. 

    
    

In the conventional radiant sintering approach, the heat is applied externally, and this method was employed in the prior mentioned research. Other research is available on the sintering of regolith by microwave energy. In microwave sintering, heat is generated directly within the material due to the presence of polar molecules, ions, impurities, and discontinuities in crystalline materials. Early investigations on the microwave susceptibility of regolith were conducted at the Los Alamos National Laboratory, USA, between 1984 and 1987 \cite{meek1985microwave}\cite{vaniman1986fusing}\cite{wright1986fusing}\cite{meek1987sintering}. Simulated Lunar regolith was used in these studies, and the susceptibility of regolith to microwave energy was attributed to the presence of the mineral ilmenite (FeTi$O_3$) and radiation defects on grain surfaces from solar winds. As mentioned in Section 3.2, nanophase Fe$^o$ can be found in Lunar soil, and its effect on sintering was neglected in the previous studies as it can not be found in the Lunar regolith simulants. Microwave sintering of real Apollo 17 soil by Taylor and Meek \cite{taylor2005microwave} revealed extreme coupling of Lunar soil with microwave energy. It was explained that the underlying reason for the distinctive behavior of Lunar soil when exposed to microwave radiation is due to the presence of nanophase Fe$^o$ which is an unusual property of real Lunar soil. The extremely small size of the nanophase Fe$^o$ particles is such that it falls below the skin depth of microwave energy. Consequently, each of these metallic Fe grains behaves as a conductor. The conductivity of the metallic Fe particles makes them efficient absorbers of microwave energy, thereby establishing "energy sinks," which result in the substantial generation of heat during microwave exposure.

Allan et al. \cite{allan2013high} investigated the high-temperature dielectric properties of the JSC-1AC Lunar simulant up to $1100$\textdegree{C} at the microwave frequency of 2.45 GHz. They observed low heating absorption below $250$\textdegree{C} and a progressively more significant absorption above $1100$\textdegree{C}. At the same microwave power, the heating speed accelerated to beyond 100/min. This rapid heating is attributed to the formation of lattice defects in crystalline materials at high temperatures \cite{farries2021sintered}. In this situation, the temperature of a sintering process increases uncontrollably, potentially leading to overheating and melting, commonly referred to as the "thermal runaway." The temperature inside the part was determined to be around $1100$\textdegree{C}, and as shown in the image, in Region A, the regolith was completely melted, forming a glassy phase. Regions B and C show the smooth fracture surface of the glassy phase and the boundary between the glassy phase and the surrounding sintered area, respectively.
Uneven heating by direct microwave sintering was addressed by combining radiant and microwave heating, which is called hybrid microwave furnaces. Within a hybrid microwave furnace, the green body is surrounded by a microwave susceptor, which absorbs a fraction of the microwave energy, generating radiant heat, while a portion of the microwave radiation penetrates the susceptor material to heat the green body directly. A more uniform condensed structure was achieved by sintering the JSC-1 Lunar simulant in a hybrid microwave furnace. Hybrid heating was used in another study to sinter MLS-1 Lunar regolith simulant. Processing parameters were optimized, and with sintering at 980\textdegree{C} for 35 minutes, an 11\% increase in density and compressive strength of 1100 psi (7.6 MPa) was achieved. Compressive strength is lower than that of samples made by furnace sintering. Fateri et al. \cite{fateri2019localized} used localized microwave irradiation for 30s (around 815\textdegree{C}) at $950$ W in air for the post-treatment of furnace-sintered JSC-2A regolith simulant. While treated samples showed a more compact structure with a 13\% decrease in porosity, no significant change in the compressive strength and Young's modulus was observed. Microwave-treated samples exhibited mostly closed porosity with microcracks, in contrast with oven-sintered samples with interconnected pores and irregular shapes. Increasing the heat with a longer irradiation time of around $45$ s (around 1100\textdegree{C}) resulted in material melting and outgassing. It is important to note that outcomes could differ under vacuum conditions where grain oxidation is avoided, and outgassing has a stronger effect than in ambient conditions.

The Pacific International Space Center for Exploration Systems (PISCES), in partnership with NASA's SwampWorks, developed interlocking tiles made from sintered basalt sourced from a quarry in Hawaii \cite{edison2019hawaiian}. This basalt, which shares geochemical properties with Lunar and Martian regolith, was processed at two different thermal profiles ($1149$\textdegree{C} and $1177$\textdegree{C}). The resulting sintered materials exhibited structural properties up to three times stronger than residential and specialty concrete. Analysis revealed that high plagioclase content, low olivine, and moderate glass levels contributed to the material's durability. While this research showed the potential for ISRU applications, further investigation was needed to fully understand the link between regolith composition and sintering quality.
Following this, the team examined the mechanical and structural properties of the sintered Hawaiian basalt aggregates, focusing on the effects of mineral composition. The basalt samples were sieved to enhance particle fusion and the cohesiveness of the sintered material. Adopting the same sintering temperature, the study found that even minor variations in the chemical composition of basalt sources could significantly affect the sintered materials. Samples with different mineral abundances, particularly varying levels of MgO, displayed differences in sinterability. The research highlighted the importance of understanding the mineral composition of basalt feedstock to identify the optimal thermal profile for producing durable and cohesive construction materials. Basalts with higher percentages of plagioclase and pyroxene, and minimal olivine content, were found to be more suitable for creating robust construction materials.

The Microwave Structure Construction Capability (MSCC) project, initiated in 2020 as part of the Moon to Mars Planetary Autonomous Construction Project (MMPACT), aims to develop payload infrastructure on the moon by sintering regolith using microwave energy \cite{effinger2020microwave, rickman2021basics}. Earth-developed Lunar regolith contains "non-Lunar" materials that produce volatiles during sintering, hindering the production of a dense part. Therefore, a heat treatment method was developed to eliminate these materials. Previous vacuum sintering studies in the literature were conducted in enclosed chambers, which benefit from reflection. To more closely replicate Lunar conditions, this project is the first to use a Lunar-like microwave applicator to sinter ceramics in a bed \cite{effinger2021microwave}. 
The latest development of the project includes microwave process advancements to minimize energy input in terms of kWh/kg. Through designed experiments, the process parameters were explored, and a controllable top-down microwave sintering process was developed. A solid paver made of 100\% in-situ resources was sintered in a vacuum chamber using CSM-LHT-1G highlands simulant, weighing up to 1.4 kg \cite{shulman2023microwave}.

In a study by Kim et al. \cite{microwave2023}, microwave sintering of the KLS-1 mare regolith simulant was performed under vacuum conditions at a frequency of 2.45 GHz and within a temperature range of 1080-1120\textdegree{C}. To ensure uniform sintered bodies, KLS-1 cylindrical specimens were preheated to 500\textdegree{C} and then sintered up to 1100\textdegree{C} at a low heating rate of 3-5\textdegree{C}/min. The sintered specimens had dimensions of 25 mm in diameter and 30 mm in height. Samples sintered in a vacuum at the same temperatures had higher density and strength compared to those sintered at atmospheric pressure, as pores can shrink under vacuum conditions. For the sample sintered at 1080\textdegree{C}, the local porosities ranged from 28.5\% to 33\%, with an average porosity of 30.87\%. It was observed that samples sintered above 1120\textdegree{C} had non-homogeneous shapes and a large number of pores due to the evaporation of some molten parts. Samples sintered at 1080\textdegree{C} and 1100\textdegree{C} under vacuum showed mechanical strengths of 17-28 MPa.

In a study by Barmatz et al. \cite{barmatz2011microwave}, microwave permittivity and permeability of four different real Lunar mare and highland regolith from Apollo 11, 14, 16, and 17 were measured to provide a more fundamental understanding of regolith properties and insights into the significance of nanophase iron in the heating of Lunar soils. The findings indicated that at room temperature and the specified frequency of 2.45 GHz, the magnetic constituents of the Lunar regolith do not primarily influence microwave absorption. Since there is a possibility that higher temperatures augment the imaginary relative permittivity and permeability,  it is imperative to conduct additional measurements at elevated temperatures. They have suggested that the high microwave absorption of Lunar soil may not be primarily attributed to the presence of nanophase-sized iron grains. Instead, the absorbed power seems to be closely related to the applied electric and magnetic fields. Consequently, the specific location of a sample within a microwave environment can significantly influence the resulting absorption \cite{barmatz2011microwave, barmatz2012microwave}. Later, their measurements revealed that fragments of Lunar rock, characterized by sharp jagged edges, exhibit improved microwave absorption \cite{barmatz2013microwave}. The presence of sharp edges in a sample alters the distribution of electromagnetic fields. Sharp edges can act as focal points, concentrating the energy and intensifying the heating in those specific areas. The sharp-edged and irregular particles were magnetically heated more efficiently than their rounder counterparts. In Lunar mare and highland regions, micrometeorite impacts can lead to the melting and subsequent recondensation of Lunar soil. This process results in the creation of sharp-edged Lunar regolith particles. These particles seem to contribute significantly to Lunar soil's high microwave absorption capabilities \cite{barmatz2013microwave}.

To enhance the properties of sintered regolith, a recent study by Wan et al. investigated the addition of Fe at concentrations of 2, 5, and 10 wt\% to the CUG-1A Lunar regolith simulant \cite{wan2024iron}. A pressureless sintering technique was employed to produce large blocks with dimensions of 101 mm \texttimes 45 mm \texttimes 20 mm. Due to Fe's significantly higher thermal conductivity compared to the Lunar regolith simulant, it facilitated heat diffusion during the sintering process. This resulted in lower required sintering temperatures and produced blocks with higher densities. The sintered block containing 10 wt\% Fe achieved a density of 90\%, compared to 82\% in samples without Fe. Post-sintering analysis revealed that a substantial amount of Fe remained, acting as a reinforcement phase. Improved mechanical properties at 10 wt\% Fe were achieved, with tensile and compressive strengths reaching 21.01 MPa and 150.92 MPa, respectively.

A recent study investigated the microstructure of microwave-sintered Lunar soil simulant CLRS-2, both with and without nanophase iron. The study confirms Taylor et al. findings that the inclusion of nanophase iron contributes to better properties in the sintered samples compared to those without it \cite{zhang2023np}. Incorporating nanophase iron in the simulant pairs well with microwaves, leading to increased temperature within the sample \cite{zhang2023np}.

\subsubsection{Regolith Concrete}
The concept of "Lunar concrete" has been proposed as an innovative way to use Lunar regolith for construction materials. Despite the distinct compositions of Lunar regolith and terrestrial concrete, researchers have investigated the feasibility of using Lunar regolith to develop concrete-like materials suitable for building structures on the Moon. Beyer et al. \cite{beyer1985Lunarcrete} came up with the idea of Lunar Cement in 1985. Lunar cement can be utilized to construct Lunar habitats and launching pads. Scientists have developed different concretes based on Lunar regolith. The first group is calcium aluminate concrete, which requires a combination of aggregate, water, and cement. Since water access is not easy on the Moon, another method, called dry-mix/steam injection, was developed by Lin et al. \cite{lin1992concrete} that reduced the required water and hydration time by 50\%. Moreover, it resulted in a low porosity and high-strength concrete. To eliminate the use of water, other concretes are developed, namely sulfur concrete, polymer concrete, and geopolymer concrete.

In sulfur concrete, sulfur is used as a binder instead of cement and water. The first step in preparing this concrete is obtaining sulfur that can be extracted from sulfates and sulfides with a purity of 85-95\%. With the assumption that sulfur can be extracted from Lunar resources, studies have been conducted on the properties of sulfur concrete for Lunar applications. Toutanji et al. \cite{toutanji2005strength} manufactured sulfur concrete with 65 wt\% regolith simulant (JSC-1) and 35 wt\% of sulfur. The mechanical properties of the samples were tested using a universal hydraulic testing machine in compression. Samples were subjected to three environmental conditions. The first samples were at room temperature. The second series was kept at -27\textdegree{C} for 50 days and the third series was subjected to 50 freeze/thawing cycles (-180\textdegree{C} to room temperature). Each cycle included 12 hours of freezing and 12 hours of thawing in water. Three samples were tested in each set and the average value of the compressive strength of the first, second, and third conditions were 2530, 2283, and 2559 psi (17.5, 15.7, and 17.6 MPa), respectively. This result showed no significant effect of severe and constant freezing on the strength of the sulfur concretes.

    

Even though sulfur concrete possesses desired mechanical properties, there are some problems associated with using it in the Lunar environment's lack of atmosphere and extreme temperature. Research on the durability of sulfur concrete has shown that it does not have proper performance in vacuum and high-temperature conditions \cite{grugel2008sulfur}. Putting sulfur concrete in a vacuum for 58 days showed sulfur sublimation. Loss of weight due to the sulfur sublimation resulted in the formation of cracks and pores in concrete. It was also observed that sulfur sublimation drastically increases with temperature. According to results, it takes 3.5 years and 2 hours to sublime a 1 cm layer at $15$\textdegree{C} and $120$\textdegree{C}, respectively. Another problem is the low melting temperature of sulfur ($120$\textdegree{C}). As the temperature goes higher than $120$\textdegree{C} near equators on the Moon, melted sulfur may lead to structural failure \cite{grugel2008sulfur}. Another issue is that sulfur is scarce on the Lunar surface, making this method even less suitable for Lunar-based manufacturing \cite{haskin1991Lunar}.

Glass fiber is largely used in terrestrial construction applications and may be a candidate for Lunar infrastructure. To improve the mechanical properties of the concrete based on Lunar regolith, glass fiber was used in a few studies \cite{toutanji2006development, toutanji2008mechanical, meyers2007analysis}. It was shown that adding only 1 wt\% of glass fiber increased the flexural strength by 40\% as well as the ductility of the concrete. Glass fibers were fabricated by melting JSC-1 at $1600$\textdegree{C} and then hand-drawing them using aluminum rods \cite{toutanji2006development}.

Geopolymer concrete is another type of concrete that can be fabricated using regolith. Geopolymer is an amorphous, refractory, inorganic polymer that was first proposed in 1979 \cite{davidovits1979synthesis}. Geopolymer forms through a polymerization reaction between alkali and aluminosilicate-rich material. The activator, alkali, is an amorphous, covalently bonded material \cite{cui2011novel} that is usually a mixture of sodium hydroxide and sodium silicate \cite{rangan2014geopolymer}. Geopolymers are a good candidate for Lunar applications due to having high melting temperature and size stability up to $800$\textdegree{C} \cite{rickard2012thermal, neves2020characterization}. They also have good vacuum stability \cite{wang2016development} as well as radiation shielding properties \cite{montes2015evaluation}. Moreover, they show superior mechanical properties and heat resistance compared to regular concretes \cite{xiong2022mechanical, van2012technical}. Despite the interesting features of the geopolymer concretes, few studies can be found on their utilization for Lunar applications \cite{montes2015evaluation, alexiadis2017geopolymers, wang2016development}.
Geopolymer concrete usually consists of 70-80\% of fine and coarse aggregates and 20-30\% of the geopolymer binder \cite{naser2019extraterrestrial}. The potential existence of alkali metals on the Moon, suggested by the detection of a Lunar sodium tail \cite{matta2009sodium}, could provide a source of alkaline solution for geopolymerization. Narendranath et. al mapped the sodium distribution on the Lunar surface for the first time and derived a global average of $1.33$ $\text{±}$ $0.03$ wt\% \cite{narendranath2022sodium}. Research studies show that geopolymer concretes can be fabricated for Lunar applications containing up to 90\% of Lunar regolith \cite{wang2016development}. Fabrication of geopolymer concrete is a low water-consuming process \cite{xiong2022mechanical}, but water content in Lunar soil is very low, and it is difficult to mine the water resources \cite{chua1998martian}. Wang et al.proposed a process for fabricating geopolymer concretes on the Moon that was equipped with a water recycling device. Using this method, water consumption would be near zero \cite{wang2016development}.

Another alternative to water for Lunar concrete is using polymers as a binder. Different thermoplastic and thermoset polymers, such as epoxy, polyester, polyurethane, polyethylene (PE), and formaldehyde, have been used to bind aggregates and develop polymer concrete. The aggregates and polymer binder volume fraction varies between 75-80\% and 20-25\%, respectively \cite{mani1987comparative}. Binding is achieved by melting or curing reaction of the thermoplastics and thermosets, respectively. Thermoplastics are usually preferred over thermosets as a binder because of the higher UV resistance \cite{alvino1971ultraviolet}, flexible molding conditions, higher toughness, and longer shelf life \cite{sezer2019polyimide}.
The effect of the type of polymer and its volume fraction, processing parameters like mixing and curing protocols, the degree of adhesion between polymer and aggregates, and reinforcing fibers on the mechanical properties of the polymer concretes have been conducted by different researchers for Earth applications. 

In terms of Lunar applications, thermoplastic polymers are preferred as well; specifically, polyethylene is the binder that has been studied the most. Thermoplastic polymers can be melted and recycled multiple times, meaning that when a structure retires, the components can be reused. Moreover, thermoplastic polymers are present on space missions for various uses. Reusing them for construction applications decreases the amount of material that needs to be transported from Earth and can also constitute a waste management approach in space. Chen et al. \cite{chen2015developing, chen2016inorganic} utilized PE at different weight fractions between 2.2 wt\% and 30.3 wt\% as the binder. It has been proved that using different sizes of aggregates, smaller particles fill the free spaces between bigger particles, leading to a lower amount of binder required and improved mechanical properties. To study the effect of the particle size on the properties of the concrete, sieving was done, and three sets of batches were utilized for manufacturing; 1. grain size ranging between \( 90\text{-}112\ \mu \)m, 2. grain size of \( 90\text{-}112\ \mu \)m and \( 20\text{-}25 \mu \)m with the weight ratio of 765:235, and 3. random grain size. Samples were made by mixing PE with JSC-1 Lunar regolith simulant and keeping it at $300$\textdegree{C} for 10 minutes, followed by cooling down to room temperature. The flexural strength results showed an exponential decrease as the PE content was reduced in the system. No specific effect of using a uniform size of regolith on the mechanical properties was observed compared to using as-received agglomerates. However, the addition of smaller particle sizes in the two-grain size particles revealed their enhancing effect on the structural integrity of the composite and, thus, the mechanical performance. It is worth mentioning that in the two-grain size system, a threshold was observed around 10-15 wt\% of PE, above which the strength was almost insensitive to the PE content, and below that, it significantly decreased as the binder content reduced. 

Lee et al. \cite{lee2015manufacture} used PE as a binder to manufacture Lunar regolith-based concrete. They used basalt as the simulant with a size below \( 75\ \mu \)m and 10 wt\% of PE. A vacuum chamber with pressure below 0.1 Torr was used to mimic the Lunar environment. The polymer was melted by attaching a hot plate on top of the mold at $230$\textdegree{C} for 5 hours. Prior to melting, the mixture was preheated at $123$\textdegree{C} for 2 hours to simulate the daily high temperatures on the Moon. Another sample was made without preheating, with a melting duration ($230$\textdegree{C}) of 24 hours. By applying heat from one side, this method of fabrication resulted in a hardened depth, meaning that the concrete was not fully hardened through the thickness (50 mm). The compressive strength of the samples was also measured. According to the results, chamber temperature significantly affected the hardening of the concrete. The non-preheated samples were only hardened in the area with direct contact with the hot plate, meaning that the heat was not conducted to other parts of the mixture within the mold. Therefore, only around 60\% of the mixture was hardened after 24h, while in pre-heated samples, 99\% of the concrete was hardened after 4-5h.

The concrete's strength typically serves as the primary criterion for its acceptance or rejection within a given structure. Specifications or codes outline the required strength, mostly compressive strength, for various components of a structure. Since the gravity of the Moon is one-sixth of the gravity of the Earth, the implemented concrete for the Moon infrastructures should resist more external loading. The compressive strength of the concrete reached 12.6-12.9 MPa after 4-5 hours of heating. This is equivalent to a compressive strength of 70 MPa under Earth's gravity, meaning that, according to the requirements mentioned in \cite{neville2012concrete}, the concrete is strong enough to resist external loading.


In order to decrease the required heating energy, a bottom-up heating method was proposed by Lee et al. instead of the top-down method \cite{lee2018bottom}. They used 10 wt\% of PE as a binder and 90 wt\% of KOHLS-1 regolith simulant with a particle size distribution of 50\% of \( 75\ \mu \)m, 20\% of \( 150\ \mu \)m, 12\% of \( 300\ \mu \)m, 10\% of \( 600\ \mu \)m, 3\% of 1 mm, and 5\% of 2.36 mm. To mimic the atmosphere and daytime temperature of the Moon, a vacuum of 5.0 \texttimes $10^{\text{-2}}$ Torr was produced, and the chamber temperature was set to $123$\textdegree{C}. The mold's width, length, and height were 50, 50, and 100 mm, respectively. The regolith/binder mixture was added to the mold and held for 5 hours with the heating pad set at $200$\textdegree{C}. In this setup, the temperature progressively decreases to $132$\textdegree{C} at the farthest point from the heating source. The compressive strength of the samples was measured by cutting them in half and measuring the strength of the bottom and top parts separately. Results of mechanical tests showed the bottom part had a higher strength due to better bonding between particles as it was closer to the heating pad, and a significant increase occurred between 3 and 4 hours of heating. SEM observations revealed physical bonding between PE and KOHLS-1 without chemical reaction. Compared to the top-down heating method, half of the strength was obtained in manufactured concretes with bottom-up heating. However, lower heating energy was consumed as the temperature decreased by $30$\textdegree{C}. Moreover, considering the weight of the Lunar roving vehicle and Lunar module from the Apollo mission, they showed the suitability of the bricks produced by bottom-up heating for practical applications.

Zhang et al. \cite{zhang2022mechanical} developed carbon nanofiber (CNF)/geopolymer nanocomposites using BH-1 Lunar regolith simulant and examined them through both theoretical and experimental methods. They evaluated a ball-milling technique for dispersing CNFs into the simulant and analyzed the mechanical properties of the resulting nanocomposites. Characterization was performed using XRD, SEM, Fourier transform infrared spectroscopy, and mercury porosimetry. The study found that CNFs significantly improved mechanical properties, with the optimal CNF ratio being 0.3 wt\%. Specifically, enhancements included a 34.8\% increase in flexural strength, a 7.5\% increase in Young’s modulus, an 83.9\% boost in flexural toughness, a 21.4\% increase in maximum displacement, and a 13.1\% rise in compressive strength. Microstructural analysis indicated that CNFs acted as nucleation sites, fillers, and bridges, reducing porosity and increasing the energy required for failure \cite{zhang2022mechanical}.

Debbarama et al. \cite{debbarma2024fiber} developed fiber-reinforced Lunar geopolymers with LMS-1 and LHS-1 regolith simulants, incorporating basalt fibers (BF) and human hair fibers (HHF). They tested these geopolymers under various curing conditions—ambient, high-temperature, and low-temperature—across different curing times and humidity levels. The study revealed that the LHS-1 geopolymer outperformed LMS-1 in both flowability and compressive strength. This was attributed to the lower Si/Al ratio and higher Ca/Si ratio in LHS-1, as shown by SEM-EDS analysis, which directly correlates with its enhanced strength. Curing conditions had a significant effect, with LHS-1 showing the best performance under high-temperature and controlled environments, while LMS-1 exhibited consistent strength across all conditions. Adding up to 1.2\% BF and 1.5\% HHF improved the flexural strength but reduced compressive strength due to increased porosity. SEM analysis further demonstrated that the fibers contributed to better flexural strength and microstructure by acting as nucleation sites and structural bridges \cite{debbarma2024fiber}.

\subsubsection{Binder-Based}
Binder-based methods for making Lunar infrastructures involve using a binding agent derived from locally available materials or brought from Earth to solidify and stabilize Lunar regolith or other materials for construction purposes. This method aims to create structures and components necessary for habitats, landing pads, or other infrastructure on the Moon. The binder acts as a glue or adhesive to hold the particles together, forming a cohesive and durable material suitable for construction in the Lunar environment. The binder that is usually used in the literature for Lunar applications is a thermoplastic or thermoset polymer. Compression molding is used in this method to reduce the amount of required binder compared to polymer concrete.

Chen et al. \cite{chen2017high} employed a high-pressure densification method to produce low-binder content parts. The JSC-1/polyester resin mixture was compression molded with a peak compaction pressure ranging from 30 to 700 MPa. The effect of the other processing parameters, e.g., compression duration, vibration, loading rate, loading number, and initial grain size distribution, was investigated as well. Compression pressure is the most critical parameter in controlling the final binder content. High pressure pushed a significant amount of the binder out of the mixture, and by applying 350 MPa, the resin content was reduced to 6.5 - 8.7 wt\%. Pressures higher than 350 MPa didn't significantly further reduce the binder amount.

In another study, Chen et al. \cite{chen2018formation} introduced a method to produce ultra-low binder content (between 2 - 5\%) parts. In regular compression molding, polymer content should be high enough to provide full wetting with agglomerates. When the interstitial space between agglomerates is filled with the binder, the portion of the binder that is in direct contact with the particles directly determines the load-bearing capacity. The rest of the binder, filling the interior space, does not contribute to load-carrying equally. A more efficient way of using a binder is the formation of polymer micro-agglomerates (PMA) in the contact area of the particles that carry the load. To achieve PMA, the compaction process requires a maximum pressure exceeding the crushing strength of the agglomerates. This ensures the breakage of large binder drops and the formation of a close-packed configuration. Capillary forces facilitate the surface diffusion of the binder at the contact points between agglomerates under increased pressure, which leads to the formation of PMA. Utilizing this method, they produced JSC-1/polyester composite Lunar cement with a 2 - 5 wt\% binder content in a two-step compression molding. SEM observations before and after the first compaction of the regolith simulant/binder mixture showed that small clusters of regolith simulant were disrupted and the large droplets of the binder became smaller. The second compaction was essential to achieve better mixing of simulant and binder. SEM images show evidence of the loosely attached simulant grains due to low compaction pressure, which caused incomplete dispersion of the binder. The flexural strength ranging between 30 - 40 MPa was achieved, comparable to typical steel-reinforced concrete \cite{qiao2007simplified}. Mimicking the Lunar large temperature gradients for flexural testing (-200\textdegree{C} to $130$\textdegree{C}) showed a satisfactory strength in the range.

To benefit from the advantages of thermoplastic polymers e.g., higher UV resistance, longer shelf life, higher toughness, and more flexible processing conditions compared to thermosets, Oh et al. \cite{oh2020ultralow} utilized polyether ketone ketone (PEKK), polyimide (PI), and polyamide nano clay composites (PAnc) to make ultralow binder content regolith-based cement. The grain size of regolith simulant, PEKK, PI, and PAnc is 100, 50, 1-10, and 50-200 \(\mu \)m, respectively. The polymer content was between 3 and 6 wt\%. One-step compression molding was used to manufacture CLS. Compaction of the mixture at low temperatures and in the solid form of the binder leads to an irreversible dispersion of the binder after removing the pressure. Short-range movements achieve dispersion of the binder through grain deformation, rotation, and sliding. During heating at temperatures higher than $T_{g}$ or $T_{m}$ of the binder, an inner pressure, arising from the volume expansion of the binder, drives the binder to the narrowest space which is the contact point between the agglomerates.

Results of flexural strength (R) as a function of binder content showed that binder content increased flexural strength. Flexural strength of 10 MPa was achieved by adding only 3 wt\% of the PEKK, which is similar to the steel-reinforced concrete (10-15 MPa) and much higher than the regular unreinforced concrete (~5 MPa) \cite{wight2016reinforced}. Maximum flexural strength of 35 MPa was obtained at 6 wt\% of PEKK and the heating temperature of 450-500\textdegree{C}. Even though PAnc binder itself has the highest strength compared to the PI and PEKK \cite{agag2001studies}, PEKK-based concrete parts showed the highest strength, which could be attributed to the larger agglomerate size of PAnc and its larger effective viscosity \cite{jang2005effect}.


\subsection{Pre-treatments} \label{Prep}

In manufacturing, pre-treatment refers to the processes applied to a material or surface before the primary manufacturing process. It can encompass a range of treatments, from surface cleaning to surface treatments like plasma spraying, chemical etching, or mechanical roughening, which are applied to the powders to improve final properties after sintering. However, in this section, we will discuss the commonly employed techniques on regolith particles and how they affect the final properties of the materials. This is particularly relevant to this paper, where we primarily focus on the manufacturing of regolith-based materials.

For any process involving sintering, the regolith particle size distribution (PSD) and morphology can significantly impact the properties of the final product \cite{popovich2020additive}. Smaller particle sizes typically result in higher post-sintering density due to improved initial packing, increased surface area, reduced sintering time, and temperature requirements. In the case of liquid-phase sintering, smaller particles enable easier particle movement driven by capillary forces. This is particularly advantageous in binder-based processes, where reducing particle size can lower the melt flow index, achieving a more homogeneous blend and allowing for a higher weight percentage of solid (regolith) content. To reduce regolith particle size or deagglomerate regolith particles, common methods include milling or grinding. The use of ultrasonic waves is less common, as it often involves liquid materials, which are not feasible in the vacuum of space. Ball milling is the most prevalent technique in the literature, but caution should be exercised when using liquid process control agents (PCAs) for in-space processes. Sieving, either alone or in combination with milling, is also highly effective \cite{rahimian2009effect, yeh1988effect, altun2021additive, heady1970analysis, walter1997outgassing}.

Particle morphology is also crucial, primarily because it influences initial packing efficiency. Before the PBF process, there might be a step to improve the spreadability of the powder. This can include coating, morphology improvement, or moisturizing treatments \cite{cordova2020measuring}. Surface treatments, such as plasma spraying a layer of metals onto regolith powder before fabrication, can modify particle behavior during sintering and enhance post-sintering properties. When possible, the powder used in PBF should consist of spherical particles \cite{zhao2020significance}. A coating can also be applied to the powder before laser processing to reduce optical reflectivity \cite{bidulsky2021coated}. Another form of material pre-treatment involves blending regolith with another powder, which may include incorporating metals into the regolith before laser-based fabrication to enhance post-sintering properties. This step can be conducted to enhance laser absorption by the ceramics, allowing for higher particle temperatures during laser processing. It can involve using molten metallic particles as a filler to seal gaps and pores between the ceramic particles \cite{azami2023laser}. Drying is a crucial pre-treatment step, as the presence of moisture during processing can lead to defects and negatively affect the final properties due to gas formation \cite{lambiase2023moisture}.

\subsection{Post-treatments}\label{Post}

Post-processing, also known as post-treatment, refers to the additional steps or procedures conducted on a product or material after the initial manufacturing or production phase. These supplementary measures are applied to modify, enhance, or refine the product's properties, appearance, or functionality. Since it requires extra time, energy, materials, or equipment, it is essential to design parts to minimize the need for post-processing. Additive manufacturing, in particular, has the potential to reduce the reliance on post-processing. This includes tasks such as removing support structures, performing secondary machining for precise dimensions, assembling and joining components, refining surface characteristics, applying surface coatings or paint, subjecting the item to heat treatment, hot isostatic pressing (HIP), and incorporating materials, among other techniques \cite{shiyas2021review, mahmood2022post}.

Heat treatment is one of the most commonly used post-processing techniques for AM parts and involves subjecting the 3D-printed part to controlled heating and cooling processes for stress relief or to alter its microstructure and properties \cite{rajan2023heat}. For metals and metal matrix composites (MMCs), it can include annealing, hardening (quenching), aging, and stress relieving, among others \cite{shiyas2021review, karamimoghadam2022comparative, mahmood2022post}. For ceramics and ceramic matrix composites (CMCs), it can involve annealing and stress relief, as well as different types of firing and sintering, such as spark plasma sintering (SPS) and Sinter-HIP (simultaneous sintering and HIP) \cite{kuang1997review, rahaman2017ceramic}. In the case of polymers and polymer matrix composites (PMCs), stress-relief annealing is less common, as residual stresses are gradually relieved upon fabrication. However, for semi-crystalline polymers and PMCs, annealing is typically conducted to improve sample crystallinity \cite{tamburrino2021post}. In the context of ISM, where regolith (a ceramic) is preferred over other raw materials, sintering is the literature's most commonly used post-processing technique. This can include polymer debinding, followed by sintering as covered in Section \ref{SLA}. Also, when no binder is used, sintering can improve regolith parts' density and overall properties since they are usually porous \cite{chavez2020influence}.

Here are a few examples of sintering and infiltration, which are the most commonly used post-processing methods for ceramics and CMCs, especially regolith. Altun et al. \cite{altun2021additive} sintered EAC-1A powders at 1000 \textdegree{C} following polymer debinding. SPS is also used to improve the properties of regolith-based materials, as demonstrated by Popovich et al. \cite{popovich2020additive}. The porosity was reduced from 23.7\text{\text{\%}} \text{±} 5.4 to 11.7\text{\%} \text{±} 3.3 when the SPS temperature was increased from 1025 \textdegree{C} to 1075 \textdegree{C}. This temperature increase also resulted in a significant increase in hardness from $443 \text{±} 56$ HV to $743 \text{±} 142$ HV \cite{popovich2020additive}. Infiltration of another material, such as an alloy or a polymer, can be another method to reduce porosity and improve the characteristics of regolith-based samples. Azami et al. \cite{azami2023laser} used a polymeric resin (Dichtol WTF 1532, Metaplastic, Germany) to infiltrate the cracks and pores of the alumina-based additively manufactured CMC specimens. According to the data provided by the resin manufacturer, Dichtol polymer has a service temperature of -40 \textdegree{C} to +300 \textdegree{C}. The polymer impregnation improved the compressive strength of the sintered samples (from 56 to 120 MPa) \cite{azami2023laser}. Metal infiltration is also a promising method for enhancing the properties of porous materials. It can be conducted through vacuum infiltration. Typically, the wettability of the ceramic with the infiltrant metallic material is critical for the success of this process, as capillary forces play a role in the infiltration \cite{vogt2010improving, li2021k}. Infiltration can also be carried out by high-pressure casting \cite{peng2004mechanical}.

\section{Construction and Manufacturing of Large-Scale Objects: The Potential of Robotics}
Traditional additive manufacturing is constrained by the workspace of the printer. The construction of large-scale objects requires either the scaling up of the workspace or novel methods to escape its confines.

There have been high profile examples of the use of 3D printing robotics for architectural applications, such as the MX3D steel bridged constructed with robotic Wire Arc Additive Manufacturing, and the HuaShang Tengda $400 m^2$ mansion constructed near Beijing ~\citep{leonard2018case}. Companies like PERI 3D Construction and COBOD are continuing to rapidly push the boundaries of this technology's present-day terrestrial applications.

The application of additive manufacturing to large-scale objects has been explored for construction since the introduction of Contour Crafting ~\citep{khoshnevis2004automated}, which was a pioneering 3D concrete printing technique, and of D-Shape~\citep{dini2008method}, which pioneered construction-scale binder jetting (using a magnesium oxychloride binder with a sand substrate). Both these techniques involve first setting up a gantry system that is larger than the object being printed. Despite the significant challenges and constraints imposed by requiring a large gantry system, Contour Crafting features as part of a NASA technology demonstration program, named Additive Construction with Mobile Emplacement (ACME), meant to demonstrate the feasibility of Lunar AM construction~\citep{mueller2017additive}. More recently, ICON has made significant advances in 3D printing structures using a large gantry, and completed 3D printing of NASA's Mars Surface Simulated Habitat in 2022 ~\citep{agnihotri20233d}.

The other main robotic configuration for positioning in general, and positioning of a printhead specifically, is a serial manipulator. Such robots have more flexible workspaces and the ability to point the printhead at varying orientations. This configuration was used by the Institute for Advanced Architecture of Catalonia's Mataerial project~\citep{finnane2013mataerial}, and by XtreeE for more advanced concrete 3D printing~\citep{gosselin2016large}. In general, there have been substantial developments in manipulator-based additive manufacturing that are highlighted in recent review papers~\citep{paolini2019additive,bhatt2020expanding}, and today most major robotic manipulator manufacturers, including ABB and KUKA, offer 3D printing application solutions, either directly or through partners. Large-scale polymer extrusion using manipulator arms includes the Oak Ridge National Laboratory's Big Area Additive Manufacturing (BAAM)~\citep{duty2017structure}, as well as commercial work by DUS Architects, BLB Industries, WASP, Thermwood, Ai Build, Branch Technologies, and others. Toward the largest end of the scale, and specifically for construction applications, Apis Cor uses a manipulator on a swivel, similar to a tower crane, with a reach of 8.5~m~\citep{puzatova2022large}, while CONPrint3D uses a truck-mounted concrete pump whose extensive boom serves as a robotic arm~\citep{mechtcherine2019large}.

The cost of launching mass from Earth is a strong driver to move away from large massive gantry structures in Lunar applications. Further, as robotic manipulators grow in scale, maintaining precision requires scaled stiffness and, thus also, mass. The ability to build large-scale objects with smaller, less massive robots is of particular interest on the Moon. This section first looks at methods to increase printed object scale by extending a robotic printer's workspace, then discusses specific ways to achieve mobile 3D printing, and finally extends this discussion to teams of mobile robots.

\subsection{Escaping Workspace Constraints: Conveyors and Mobile 3D Printing}
To print large-scale objects without utilizing large massive printing robots, methods must be used that extend printing beyond the confines of a small printer's workspace. This can be achieved either by moving the part being printed through the workspace (i.e., on a conveyor) or moving the workspace itself along the part being printed (i.e., Mobile 3D Printing).

Voxeljet developed continuous powder-based 3D printing with binder jetting, based on silica sand with a furan resin binder~\citep{gunther2014continuous}, that uses a conveyor to pass material continuously through the printer to manufacture large parts. Various university groups have subsequently also developed conveyor belt-based 3D printer projects ~\citep{mccarthy2018conveyor,ozsoy2020design}.

Early mobile 3D printing was attempted with extrusion printheads mounted to simple mobile robot platforms, by De Sa Bonelli~\cite{de2016pratica} followed by Marques et al.~\cite{marques2017mobile}. Any horizontal printhead motion was achieved by moving the robot, and as no sensor was applied other than wheel odometry, the accuracy of these initial attempts was low, and the error continuously accumulated. Tiryaki et al.~\cite{tiryaki2019printing} incorporated computer vision-based localization using ArUco markers along with a 6DOF manipulator-arm-mounted printhead to achieve "printing-while-moving".

Another related pioneering mobile robotic additive manufacturing approach was the Minibuilders system developed at the Institute for Advanced Architecture of Catalonia~\cite{IAAC2013}, which includes a "grip robot" that climbs a structure that is actively in the process of printing via extrusion of cementitious material.

Keating et al. \cite{keating2017toward} demonstrated the Digital Construction Platform (DCP), which consists of a compound arm system (manipulator arm at the end of the rotating boom) carried on a tracked mobile platform. "Printing while driving" and "print from a stationary position" strategies are both mentioned, but only the latter strategy (with all printing done from a single position) is conducted as a case study in their work. They mention requiring additional sensor feedback for the full-body motion planning required in the "Printing while driving" strategy. CyBe and InSitu Fabricator systems also involve manipulator arms on tracked bases, but in practice are used in a quasi-stationary manner~\cite{dorfler2022additive}.

Localization and printhead positioning are crucial to robotic mobile 3D printing~\cite{dorfler2022additive}. Li et al. developed Simultaneous Localization and Additive Manufacturing (SLAAM) to fuse local (3D scanner) and global (onboard laser total station) sensing to achieve print-drive-print mobile 3D printing of large looped objects with high dimensional accuracy~\cite{li2020slaam}; performance was demonstrated in the lab with a mobile Delta-configuration printer.

\subsection{Extensions to Teams of Mobile Robots}
One of the advantages of mobile 3D printing is its potential for parallelization. Once an object is being segmented in a printing process, it is conceivable to assign different segments to different robots for more rapid completion.

Zhang et al.~\cite{zhang2018large} demonstrated pre-coordinated printing of 2 cement segments by 2 robots that interfaced at the union of the segments such that it resulted in a unitary piece. On the other hand, Sustarevas et al.~\cite{sustarevas2019youwasps} have focused research on task decomposition and allocation. Still, in its very early stages, mobile 3D printing by teams of mobile robots is expected to play an important role in the future of this research area.

\section{The Potential of Artificial Intelligence for LBMC}

The integration of machine learning in manufacturing has been well-established in recent years, aiming to enhance efficiency and precision in the manufacturing process \cite{rai2021machine, qin2022research, razvi2019review, qi2019applying, wuest2016machine}. Progress in data acquisition and storage technologies has led to the widespread adoption of machine learning-based data-driven approaches to streamline manufacturing processes. Machine learning models can address challenges in the manufacturing process in terms of product design, materials, process optimization, quality control, and \textit{in situ} monitoring during manufacturing operations. The application of machine learning is not limited to terrestrial operations; machine learning could also potentially be implemented in construction and manufacturing endeavors on the Moon's surface.

The selection of an additive manufacturing method depends on factors such as the regolith's properties, including particle size, composition, mineralogy, and physical attributes like thermal conductivity. These characteristics vary across various Lunar sites \cite{mckay1991Lunar, noble2009Lunar, zhao2023variations}.
One significant area of machine learning application involves the analysis of Lunar soil samples. Machine learning algorithms can be specifically trained to identify minerals and chemical compositions present in Lunar soil samples, as well as their physical characteristics. In a study conducted by Kodikara et al. \cite{kodikara2020machine}, machine learning algorithms were implemented to categorize mineralogical and physical attributes of Lunar soil. The classification was based on criteria such as soil type (mare soil and highland soil), particle size, maturity, and the predominant type of pyroxene (High-Ca and Low-Ca), utilizing reflectance spectra. The researchers used data from the Lunar Soil Characterization Consortium (LSCC) database, which includes compositional and spectral information derived from Lunar highland and mare soil samples collected during the Apollo missions for training and validating their models. In another study, the morphological characteristics, such as size, shape, and distribution of regolith particles, as well as the physical attributes like density, porosity, and composition of the Lunar regolith sample retrieved by the Chang'e-5 mission, were evaluated through the integration of high-resolution micro-CT imaging and an image processing approach based on machine learning \cite{wu2023micro}.

Machine learning has the potential to mitigate challenges in other aspects of Lunar-based additive manufacturing, leading to improved designs, optimized processes, real-time monitoring, and enhanced properties of the final print. Exploring these areas opens up opportunities for future research, pushing the limits of what we can achieve in space exploration and Lunar construction. Before the AM process begins, designers must define parameters to optimize the design and manufacturing process and ensure the desired production quality. These parameters include production path strategies, part locations, and build orientations to enhance the quality and characteristics of the final printed products \cite{zhang2018statistical}. Support structures are also important in certain processes in which the designed models contain overhang parts with no solid material underneath. A machine learning model can be developed that enables designs for which a minimum amount of support is required \cite{huang2019surfel}. For printing composite materials, factors such as fiber size, volume fraction, and direction must be optimized to secure the desired properties for the printed part \cite{yanamandra2020reverse}.

Efficient quality assurance and monitoring technologies to detect the onset of defects and control the printing process are also crucial \cite{everton2016review, grasso2017process}. Given the process and material used, a wide range of defects and quality issues may arise during the additive manufacturing processes. Powder-based processes are prone to issues of porosity, lack-of-fusion, balling, and cracks \cite{everton2016review, zhang2017defect, taheri2017powder}, while geometry deviation, shape shrinkage, surface imperfections, layer misalignment, and adhesion issues \cite{lee2014development, wang2016plane, xia2022modelling, gerdes2021hyperspectral} are the main concerns in the FFF processes. The timely detection and prompt rectification of these defect issues during the process are of utmost importance. With the advancement of data acquisition, machine learning algorithms can be applied to address this aspect \cite{bisheh2021layer, bugatti2022towards}. The fusion process and consequent properties and characteristics of the printed material are determined by the process parameters. There is a complex interaction between the process parameters as well as nonlinear correlations between the input parameters and the output quality of the fusion. Machine learning models can be developed to predict mechanical properties of materials fabricated through additive manufacturing as they can learn and model highly complex relationships \cite{herriott2020predicting, yan2018data, koeppe2018efficient, zhou2019accelerating, zhan2021novel, demir2021laser, muhammad2021machine, nasiri2021machine}. In particular, the process conditions can be linked with properties of the final products, such as strain rates \cite{mehrpouya2019prediction} and compressive strength \cite{sood2012experimental}, tensile strength \cite{zhang2019deep}, tensile modulus, nominal stress, and elongation \cite{baturynska2019application} using artificial neural networks. Machine learning models can also be trained to quantify spatial variations within material properties and measure local mechanical properties of printed materials based on digital image correlation (DIC) strain fields as input \cite{10.1115/IMECE2022-95195, pitz2023estimation}.

\section{Outlook and Conclusion}

\subsection{Summary of the Current State of the Art}
This review encompasses various aspects of Lunar-based manufacturing and construction (LBMC). It discusses the necessity of advancing LBMC for future space exploration and its synchrony with the ongoing research and roadmaps of space agencies, including NASA. It describes how different environmental characteristics of the Moon, including vacuum conditions, low gravity, energy constraints, and large temperature fluctuations, among others, affect procedures and material choices and impose limitations. It includes a description of what Lunar regolith offers for in-situ resource utilization, including mineralogical and morphological characteristics of the regolith and how it varies from one region to another on the Lunar surface. Additionally, it addresses Lunar regolith simulants developed to represent Lunar regolith and facilitate potential research on LBMC and space exploration as a whole. 

The review also explores the potential for extracting metals and oxygen from Lunar regolith, a technology that remains at a low to medium maturity level but could revolutionize space exploration, particularly in Lunar-based manufacturing. While this technology appears energy-intensive—especially since many extraction processes require high temperatures—its necessity is clear due to the crucial need for oxygen production on the Moon. For instance, molten regolith electrolysis requires melting regolith at a minimum of around $1200$\textdegree{C}, while molten salt electrolysis requires salts to be melted at around $700$\textdegree{C} or above. Furthermore, some acid-leaching processes with common acids require heating to high temperatures. Despite these challenges, the use of Lunar regolith, along with ice water, as primary sources for oxygen extraction is indispensable. Oxygen is vital for life support and propulsion, making it crucial for the expansion of Lunar habitation and exploration. So, metals recovered as byproducts in these processes could further support Lunar infrastructure.

The primary focus of this review is on various manufacturing and construction techniques for LBMC, encompassing both Additive Manufacturing (AM) and traditional (non-AM) methods. The review evaluates their current state, the potential advantages they offer, and the specific challenges and limitations they encounter or impose. A brief assessment for each AM or traditional technique is included in Table \ref{AMTechniquesassessment} and Table \ref{NON-Techniquesassessment}, respectively. Please note that these tables provide an evaluation of technological maturity stages relevant to outdoor ISRU-based fabrication on the Lunar surface, regardless of the weight percentage of regolith involved. In this context, a low level of technological maturity implies the presence of inherent issues requiring attention for feasibility or there has been an insufficient amount of research to facilitate a comprehensive assessment. Conversely, a medium stage of maturity indicates the absence of inherent issues, with successful research pieces available for consideration. A high stage of maturity would signify that the technology has undergone necessary testing and seems to be immediately ready for deployment in the specified context; no technology has yet to meet this threshold. When considering the applicable size ranges for each technology, small-scale refers to fabricating components with relatively compact dimensions (typically a few centimeters) and involving intricate details. Medium-scale implementation occurs when the technology is employed for the fabrication of components with dimensions that fall between small and large, striking a balance between detail and overall size. On the other hand, large-scale technologies are harnessed for the fabrication of components or products with substantial dimensions (typically around 1 meter or larger), with specific size-influenced material requirements. Items fabricated on a large scale may find suitability in applications such as construction, structural components of spacecraft, or other macro-scale uses. It is noteworthy to highlight that certain techniques, such as regolith sintering, may not facilitate the creation of large structures in a single operation. However, the bricks produced through these methods can be utilized for constructing sizable structures.

Early investigations on Lunar-based manufacturing date back to the 1980s, focusing initially on regolith casting and sintering methods. Both methods involve the application of high temperatures, reaching up to around $1200\text{-}1300$\textdegree{C}, making them energy-intensive processes. Sintered regolith has been shown to meet the structural needs in different studies, even though the requirement for a furnace or microwave chamber for sintering imposes constraints on the size of the produced components. Despite the energy-intensive nature of melting regolith in regolith casting, this method holds potential benefits for LBMC, particularly in the fabrication of glass fibers from regolith, an area where knowledge is currently limited. 
Regolith concrete also holds great potential for LBMC. Geopolymer concrete can maintain stability under elevated temperatures and vacuum conditions. However, its reliance on water poses a constraint on its application. Polymer concrete and binder-based techniques that use polymers as binders have the advantage of being available in the waste generated during space missions, offering potential for recycling and reuse. One limitation involves the substantial compression pressure required for producing parts with ultra-low binder content, reaching up to 350 MPa.

To compare the potential of AM and non-AM techniques for LBMC, it should be noted that as awareness and adoption of AM saw a significant increase in the 2010s, it also gained attention in LBMC. Due to its ability to create complex geometries, reduce material waste, and enable rapid prototyping, it continued to grow and became more popular for extra-terrestrial manufacturing compared to traditional manufacturing methods. The larger number of publications and ongoing projects shows the focus on AM methods. However, traditional methods have been complementary to AM methods as the knowledge developed in these methods can benefit advancing AM methods.

To conclude a summary of the current state of each AM technology for LBMC, SLA has the highest maturity level for indoor manufacturing of small precise parts. However, a significant challenge arises for fabrication in outdoor vacuum conditions due to outgassing issues. SLA also encounters recycling challenges and curing/debinding/sintering issues, particularly for large-scale parts. PBF technologies demonstrate great potential, enabling the use of a high ratio of regolith in fabrication. However, they are still at a relatively low maturity level, and there is a shortage of comprehensive publications on the subject. Most available literature consists of conference papers or projects conducted by companies and space agencies. Printing large-scale parts using these technologies proves challenging due to the nature of the technologies and the behavior of regolith as a ceramic (as opposed to metal), making this approach currently more applicable for parts manufacturing. The feasibility of achieving higher process control during SS-PBF could be especially promising, as it offers great potential in energy requirements. DED-based technologies share similar advantages and challenges with PBF. They have a more challenging powder feeding system compared to PBF but exhibit better compatibility with repairing processes. EMR technology holds significant promise as one of the few remaining options for utilizing a high wt\% of regolith without relying on polymeric binders as off-site materials, enabling the fabrication of large-scale structures and constructions. There is potential to source molten regolith from the electrolysis of molten regolith, should this process be employed for oxygen extraction. However, process control poses challenges, and the technology is still at a low maturity level. Among other ME-based technologies, FFF and FGF show great potential due to their simplicity, low energy consumption, process speed, and recyclability of the thermoplastics used. CC emerges as the most promising technology for large-scale Lunar-based construction, although challenges exist, including evaporation, outgassing, and viscosity increase in a vacuum condition. DIW offers good precision but faces binder-related issues such as outgassing, boiling, and freezing in a vacuum, along with recycling challenges. BJ presents similar promises and challenges, with the nature of its feeding system making it even more challenging in a vacuum. 

Despite a significant portion of the research on LBMC focusing on polymeric materials, a major challenge remains: polymers are inherently off-site materials. This, first of all, makes these approaches most attractive for smaller-scale specialty parts rather than large-scale construction. While manufacturing polymer-based parts on Earth and transporting them to the Moon may be a cost-effective option in some cases, several factors underscore the importance of on-site manufacturing, particularly for polymer/regolith composites. Firstly, in cases where a part is urgently needed on-site, the time required for fabrication and transportation from Earth can be prohibitive. This underscores the need to enhance the capability for on-site manufacturing. Secondly, ongoing research worldwide aims to enable the fabrication of polymer/regolith composite parts through material extrusion, with in situ additives potentially making the process significantly more economical. Finally, FFF or FGF rely on thermoplastics or their composites. These methods can use raw materials repurposed from failed/obsolete parts and scraps, offering significant potential to make the process more economical.

All in all, among various AM techniques, none takes precedence over others, although the current maturity level for the techniques is different. One can prioritize them considering equipment and material accessibility, energy availability, energy usage deliberations, environment (including whether it is an indoor or outdoor process), costs and budget considerations, material characteristics requirements, size, and geometry of the parts, among other factors, summarized in the tables.

This paper also encompasses applicable pre-treatment (\ref{Prep}) and post-treatment (\ref{Post}) strategies aimed at enhancing fabrication quality. This offers readers various potential approaches to improve quality, enabling them to make informed choices.

Another section of the review explores the potential of robotic additive manufacturing, employing either a single robotic 3D printer or a team of such robots, to overcome chamber-size constraints and fabricate large-scale structures. This section delves into the advantages of using small mobile robots in additive manufacturing, highlighting their advantages over large gantry-based AM machines when printing large-scale objects, particularly considering the significant mass/volume constraints on the Moon.



This review also covers the potential of AI in LBMC. This section explores the existing literature, addressing how AI can impact various aspects of LBMC. This includes the analysis of soil characterization, such as morphological features, chemical compositions, mineralogy, and physical attributes. These characteristics vary across Lunar sites and affect the selection of manufacturing processes. Additionally, machine learning models can optimize manufacturing by fine-tuning path strategies, determining part locations and build orientations, and minimizing support structures for overhang parts. The significance of efficient quality assurance and monitoring technologies for detecting defects and controlling printing is also emphasized. Timely detection and prompt rectification of defects during the manufacturing process are crucial. With advancements in data acquisition, machine learning algorithms can address these challenges. Models can also be developed to predict the mechanical properties of materials fabricated through additive manufacturing processes based on input process parameters, taking into account complex and nonlinear correlations.




\begin{landscape}
\begin{table}[htp]
\centering
\caption{Assessment of Manufacturing and Construction Techniques for LBMC - AM Techniques}  
\label{AMTechniquesassessment}
\scriptsize 
\setlength{\tabcolsep}{3pt} 
\begin{tabularx}{\linewidth}{l l X X X X X X X X}
\toprule
\textbf{Category} & \textbf{Technology} & \textbf{Outdoor ISRU LBMC TMS} & \textbf{Applicable Materials} & \textbf{Max \% of Regolith} & \textbf{Applicable Size Range} & \textbf{Indoor/Outdoor} & \textbf{Energy Consumption} & \textbf{Current Advantages} & \textbf{Current Challenges} \\
\midrule
\multirow{4}{*}{PBF} & L-PBF & Low to Medium & Metals, Polymers, and Ceramics & High to Very High & Small to Medium & Both & Moderate to High & High Precision, No Additives or Chemicals Needed & Low Optical Absorption of Regolith Causes Voids and Pores\\
& EB-PBF & Low & Metals Mainly & High to Very High & Small to Medium & Both & Moderate to High & High Precision, Less Influenced by Material's Absorption Characteristics & Less Effective for Non-metals, Powder Emission in Microgravity \\
& SS-PBF & Low & Regolith Mainly & Very High & Small to Medium & Outdoor & Uses Solar Energy Directly & Energy Efficient & Difficulties in Process Control, Low Strength, Geometric Inaccuracies \\
& MWS-PBF & Low to Medium & Various & Very High & Small to Medium & Indoor & Low & Energy and Time Efficient, High MW Penetration into Regolith, Consistent Sintering & Difficulties in Process Control and Resolution.\\
\midrule
\multirow{5}{*}{ME} & FFF & Medium & Polymer Matrix, Potentially Plus Solid Materials, including Regolith & Medium & Small to Large & Both & Low & Simplicity, Cost-effectiveness & Layered Structures\\
& FGF & Low to Medium & Polymer Matrix, Potentially Plus Solid Materials, including Regolith & Medium & Small to Large & Both & Low & Eliminates Filament-making Step & Less Control Compared to FFF\\
& EMR & Low & Regolith & Very High & Medium to Large & Outdoor & Very High & ISRU, Molten Regolith Provided from Oxygen Extraction Process &  Hard to Control\\
& DIW & Low & Polymer Matrix, Potentially Plus Solid Materials, including Regolith & Medium & Small to Medium & Indoor & Low & Customizable, Precision & Recycling Challenges, Binder Melting/Freezing and Volatility.\\
& CC & Medium & Regolith-based Cement & Medium to High & Large & Outdoor & Low & Construction Speed & Water Evaporation \\
\midrule
\multirow{2}{*}{SLA} & & Low & Photopolymers, Potentially Containing Regolith & Medium & Small & Indoor & Low & High Precision, Complex Shapes & Indoor Only, Recycling Issues\\
\midrule
\multirow{2}{*}{BJ} & & Low & Binder + Various Materials, including Regolith & Medium & Small to Medium & Indoor & Moderate & Fast, Multi-material & Recycling Concerns, Imprecise Droplet Ejection in Low-g, Ink-related Issues: Outgassing, Boiling/Freezing\\
\midrule
\multirow{3}{*}{DED} & L-DED & Low & Metals, Ceramics, and Polymers & High to Very High & Small to Large & Both & Medium to High & Build on Existing Structures & Low Optical Absorption of Regolith Results in Porosity\\
& EB-DED & Low & Metals Mainly & High to Very High & Small to Large & Both & High & Build on Existing Structures & Difficulties in Working with Non-metals \\
& WAAM & Low & Metals & Low & Medium to Large & Indoor & High & Build on Existing Structures & A Metallic Matrix is Essential\\
\bottomrule
\end{tabularx}
\end{table}
\end{landscape}


\begin{landscape}
\begin{table}[htp]
\centering
\caption{Assessment of various manufacturing and construction techniques for LBMC - Non-AM Techniques}  
\label{NON-Techniquesassessment}
\scriptsize 
\setlength{\tabcolsep}{2pt} 
\begin{tabularx}{\linewidth}{@{} l l X X X X X X X X X @{}}
\toprule
\textbf{Category} & \textbf{Technology} & \textbf{Outdoor ISRU LBMC TMS} & \textbf{Applicabale Materials} & \textbf{Max \% of Regolith} & \textbf{Applicable Size Range} & \textbf{Indoor/Outdoor} & \textbf{Energy Consumption} & \textbf{Current Advantages} & \textbf{Current Challenges} \\
\midrule
\multirow{1}{*}{Regolith Casting} & Casting & Low & Regolith & Very High & Medium to Large & Both & High & ISRU & High Energy Consumption \\
\bottomrule
\multirow{1}{*}{Regolith Sintering} & Furnace/MW Sintering & Low to Medium & Regolith & Very High & Small to Medium & Both & High & Meets Structural Needs & Needs Furnace/MW Chamber \\
\midrule
\multirow{3}{*}{Regolith Concrete} & Sulfur Concrete & Low & Regolith and Sulfur & Medium to High & Medium to Large & Indoor & Moderate & Stable under Freezing Condition & Sulfur Instability under Vacuum and Lunar Days \\
& Geopolymer Concrete & Low & Regolith and Geopolymer Binder & High & Medium to Large & Outdoor & Moderate & Stable During the Lunar Days after Curing & Needs Water \\
& Polymer Concrete & Low to Medium & Regolith + Polymer Binder & High & Medium to Large & Both & Moderate & Water-Free Process & Poor Mechanical Performance \\
\midrule
\multirow{1}{*}{Binder-Based} & & Low & Regolith + Polymer Binder & High to Very High & Small to Medium & Both & Moderate & Binder Recycling and Reuse Feasible & Requires High Compression Pressure \\
\midrule

\end{tabularx}
\end{table}
\end{landscape}

\subsection{Implications of the Review}
\normalsize
This review focuses on the significance of LBMC as humanity prepares for more extensive Lunar exploration, necessitating infrastructure development on the Moon. It emphasizes the potential of ISRU and different fabrication techniques, particularly AM, to reduce costs and time associated with transporting parts or materials from Earth. Sustainability, resource efficiency, and technological advancements are key themes, highlighting the need for evolving manufacturing techniques and adapting to Lunar conditions. The review considers energy efficiency and structural characteristics, as well as the role of automation and AI, offering insights into both opportunities and challenges. The review aims to guide future research, policy decisions, and investments for LBMC to achieve sustainable Lunar exploration.

\subsection{Future Directions for Research and Development}
While a more detailed assessment of current research gaps and potential future research directions to address them is discussed in the previous sections, the main remaining research gaps are summarized in Table \ref{FutureResearch} for different fabrication techniques. SLA demonstrates advantages within the context of indoor manufacturing of small precise parts. FFF is attractive due to its simplicity, process speed, low energy consumption, and recyclability. CCC demonstrates advantages in the context of large-scale construction. All in all, among various AM techniques, none takes overall precedence over others, as they all still present challenges and open research directions, as summarized in Table \ref{FutureResearch}, and breakthroughs in any of the listed directions could still significantly alter the path forward. A concerted effort along the various directions outlined will result in the highest chance of success for this emerging technological field.


\begin{table}[htp]
\centering
\caption{Recommended Future Directions for Research and Development Based on Identified Current Research Gaps.}  
\label{FutureResearch}
\scriptsize 
\setlength{\tabcolsep}{3pt} 
\begin{tabularx}{\linewidth}{l l X}
\toprule
\textbf{AM Techniques} & \textbf{Technology} & \textbf{Future Research Directions} \\
\midrule
\multirow{4}{*}{PBF} & L-PBF & Addressing optical absorption challenges in regolith, enhancing understanding of powder morphology and PSD effects.\\
& EB-PBF & Optimizing electron-beam interaction with regolith to enhance sintering, addressing powder emission challenges in microgravity. \\
& SS-PBF & Improve control over the process to enhance geometrical precision and overall print quality, addressing outgassing challenges at the melting point. \\
& MWS-PBF & Investigate methods to improve structural strength.\\
\midrule
\multirow{5}{*}{ME} & FFF & Increasing the weight percentage of regolith to enhance interlaminar bonding and thermal control. Investigating the effect of recycling on material characteristics. \\
& FGF & Explore actions for better process control and precision. \\
& EMR & Investigate approaches for better process control and precision, examining the effect of different compositions on process parameters. \\
& DIW & Address challenges in polymer recycling, outgassing concerns, and issues associated with binders, including those related to low melting and freezing points. \\
& CCC & Research on reducing water evaporation, for example, by adding additives or increasing construction speed. \\
\midrule
\multirow{1}{*}{SLA} &  & Address outgassing challenges, investigate the impact of PSD and composition on post-sintering characteristics, and tackle recycling issues for the binder.\\
\midrule
\multirow{1}{*}{BJ} &  & Address challenges related to liquid binders, including outgassing concerns, precise droplet deposition in microgravity, ink boiling and freezing, minimizing dependence on significant amounts of chemicals and water, exploring the influence of PSD and composition on post-fabrication characteristics, and tackling recycling issues associated with the binder. \\
\midrule
\multirow{3}{*}{DED} & L-DED & Develop strategies to mitigate optical absorption challenges in regolith. \\
& EB-DED & Improve interaction of EB with regolith for enhanced strength. \\
& WAAM & Research ways to expand WAAM for applications beyond metals, such as metal-regolith composites. \\
\midrule
\textbf{Non-AM Techniques} & & \textbf{Future Research Directions} \\
\midrule
\multirow{1}{*}{Regolith Casting} & & Conducting research on fabrication of glass fiber from regolith. Addressing losses of the susceptor-assisted heating.  \\
\midrule
\multirow{1}{*}{Regolith Sintering} & & Investigating energy-efficient and solar sintering methods.  Addressing temperature control and thermal runaway challenges\\
\midrule
\multirow{1}{*}{Regolith Concrete} & & Investigating methods to enhance strength in polymer concrete. Conducting research on reducing water consumption in geopolymer concrete and exploring the effects of water recycling on concrete properties. \\
\midrule
\multirow{1}{*}{Binder-Based} & & Researching the outgassing of binders. Exploring methods to reduce compression pressure requirements.\\
\midrule

\bottomrule

\end{tabularx}
\end{table}
\normalsize

\section{Declaration of Generative AI and AI-assisted technologies in the writing process}
While preparing this work, the authors used ChatGPT and Grammarly to enhance the manuscript's language and readability. After utilizing these tools, the authors reviewed and edited the content as necessary and take full responsibility for the publication's content.

\section{Acknowledgements}
We gratefully acknowledge the financial support provided by the Natural Sciences and Engineering Research Council of Canada (NSERC) for this project.


\bibliographystyle{unsrt}


\bibliography{Main}

\end{document}